\documentclass[prd,aps,a4paper,twocolumn,nofootinbib]{revtex4-2}

\usepackage[utf8]{inputenc}

\usepackage{graphicx,psfrag}
\usepackage{mathrsfs}
\usepackage{amsmath,amsfonts,amssymb}
\usepackage{multirow}
\usepackage{bm}
\usepackage{color}
\usepackage{mathrsfs}
\usepackage{hyperref}
\usepackage{diagbox}
\usepackage{amsthm,wasysym}
\usepackage{physics}
\usepackage{orcidlink}

\def\mbar{\overline{m}}

\definecolor{gray}{rgb}{0.5,0.5,0.5}
\definecolor{cyan}{rgb}{0,0.9,0.9}
\definecolor{orange}{rgb}{0.9,0.5,0}
\definecolor{magenta}{rgb}{1,0,1}
\definecolor{purple}{rgb}{0.8,0.4,0.8}
\definecolor{darkgreen}{rgb}{0,.8,0}
\definecolor{turquoise}{rgb}{0.25,0.88,0.82}

\newcommand{\Omi}{\omega_{\mathrm{i}}}

\begin{document}
\interfootnotelinepenalty=10000
\raggedbottom

\title{Post-Newtonian expansions of extreme mass ratio inspirals of spinning bodies into Schwarzschild black holes} 

\author{Viktor Skoup\'y \orcidlink{0000-0001-7475-5324}}
\email{viktor.skoupy@matfyz.cuni.cz}

\author{Vojt{\v e}ch Witzany \orcidlink{0000-0002-9209-5355}}
\email{vojtech.witzany@matfyz.cuni.cz}

\affiliation{Institute of Theoretical Physics, Faculty of Mathematics and Physics, Charles University, CZ-180 00 Prague, Czech Republic}

\begin{abstract}
Space-based gravitational-wave detectors such as LISA are expected to detect inspirals of stellar-mass compact objects into massive black holes. Modeling such inspirals requires fully relativistic computations to achieve sufficient accuracy at leading order. However, subleading corrections such as the effects of the spin of the inspiraling compact object may potentially be treated in weak-field expansions such as the post-Newtonian (PN) approach. 

In this work, we calculate the PN expansion of eccentric orbits of spinning bodies around Schwarzschild black holes. Then we use the Teukolsky equation to compute the energy and angular momentum fluxes from these orbits up to the 5PN order. Some of these PN orders are exact in eccentricity, while others are expanded up to the tenth power in eccentricity. Then we use the fluxes to construct a hybrid inspiral model, where the leading part of the fluxes is calculated numerically in the fully relativistic regime, while the linear part in the small spin is analytically approximated using the PN series. We calculate LISA-relevant adiabatic inspirals and respective waveforms with this model and a fully relativistic model. Through the calculation of mismatch between the waveforms from both models we conclude that the PN approximation of the linear-in-spin part of the fluxes is sufficient for lower eccentricities.

\end{abstract}

\maketitle

\section{Introduction and summary}
\label{sec:intro}
\subsection{Extreme mass ratio inspirals}
Forthcoming space-based gravitational wave (GW) detectors such as LISA, TianQin, or Taiji \cite{LISA,LISA:2024,TianQin,Taiji} will be able to detect signals from various sources, including extreme mass ratio inspirals (EMRIs) \cite{Babak:2017}. These systems consist of a stellar mass black hole or neutron star in orbit around a massive black hole with the mass ratio $\epsilon = \mu/M$ of the small (secondary) mass $\mu$ and large (primary) mass $M$ between $10^{-7}$ and $10^{-4}$. Because of gravitational radiation reaction, the orbit of the small body decays, and it inspirals into the primary body while radiating GWs to infinity. Because the secondary body completes many orbits in the strong gravitational field of the primary body, the detection of GWs from such systems will give a unique insight into the strong-field regime around massive black holes, which will also allow us to test general relativity to high precision \cite{Barack:2006pq,LISAFunWG:2022}. Furthermore, the study of EMRI populations will provide new insights in cosmology and astrophysics \cite{Babak:2017,LISAAstroWG:2022}.

To achieve the aforementioned goals, the parameters of the detected systems must be estimated with high precision. Because signals from EMRIs and other astrophysical sources will overlap, detection and parameter estimation will be done through matched filtering, which is based on the comparison of the received signal with many waveform templates. For this purpose, the waveforms must be generated for a wide range of parameters with the phase accurate to fractions of radian \cite{LISAWavWG:2023}. There are several methods for modeling binary systems, and the choice of the most suitable method depends on the parameters of the system, such as the mass ratio and compactness.

\subsection{Black hole perturbation theory}

In particular, for the modeling of EMRIs, black hole perturbation theory (BHPT) is often employed, where the spacetime is expanded in the mass ratio around a background spacetime of the primary \cite{Pound:2021}. Then, the system can be effectively described as a point particle moving in the background spacetime while inducing a perturbation of this spacetime. This perturbation acts on the particle with the so-called self-force, which can be expanded in the powers of the mass ratio. Because the mass ratio and, therefore, the perturbation is small, the inspiral timescale is much slower than the orbital timescale. Thus, to efficiently solve the problem, two-timescale expansion is often used, where the system is described using a set of orbital parameters $\mathcal{J}_i$, which evolve slowly ($\dot{\mathcal{J}}_i = \order{\epsilon}$) and a set of phases $\psi^i$ that evolve quickly ($\dot{\psi}^i = \order{1}$) \cite{Hinderer:2008}. The phases are directly related to the phase of the GW. When we consider an inspiral that sweeps through some finite range of frequencies such as a GW detector band, we can use the separation of scales to expand the phase elapsed during this process as
\begin{equation}
    \Phi_\text{GW} = \epsilon^{-1} \Phi_0(t) + \Phi_1(t) + \order{\epsilon} \, .
\end{equation}
The first term, which is called the adiabatic term, is of the order of $\epsilon^{-1}$ radians, while the second, postadiabatic term, is in the order of radians and cannot be neglected to achieve subradian accuracy. The adiabatic term consists of the contribution from the time-averaged dissipative (time-antisymmetric) part of the first-order in the mass ratio self-force, while the postadiabatic term consists of a number of contributions from different physical effects \cite{Pound:2021}. Namely, the postadiabatic term requires the inclusion of the oscillating part of the dissipative and conservative (time-symmetric) first-order self-force, time-averaged dissipative part of the second-order self-force, the force caused by the spin-curvature coupling of the secondary, and corrections to the dissipative self-force caused by the secondary spin. 

To find all the contributions up to the postadiabatic term, one in principle has to find the metric perturbation up to the second order in the mass ratio, regularize it near the particle, and calculate the self-force from the regular part. However, thanks to flux-balance laws, the averaged dissipative part of the self-force can often be obtained from the asymptotic GW fluxes to infinity and through the horizon of the primary black hole \cite{Mino:2003,Sago:2006,Akcay:2020,Warburton:2021}. The first-order flux-balance laws for nonspinning secondaries were obtained by \citet{Mino:2003} and \citet{Sago:2006}. For spinning secondaries, the flux-balance law was proven only for the evolution of energy and azimuthal angular momentum \cite{Akcay:2020}. Second-order flux-balance laws for the energy and azimuthal angular momentum of nonspinning secondaries on quasicircular orbits in Schwarzschild spacetime were derived by \citet{Miller:2020bft}; these derivations are expected to straightforwardly generalize to generic orbits. The currently open question is a concrete formulation of some sort of flux-balance law for the so-called Carter constant at second order in the mass ratio and under secondary spin corrections to the motion (see Ref. \cite{Grant:2024ivt} for some recent effort in this direction). A less obvious quantity that did not have a flux-balance law to date was the aligned component of the secondary spin $s_\parallel$; this question is resolved by us in Section \ref{sec:evolution}.

The first-order adiabatic fluxes must be calculated with high accuracy since the error will be enhanced by $\epsilon^{-1}$ compared to the postadiabatic terms. As a general rule, the error of the adiabatic term must be $\order{\epsilon}$ smaller than the error of the postadiabatic term. This opens up the possibility of using various approximations for the calculation of the postadiabatic effects. 

\subsection{Post-Newtonian expansions}
As mentioned above, there are other techniques to model binary systems with different mass ratios and separations. One such technique is the post-Newtonian (PN) theory \cite{Blanchet:2014}, which is valid for systems with large separations and low relative velocities. This method relies on expanding the Einstein equations in the inverse square of the speed of light in vacuum $1/c^2$, thus assuming that quantities such as the dimensionless speed squared $v^2/c^2$ or the dimensionless gravitational potential $GM/(r c^2)$ are small. Currently, the state-of-the-art results for the dissipative effects in comparable-mass spinning binaries on eccentric orbits are expansions of the energy and angular momentum fluxes to 3PN order \cite{Henry:2023} beyond the Newtonian quadrupole formula \cite{Peters:1963} and for nonspinning objects on circular orbits to 4.5PN \cite{Blanchet:2023bwj}.

The regime in which both PN theory and BHPT are valid offers the possibility of cross-validating the results of both theories. In particular, when the results of BHPT are analytically expanded in a PN parameter (see a review of older results in \cite{Mino:1997,Sasaki:2003}), direct comparisons can be made with pure PN computations truncated at the first order in the mass ratio. Furthermore, the BHPT computations can typically be expanded to higher PN orders than the existing PN computations at finite mass ratio. Finally, careful considerations of the symmetries of the mass ratio expansion of the PN series reveal that the BHPT results can often have a ``strategic'' importance for obtaining unknown pieces of the equations of motion of binaries at any mass ratio \cite{Bini:2019nra,Blumlein:2020pyo,Blumlein:2021txe,Bini:2024tft}.   

Such results can also be utilized to calibrate effective-one-body models, which is an approach to binary modeling that takes input from numerical relativity, PN theory, and BHPT \cite{Buonanno:1998gg,Albertini:2022rfe,vandeMeent:2023ols}. In particular, the dynamics of spinning test particles in black hole spacetimes proved to also be useful in the development of effective-one-body models (see, e.g., \cite{Barausse:2009xi,Nagar:2019,Albertini:2024rrs}).

The PN expansion of the BHPT results was first used by Poisson \cite{Poisson:1993}, where the energy fluxes to infinity from circular orbits in the Schwarzschild spacetime were expanded to 1.5PN orders beyond the Newtonian order. These results were then extended to higher PN orders, to fluxes through the horizon, and to the Kerr spacetime \cite{Tagoshi:1994,Poisson:1995,Tanaka:1996,Shibata:1995,Tagoshi:1996,Tagoshi:1997jy}. The latest results are infinity fluxes and horizon fluxes from circular orbits in the Schwarzschild spacetime to 22PN \cite{Fujita:2012} and in the Kerr spacetime to 11PN order \cite{Fujita:2015}.

The effects of the spin of the secondary body were first incorporated into the fluxes from circular orbits around a Kerr black hole by \citet{Tanaka:1996b} to 2.5PN order, and later by \citet{Nagar:2019} and \citet{Akcay:2020} for circular orbits in the Schwarzschild spacetime to the 5.5PN order.

The formalism was also extended to generic orbits of nonspinning bodies in Kerr spacetime, where one needs the evolution of three constants of motion, namely the energy, angular momentum, and the Carter constant \cite{Sago:2006,Ganz:2007,Sago:2015}. 
The latest results, i.e. 5PN fluxes with expansions in eccentricity to $e^{10}$ were used by \citet{Isoyama:2022} to generate generic adiabatic inspirals and waveforms.

Another direction in which this technique was utilized was to calculate PN expansions of energy and angular-momentum fluxes from highly eccentric orbits in the Schwarzschild spacetime. In Ref. \cite{Forseth:2016} the authors identified singular factors in the form $(1-e^2)^{-k}$ yielding convergent series in eccentricity after the factorization of such terms. In addition, they used highly accurate numerical calculations to find the coefficients of the series in eccentricity and the PN series to 7PN order. Later, an analytical form for the leading and subleading logarithmic terms was found by \citet{Munna:2019,Munna:2020a}. Finally, the energy and angular momentum fluxes to infinity and through the horizon were found up to the 19PN order \cite{Munna:2020,Munna:2020b,Munna:2023}. 

These expansions not only provided validation of the results of the PN theory, but were used by \citet{Burke:2023} in the calculation of adiabatic inspirals, where the authors also tested the possibility of using the waveforms derived from such inspirals for accurate parameter estimation. It was found that the 9PN adiabatic fluxes from eccentric orbits in Schwarzschild spacetime introduce bias on the system parameters and, therefore, cannot be used instead of the fully relativistic fluxes. However, in the same work, a hybrid model with fully relativistic adiabatic (first order in the mass ratio) fluxes and 3PN expansion of postadiabatic fluxes (second order in the mass ratio) was used, which was proven to be sufficient for accurate parameter estimation in some cases. 

The secondary spin corrections to the fluxes are of the order of the mass ratio and, consequently, contribute at postadiabatic order, the same as the PN-expanded pieces used in \citet{Burke:2023}. Therefore, it may be possible to approximate them using PN expansion and avoid computationally expensive numerical calculations of the fully relativistic fluxes such as was done in Refs. \cite{Harms:2016ctx,Lukes-Gerakopoulos:2017vkj,Akcay:2020,Skoupy:2021,Mathews:2022,Skoupy:2023}. 

\subsection{Summary of results}

\begin{itemize}
\item In this work, we PN-expanded the analytical expressions for eccentric, precessing trajectories of spinning bodies in Schwarzschild spacetime that were recently found by \citet{Witzany:2023}. The expanded trajectories and other relevant quantities can be found in the Supplemental Material \cite{SupMat}.

\item Then, we employed the Teukolsky equation to find the energy and angular momentum fluxes from these orbits as a closed-form series in the PN parameter and eccentricity. We linearized the fluxes in the secondary spin and found the linear-in-spin correction up to 5PN and at least tenth power in eccentricity. We were able to fully factorize and resum the fluxes as a finite series in eccentricity up to 2.5 PN with partial resummation results also at higher orders. We demonstrated that the resulting eccentricity series converges even up to $e\to 1$ in Figures \ref{fig:deltaFE_coefs} and \ref{fig:deltaFJz_coefs}. The resulting spin corrections to fluxes are in Eqs.~\eqref{eq:deltaFEn} and\eqref{eq:deltaFJz} and Appendix \ref{app:PNseries} as well as the Supplemental Material. 

\item We tested the convergence of the PN series by analytically integrating the phase evolution of quasicircular inspirals with the result in Eq.~\eqref{eq:Deltaphi}. Using this general result, we computed the phase contributions of LISA-band inspirals of $100 M_\odot$ spinning black holes into massive black holes of mass $10^6 M_\odot$ in Table \ref{tab:PNcirc}. This demonstrated that the 5PN expansion is not sufficiently accurate for the nonspinning part of the flux, but it is sufficient for the spin correction in the case of quasicircular inspirals. 

\item Hence, we then used these flux corrections in a hybrid inspiral model, where the nonspinning part was calculated numerically in a fully relativistic regime and the linear-in-spin part is expressed analytically as a PN series. To be able to do so, we also derived that the aligned component of the secondary spin $s_\parallel$ will stay conserved during generic EMRIs.

\item Using the hybrid model, we computed fiducial LISA-band eccentric inspirals using the same binary masses as for the quasicircular case. Additionally, we used the FEW package \cite{Katz:2021,Chua:2020stf,michael_l_katz_2020_4005001,Chua:2018woh} to generate relativistic waveforms corresponding to the inspirals. As a test of the formalism, we computed mismatches of the hybrid-model waveforms with waveforms corresponding to fully relativistic inspirals. The mismatches presented in Figure \ref{fig:mismatch} imply that the hybrid model should be adequate for the detection of the vast majority of LISA EMRIs. Additionally, it should not introduce significant biases for parameter estimates of less eccentric events.
\end{itemize}


\subsection{Organization of paper}

This paper is organized as follows. Section \ref{sec:spinning_trajectory} reviews the motion of spinning bodies in Schwarzschild spacetime and introduces PN and eccentricity expansions of these trajectories. This is followed by Section \ref{sec:evolution} where the self-torque acting on the spin vector is presented, which is then used to derive the adiabatic evolution of the parallel component of the secondary spin. Next, Section \ref{sec:GW_fluxes} examines the GW fluxes from orbits described in the previous Section. First, the Teukolsky formalism is presented, which is then used to calculate the PN expansions of the fluxes. Next, Section \ref{sec:inspirals} presents the hybrid model for the adiabatic inspirals that includes the PN expansion of the fluxes and the calculation of inspirals using this model. Finally, Section \ref{sec:Concl} provides a discussion of the importance of the results and outlooks.

\subsection{Notation} 

Geometrized units, where the gravitational constant and the speed of light in vacuum are set to unity ($G=c=1$), are used throughout this paper. Spacetime indices are denoted with Greek letters, while tetrad indices are denoted with Latin letters. The signature of the metric is $(-,+,+,+)$, while the Riemann tensor is defined as $a_{\nu;\kappa\lambda} - a_{\nu;\lambda\kappa} \equiv R^\mu{}_{\nu\kappa\lambda} a_\mu$, where the semicolon denotes the covariant derivative and $a_\mu$ is a covector. The sign of the Levi-Civita pseudotensor is defined as $\epsilon_{tr\theta\phi}/\sqrt{-g} = - \epsilon^{tr\theta\phi} \sqrt{-g} = 1$.

\section{PN expansion of eccentric equatorial trajectories of spinning bodies}
\label{sec:spinning_trajectory}

\subsection{Spinning-particle trajectory}
Let us briefly summarize the properties of the closed analytical solution of the bound motion of spinning particles near Schwarzschild black holes as presented in Ref. \cite{Witzany:2023}. We consider the motion in Schwarzschild spacetime given as
\begin{align}
    \dd s^2 = -f(r) \dd t^2 + \frac{1}{f(r)} \dd r^2 + r^2( \dd \theta^2 + \sin^2\theta \dd \phi^2 )\,,
\end{align}
where $f(r) = 1 - 2M/r$. The motion of the spinning particle is described by Mathisson-Papapetrou-Dixon equations under the Tulczyjew-Dixon or covariant spin supplementary condition $s^{\mu\nu}P_\nu = 0$, where $P_\mu$ is the particle momentum and $s^{\mu\nu}$ is the spin tensor per unit particle mass. The solution is valid up to $\mathcal{O}(s)$ corrections to the orbital motion, and to leading order in the spin sector. In this truncation one has $P_\mu = \mu u_\mu + \mathcal{O}(s^2)$, where $\mu$ is the particle mass and $u_\mu$ is the four velocity. One then equivalently parametrizes the spin by the spin vector and the spin tensor
\begin{align}
    & s^\mu = -\frac{1}{2} \epsilon^{\mu\nu\kappa\lambda} s_{\nu\kappa} u_\lambda \,, \label{eq:sVectorFromTensor}\\
    & s^{\mu\nu} = \epsilon^{\mu\nu \kappa \lambda}  u_\kappa s_\lambda \,, \label{eq:sTensorFromVector}
\end{align}
where $\epsilon^{\mu\nu\kappa\lambda}$ is the Levi-Civita pseudotensor.

Note that the definition of $s^\lambda$ depends on the orientation of the basis. This is further complicated by the fact that raising or lowering indices formally changes the orientation. Here we fix the convention by
\begin{align}
\epsilon_{tr\theta\phi} = -(-g) \epsilon^{tr\theta\phi} = \sqrt{-g}\,,
\end{align}
which yields a right-handed basis for upper-index quantities under the assumption of a conventional transformation from $r,\theta,\phi$ to Cartesian coordinates. However, this also means that our formulas have a relative minus sign in front of any spin correction as compared to Ref. \cite{Witzany:2023}.

The 3 rotational symmetries of the Schwarzschild spacetime around the $x,y,z$ axes generate a conserved total angular momentum vector of the generically inclined spinning particle. Upon rotation of the coordinate equator $\theta = \pi/2$ into the plane perpendicular to this vector, the generic motion becomes near equatorial in the resulting frame, $\theta = \pi/2 + \delta \theta + \mathcal{O}(s^2)$. As a result, the Mathisson-Papapetrou-Dixon equations fully separate and the solutions are parametrized by the three constants of motion
\begin{align}
    & E = - u_{\mu} \xi_{(t)}^\mu + \frac{1}{2} \xi^{(t)}_{\mu;\nu} s^{\mu\nu} \,,\\ 
    & J_z = u_{\mu} \xi_{(\phi)}^\mu - \frac{1}{2} \xi^{(\phi)}_{\mu;\nu} s^{\mu\nu} \,, 
    \\ 
    & s_{\parallel} = \frac{s^\mu l_\mu}{\sqrt{l_\alpha l^\alpha}} = \frac{Y_{\mu\nu} s^\mu u^\nu}{2\sqrt{K_{\alpha\beta} u^\alpha u^\beta}} \,, \label{eq:spara}
    \\
    & l^\mu \frac{\partial}{\partial x^\mu} \equiv \frac{r \dot{\theta}}{\sin \theta} \frac{\partial}{\partial \phi} - r \sin\theta \dot{\phi}\frac{\partial}{\partial \theta}\,,
\end{align}
where $E$ has the meaning of total orbital and spin-orbital energy per unit mass, $J_z$ the orbital and spin-orbital angular momentum, and $s_\parallel$ is the component of spin aligned with the orbital angular momentum. Furthermore, $Y_{\mu\nu} = -Y_{\nu\mu}, Y_{\mu\nu;\kappa} = -Y_{\mu\kappa;\nu}$ is the Killing-Yano tensor of the Schwarzschild spacetime, which means that $s_\parallel$ is related to the more general R{\"u}diger constants in Kerr spacetime and the separation constant for spinning particles in Kerr found by separation of the Hamilton-Jacobi equation \cite{Rudiger:1981c,Witzany:2019}. It should also be noted that in the aligned frame the magnitude of the total angular momentum is by construction equal to the single component $J_z$.

The solution is parametrized by Carter-Mino time $\dd \lambda = \dd\tau/r^2$, where $\tau$ is proper time. The radial solution is then expressed in the form
\begin{align}
    & r(\lambda) = \frac{r_3(r_1 - r_2) \mathrm{sn}^2\left(\frac{K(k)}{\pi}  q^r, k\right) - r_2 (r_1 - r_3)}{(r_1 - r_2)\mathrm{sn}^2\left(\frac{K(k)}{\pi} q^r, k\right) - (r_1 - r_3)}\,, \label{eq:rlambda}\\ 
    & k^2 = \frac{(r_1 - r_2)r_3}{(r_1 - r_3)r_2} \,, \label{eq:k} \\
    & q^r \equiv \Upsilon^r \lambda + q^r_0\,,
\end{align}
where $K(k)$ is the complete elliptic integral of the first kind, $\mathrm{sn}()$ is the Jacobi sn function, and $q^r_0$ is an integration constant determined by initial conditions. $\Upsilon^r$ is given in Ref. \cite{Witzany:2023} and represents the fundamental frequency of motion with respect to Mino time. The radii $r_1>r_2>r_3$ are then the nonzero roots of the polynomial $R(r)$ appearing in the radial equation of motion
\begin{align}
    & \frac{\dd r}{\dd \lambda} = \pm \sqrt{R(r)} \,, \\
    \begin{split}
    & R(r) \equiv r\Big[r^3(E^2 - 1) - r J_z (J_z - 2 s_\parallel E) 
    \\ 
    & \phantom{R(r) =} + 2M (r^2 + J_z(J_z - 3 s_\parallel E))\Big]\,
    \\
    & \phantom{R(r)} = (1-E^2)r(r_1-r)(r-r_2)(r-r_3)\,.
    \end{split} \label{eq:Rpoly}
\end{align}
The roots $r_1,r_2$ are the physical turning points of the bound motion and are conventionally parametrized by the orbital parameters dimensionless semilatus rectum $p$ and eccentricity $e$ defined through the relation
\begin{align}
    r_1 = \frac{M p}{1-e} \; , \qquad r_2 = \frac{M p}{1+e}\,.
\end{align}
One can then express $E(p,e,s_\parallel),J_z(p,e,s_\parallel),r_3(p,e,s_\parallel)$ in closed form by examining Eq.~\eqref{eq:Rpoly}.

The $t,\phi$ degrees of freedom are then given as
\begin{align}
    & t(\lambda) = q^t + \Delta t(q^r)\,,   \; \phi(\lambda) = q^\phi, \label{eq:vphl}\\
    & q^t \equiv \Upsilon^t \lambda + q^t_0\,, \;
    q^\phi \equiv \Upsilon^\phi \lambda + q^\phi_0,\,\\
    & \Delta t(q^r) = \tilde{T}_{r}\left(\mathrm{am}\left(\frac{q^r}{\pi} K(k),k\right)\right) - \frac{\tilde{T}_{r}\left(\pi\right)}{2 \pi} q^r \,,
\end{align}
where $\mathrm{am}()$ is the Jacobi amplitude, $\tilde{T}_r$ is given in Eq.~(49) of Ref. \cite{Witzany:2023}, $\Upsilon^t,\Upsilon^\phi$ are the $t,\phi$ Mino frequencies, and $q^t_0,q^\phi_0$ are integration constants. It is interesting to note that unlike in Kerr, in Schwarzschild spacetime the Carter-Mino time $\lambda$ is simply proportional to $\phi$, which means that the $\phi$ counterpart of $\Delta t(q^r)$ vanishes in Eq.~\eqref{eq:vphl}. 

Finally, the spin degree of freedom and $\delta \theta$ depend on a precession angle $\psi$ with the evolution
\begin{align}
    & \psi(\lambda) = q^\psi + \tilde{\Psi}_{r}\left(\mathrm{am}\left(\frac{q^r}{\pi} K(k),k\right)\right) - \frac{\tilde{\Psi}_{r}\left(\pi\right)}{2 \pi} q^r,\\
    & q^\psi \equiv \Upsilon^\psi \lambda + q^\psi_0,\,
\end{align}
where $\Upsilon^\psi, q^\psi_0$ again have analogous meanings as above and $\tilde{\Psi}_{r}$ is a known function.

The deviation from the equatorial plane is then given as
\begin{align}
    \delta \theta = -\frac{\sqrt{(s^2 - s_\parallel^2)(J_z^2 + r^2)} \sin \psi}{J_z r}\,,
\end{align}
and the spin vector can be expressed as
\begin{subequations}
\label{eq:svec}
\begin{align}
    & s^t = -\sqrt{\frac{s^2 - s_\parallel^2}{f(J_z^2 + r^2)}}\left(\frac{E J_z \cos \psi}{\sqrt{f}} + \dot{r} r \sin \psi\right) \,, 
    \\
    & s^r = -\sqrt{s^2 - s_\parallel^2} \left( \frac{J_z \dot{r} \cos \psi}{r} + \frac{E r \sin \psi}{\sqrt{ J_z^2 + r^2}} \right)\,,
    \\
    & s^\phi = -\frac{\sqrt{(s^2 - s_\parallel^2)(J_z^2 + r^2)} \cos \psi}{r^2} \,,
    \\
    & s^\theta = \frac{s_\parallel}{r}\,,
\end{align}
\end{subequations}
where $\dot{r}$ is the proper-time derivative of $r$ expressed as $r^2\sqrt{R(r)}$. Note that even though the spin is parametrized by $s_\parallel$, the orientation of the spin vector is generic, and we are thus dealing with absolutely generic bound orbits of spinning test particles in Schwarzschild spacetime in this paper.

\subsection{PN expansion of the trajectory}\label{sec:PNtrajectory}

The constants of motion, orbital frequencies, and trajectory $(t,r,\phi)$ as a function of the phase $q^r$ can be expanded in a formal PN parameter. In this work we use the parameter
\begin{align}
    v = \sqrt{\frac{1}{p}} \; .
\end{align}
Other choices include, e.g., the gauge independent parameter $x=(M \Omega_\phi)^{2/3}$; however, the parameter $v$ is convenient when the orbit is parametrized with $p$ and $e$ and one can reexpress the final result in different PN parameters. Every order in $v$ corresponds to one half of the PN order, that is, the expansion to 7 orders in $v$ next to the leading order (NLO) corresponds to $3.5$PN orders NLO.

Since the expansions of the geodesic quantities were calculated before and are available in the literature \cite{Sago:2015}, here we present only the PN expansion of the linear-in-spin correction of any given quantity. We write the expansion as $E = E_{\rm (g)} + s_\parallel \delta E/M$ and $J_z = J_{z \rm (g)} + s_\parallel \delta J_z/M$, where $E_{\rm (g)}, J_{z \rm (g)}$ are the geodesic expressions at fixed orbital parameters $e,v$. Then it is straightforward to expand the linear-in-spin part of energy $\delta E$ and angular momentum $\delta J_z$ from Eqs.~(32)--(33) in Ref.~\cite{Witzany:2023}. The results read as
\begin{align}
    & \delta E = -\frac{(1-e^2)^2 v^5}{2} \sum_{k=0}^{\infty} \binom{-3/2}{k} (-(3+e^2) v^2 )^k \,,
    \\
\begin{split}
    & \frac{\delta J_z}{M} = 
        1 - 2v^2 - \frac{27 + 34e^2 + 3e^4}{8} v^4 
        \\ 
        & \phantom{\delta J_z =}
        - \frac{81 + 131e^2 + 39e^4 + 5e^6}{8} v^6 + \order{v^8} \, .
\end{split}
\end{align}
$\delta E$ was expanded to the order $v^{14}$ which corresponds to 9 orders NLO, while $\delta J_z$ was expanded to $v^{11}$ (here we show only the expansion to $v^6$ for simplicity; the full expressions can be found in the Supplemental Material \cite{SupMat}). 

Since the parameter of the elliptic integrals $K$, $E$, and $\Pi$ in Eqs.~(S22)--(S24) in Ref.~\cite{Witzany:2023} is $k^2 \sim \order{e v^2}$, we can expand the expression in $k^2$. We then write the linear-in-spin parts of the orbital frequencies in Carter-Mino time as $\Upsilon = \Upsilon_{\rm (g)} + s_\parallel \delta \Upsilon/M$ and obtain
\begin{align}
    \begin{split}
        \frac{\delta \Upsilon^t}{M} &= \frac{e^2}{(1-e^2)^{3/2}} v^{-1} + \qty( \frac{9 + 8e^2}{2 (1-e^2)^{3/2}} - 6 ) v 
        \\ 
         & \phantom{=} - \qty( 48 + \frac{9e^2}{2} - \frac{330 + 255e^2 - 88 e^4 + 9 e^6}{(1-e^2)^{3/2}} ) v^3 
        \\ 
        \phantom{\delta \Upsilon^t} & \phantom{=} + \order{v^5} \,,
    \end{split}
    \\
    \begin{split}
        \delta \Upsilon^r &= \frac{1}{2} \left(3-e^2\right) v^2 + \frac{1}{4} \left(-3 e^4+4 e^2+33\right) v^4 
        \\ 
         & \phantom{=} +\frac{1}{16} \left(-15 e^6-23 e^4+357 e^2+693\right) v^6 
         \\ 
         & \phantom{=} + \order{v^8} \,,
    \end{split}
     \\
     \begin{split}
        \delta \Upsilon^\phi &= (e^2+3) \bigg(-\frac{1}{2} v^2 - \frac{1}{4} \left(3 e^2+5\right) v^4 
         \\ 
         &\phantom{=} -\frac{1}{16} \left(15 e^4+50 e^2+63\right) v^6 + \order{v^8} \bigg) \, .
     \end{split}
\end{align}

From the Mino time frequencies, we can calculate the coordinate time frequencies
\begin{align}
    \Omega_i = \frac{\Upsilon^i}{\Upsilon^t} \,,
\end{align}
and their linear-in-spin parts 
\begin{align}
    \delta \Omega_i = \frac{\delta\Upsilon^i \Upsilon^t_{\rm (g)} - \Upsilon^i_{\rm (g)} \delta \Upsilon^t}{(\Upsilon^t_{\rm (g)})^2} \, ,
\end{align}
where $i = r,\phi$. A nice special formula is that $\delta \Omega_\phi = -3 v^6/2$ to all orders in $v$ for $e=0$. 

We expand the expressions to 9 orders in $v$ beyond the leading order while keeping the eccentricity dependence exact. The results are then expanded to $e^{10}$ for later calculations. Similarly to $\delta J_z$, here we show only the expansion to 4 orders NLO. The full expressions can be found in the Supplemental Material \cite{SupMat}.

By expanding the Jacobi elliptic function ${\rm sn}(u,k)$ in $k^2$ in Eq.~\eqref{eq:rlambda}, we find the radial coordinate parametrized with $q^r$ as a series in $v$ and $e$ and extract the linear-in-spin part $\delta r(q^r)$. Again, due to the length of the expression, it is not included here but can be found in the Supplemental Material \cite{SupMat}.

Next, we focus on $\Delta t(q^r)$, which is the oscillating part of $t = (\Upsilon^t/\Upsilon^r) q^r + \Delta t(q^r)$. Note that the oscillating part of $\phi = (\Upsilon^\phi/\Upsilon^r) q^r$ is zero. Since the expression for $t(q^r)$ is too long, its PN expansion is computationally expensive. Instead, we expand the equation
\begin{align}
    \dv{t}{q^r} = \dv{t}{\lambda} \dv{\lambda}{q^r} = \dv{t}{\lambda} \Upsilon_r^{-1}
\end{align}
in $v$ and $e$, where
\begin{align}
    \dv{t}{\lambda} = \frac{r^2 E}{f} - s_\parallel \frac{J_z f' r^2}{2f} 
\end{align}
(cf. Eq.~(29) in Ref.~\cite{Witzany:2023}). In this way, we obtain a Fourier series of $\cos(n q^r)$ that is trivial to integrate to obtain $t(q^r)$. Then we extract the linear-in-spin part $\delta \Delta t(q^r)$, which is available in the Supplemental Material.

\section{Adiabatic evolution of the constants of motion}
\label{sec:evolution}

The metric perturbations sourced by the spinning secondary will lead to a self-torque and a self-force, which will drive its motion away from the motion of the spinning test body in the Schwarzschild metric. In this section, we derive the equations governing the leading-order secular evolution of the spinning-secondary orbit under this perturbation.

The evolution of the secondary under self-force and self-torque can be cast in the form of MPD equations in the effective regularized metric $\hat{g}_{\mu\nu} = g_{\mu\nu} + h_{\mu\nu}$ \cite{Detweiler:2002mi,Harte:2011ku,Akcay:2020,Mathews:2022} (we drop the conventional ``R" index on the regularized metric perturbation $h_{\mu\nu}$ here for notational simplicity). As a result, under the assumption that the spin tensor $s^{\mu\nu}$ is unchanged in the perturbed metric, using Eq.~\eqref{eq:sVectorFromTensor} we obtain different definitions of the spin vector $s^\mu$ with respect to the Schwarzschild metric, and $\hat{s}^\mu$ with respect to the effective metric. The two definitions are related as follows
\begin{align}
    & s^\mu = \hat{s}^\mu - \epsilon \qty( \frac{1}{2} h_\alpha{}^\alpha \hat{s}^\mu + \frac{1}{2} h_{\alpha\beta} u^\alpha u^\beta \hat{s}^\mu - h^\mu{}_\nu \hat{s}^\nu )\,. \label{eq:sshat}
\end{align}
Because the spin tensor $\hat{s}^\mu$ is parallel transported in the effective metric, the spin vector on the Schwarzschild metric $s^\mu$ then experiences the self-torque 
\begin{align}
    & \frac{{\rm D} s^\mu}{\dd \tau} = -\frac{1}{2} h_{\alpha\beta;\rho} N^{\mu\alpha\beta\rho} \,, \label{eq:selftorque}
    \\
    \begin{split}
    & N^{\mu\alpha\beta\rho} = 2 g^{\mu\alpha} u^{[\beta} s^{\rho]} - g^{\mu\rho}  u^\alpha s^\beta 
    \\
    & \phantom{N^{\mu\alpha\beta\rho} =} + g^{\alpha\beta} u^\rho s^\mu + u^\alpha u^\beta u^\rho s^\mu \,.
    \end{split}
\end{align}
The Tulczyjew-Dixon SSC in the effective metric $u^\mu \hat{s}_\mu = 0$ is conserved up to $\mathcal{O}(s^2)$ due to the general properties of the MPD equations in any metric. From Eq.~\eqref{eq:sshat} it can be seen that $\hat{s}^\mu$ is always $\mathcal{O}(\epsilon s)$ close to $s^\mu$ without secularly growing terms. As a result, $u^\mu s_\mu = \mathcal{O}(\epsilon s, s^2)$ will also hold during evolution. Similarly, the spin magnitude with respect to the effective metric $\hat{s}^\mu \hat{s}_\mu$ is conserved up to higher-order terms. The background spin magnitude will then also be conserved up to $\mathcal{O}(s \epsilon,s^2)$ terms at all times. 

The energy and angular momentum of the spinning secondary are generated by the Killing symmetries of the background, so it is not surprising that their evolution averaged over the orbital timescale balances the corresponding gravitational-wave fluxes \cite{Akcay:2020,Mathews:2022}
\begin{align}
    & \left\langle\frac{\mathrm{d} E}{\mathrm{d} \tau} \right\rangle = - \mathcal{F}^E \,, \;\, \left\langle\frac{\mathrm{d} J_z}{\mathrm{d} \tau} \right\rangle = - \mathcal{F}^{J_z}\,. \label{eq:EJdecay}
\end{align}
In other words, to obtain the leading secular order, we do not need to go through the computationally demanding procedure of sourcing the full metric perturbation by the motion of the pole-dipole particle and then regularizing it to obtain the local equations of motion. Instead, we can use the pole-dipole stress energy tensor of the spinning particle evolving along the orbit described in Section \ref{sec:spinning_trajectory} as a source of \textit{curvature} perturbations in the Teukolsky equation. From these we can then compute the fluxes by reconstructing the metric at infinity (and, starting from a certain PN order, also at the horizon), as we will do in Section \ref{sec:GW_fluxes}. This should allow us to drive the evolution of the large mass ratio binary and obtain the full spin contribution to the 1PA phase.

However, the generic bound motion of the spinning particle has an additional degree of freedom in the form of the direction of the spin vector $s^\mu$. Specifically, it is conceivable that the self-torque drives the spin vector toward a more aligned, counteraligned, or orthogonal configuration with respect to the angular momentum of the orbit. In other words, we need to derive the evolution of the constant $s_\parallel$. 

Using the definition of $s_\parallel$ from Eq.~\eqref{eq:spara} we express the time derivative as
\begin{align}
    \dot{s}_\parallel = \frac{\dot{l}_\nu s^\nu}{\sqrt{l_\alpha l^\alpha}} + \frac{l_\nu \dot{s}^\nu}{\sqrt{l_\alpha l^\alpha}} - \frac{l_\nu s^\nu}{(l_\alpha l^\alpha)^{3/2}} l_\sigma \dot{l}^\sigma \,.
\end{align}
We now average the relation above over orbital timescales to obtain the secular contribution to the evolution. We also discard terms of order $\mathcal{O}(s^2,\epsilon^2)$ but keep terms of order $\mathcal{O}(s,\epsilon s, \epsilon)$ as is consistent with the order of the scheme. 

After averaging, only the parallel part of spin remains since it can be seen from Eq.~\eqref{eq:svec} that all the other components of $s^\mu$ are fully oscillating on the orbital timescale. Additionally, the third term can be written using $s_\parallel$
\begin{align}
    \langle \dot{s}_{\parallel} \rangle = \left\langle \frac{\dot{l}_\nu s_{\parallel}^\nu}{\sqrt{l_\alpha l^\alpha}} + \frac{l_\nu \dot{s}^\nu}{\sqrt{l_\alpha l^\alpha}} - \frac{s_\parallel}{l_\alpha l^\alpha} l_\sigma \dot{l}^\sigma \right\rangle \,.
\end{align}
The parallel part of the angular-momentum vector is expressed as $s_\parallel^\mu = s_\parallel l^\mu/\sqrt{l_\alpha l^\alpha}$, which yields
\begin{align}
    \langle \dot{s}_{\parallel} \rangle = \left\langle \frac{s_\parallel \dot{l}_\nu l^\nu}{l_\alpha l^\alpha} + \frac{l_\nu \dot{s}^\nu}{\sqrt{l_\alpha l^\alpha}} - \frac{s_\parallel}{l_\alpha l^\alpha} l_\sigma \dot{l}^\sigma \right\rangle\,.
\end{align}
The first and third terms above subtract. Finally, we substitute the self-torque $\dot{s}^\mu$ from Eq.~\eqref{eq:selftorque} to obtain
\begin{align}
    \begin{split}
        &\langle \dot{s}_{\parallel} \rangle = \left\langle \frac{l_\nu \dot{s}^\nu}{\sqrt{l_\alpha l^\alpha}} \right\rangle 
        \\ & = - \frac{1}{2} \left\langle \frac{- l^\mu \frac{{\rm D}}{\dd \tau} b_\mu + l^\mu h_{\mu\beta;\rho} u^\beta s^\rho - l^\rho h_{\alpha\beta;\rho} u^\alpha s^\beta}{\sqrt{l_\alpha l^\alpha}} \right\rangle,
    \end{split}
    \\
    & b^\mu \equiv \qty( g^{\mu\alpha} s^\beta - s^\mu \qty(g^{\alpha\beta} + u^\alpha u^\beta) ) h_{\alpha\beta} \,.
\end{align}
Because $\dot{l}^\mu = \order{\epsilon} + \order{s}$ and $b^\mu = \order{\epsilon s}$, the first term together with the denominator can be written as a total derivative that does not contribute to the average. Additionally, since $s_\parallel^\mu$ and $l^\mu$ are colinear, the second and third terms cancel under the average and we obtain $\langle \dot{s}_\parallel \rangle =0+ \mathcal{O}(\epsilon^2,s^2)$. 

In conclusion, the leading-order adiabatic evolution of the spinning secondary orbit will be only due to the decay of $E$ and $J_z$ as given by Eq.~\eqref{eq:EJdecay}, and $s$ and $s_\parallel$ can be treated simply as constants for the purposes of 1PA inspirals. 

This derivation holds for generic orbits in the Kerr spacetime and extends the same result for circular orbits in Schwarzschild spacetime in \cite{Mathews:2022}. This is because $l_\mu \equiv Y_{\mu\nu} u^\nu$ is parallel transported also along Kerr geodesics and thus all the derivation steps above apply without any change also for the motion near spinning primary black holes.

\section{Gravitational-wave fluxes}
\label{sec:GW_fluxes}

\subsection{Teukolsky formalism}

For the calculation of the PN expansion of GW fluxes in the framework of black hole perturbation theory, we use a similar approach to the one we used in Refs.~\cite{Skoupy:2021,Skoupy:2022,Skoupy:2023} where we solved the Teukolsky equation in the frequency domain. Because the radial motion is periodic, the strain at infinity $h = h_+ - i h_{\cross}$ can be written as a sum over $l$, $m$ multipoles and harmonic modes $n$ and $j$ as
\begin{align} \label{eq:strain}
    h = \frac{2}{r} \sum_{lmnj} \frac{C^+_{lmnj}}{\omega_{mnj}^2} {}_{-2}Y_{lm}(\theta) e^{-i\omega_{mnj}(t-r^\ast) + im\phi} \; ,
\end{align}
where we sum over $2 \leq l < +\infty$, $-l\leq m \leq l$, $-\infty < n < +\infty$, and $-1 < j < 1$, $C^+_{lmnj}$ are the Teukolsky amplitudes at infinity, $\omega_{mnj} = m \Omega_\phi + n \Omega_r + j \Omega_\psi$ is the frequency of the given mode, ${}_{-2}Y_{lm}(\theta)$ is the spin-weighted spherical harmonic, $(t, r, \theta, \phi)$ are the coordinates of the observer, and $r^\ast = r + 2M \log(r/(2M)-1)$ is the tortoise coordinate.

The orbit-averaged energy and angular momentum fluxes to infinity can be expressed as sums over the $l$, $m$, $n$, $j$ modes in the form 
\begin{subequations}\label{eq:fluxes}
\begin{align}
    \mathcal{F}^{E} &= \sum_{lmnj} \frac{\abs{C^+_{lmnj}}^2}{4\pi \omega_{mnj}^2} \; , \\
    \mathcal{F}^{J_z} &= \sum_{lmnj} \frac{m\abs{C^+_{lmnj}}^2}{4\pi \omega_{mnj}^3} \; .
\end{align}
\end{subequations}
Because the amplitudes for $j=\pm 1$ are proportional to $s_\perp = \sqrt{s^2-s_\parallel^2}$ and for $j=0$ are independent of $s_\perp$, to linear order in spin the fluxes are independent of $s_\perp$ and we can sum only over $l$, $m$, and $n$ with $j=0$ \cite{Tanaka:1996b,Skoupy:2023}. Therefore, we will write $C^+_{lmn} \equiv C^+_{lmn0}$ and $\omega_{mn} \equiv \omega_{mn0}$. Furthermore, as discussed later in this Section, the horizon fluxes are of higher PN order, and we do not consider them here.

The asymptotic amplitudes can be found from the integral over the radial phase $q^r$
\begin{align}\label{eq:Cpm_lmn}
    C^{+}_{lmn} = \frac{1}{W_{lmn} \Upsilon^t} \int_0^{2\pi} I^{+}_{lmn}(q^r) e^{i \psi_{mn}(q^r)} \dd q^r \, ,
\end{align}
where $I^{+}_{lmn}(q^r)$ is a certain moment of the spinning particle source
\begin{align}
    I^+_{lmn} &= r^2 \sum_{ab} \qty( B^0_{ab} F^{ab}_{lmn} + B^r_{ab} \pdv{F^{ab}_{lmn}}{r} ) \, , \\
    \psi_{mn}(q^r) &= \omega_{mn} \Delta t(q^r) - m \Delta \phi(q^r) + n q^r \,,
\end{align}
with the sum over Kinnersley tetrad legs $\lambda^\mu_a$, $ab = nn, n\mbar, \mbar\mbar$. Note that we have rearranged the expression for $I^+_{lmn}$ in Eq.~(52) from \cite{Skoupy:2023} and introduced quantities
\begin{align}
    F^{ab}_{lmn} &= \sum_{i=0}^{I_{ab}} (-1)^i f^{(i)}_{ab} \dv[i]{R^-_{lmn}}{r} \, , \\
    B^0_{ab} &= A^{\rm m}_{ab} + A^{\rm d}_{ab} + i \qty( \omega_{mn} B^t_{ab} - m B^\phi_{ab} ) \, .
\end{align}
The functions $F^{ab}_{lmn}$ depend on the spin-weighted spherical harmonics $_{2}Y_{lm}$ through the functions $f^{(i)}_{ab}$ defined in Eqs.~(B4) in \cite{Skoupy:2023}, and on the solution of homogeneous radial Teukolsky equation $R^-_{lmn} \equiv R^-_{lm\omega_{mn}}$ satisfying a purely outgoing boundary condition at the horizon (sometimes called the ``in'' solution). The quantities $A^{\rm m,d}_{ab}$ and $B^{\mu}_{ab}$ are calculated from the dynamical stress energy tensor of the spinning particle to the pole-dipole order as
\begin{align*}
    A^{\rm m}_{ab} &= \mu u_{a} u_{b} \; ,
\end{align*}
\begin{align*}
    A^{\rm d}_{ab} &= \mu s^{c d} u_{(b} \gamma_{a)dc} + \mu s^{c}{}_{(a} \gamma_{b)dc} u^d \; , \\
    B^\rho_{ab} &= \mu s^{\rho}{}_{(a} v_{b)}
\end{align*}
with the spin coefficients
\begin{equation*}
    \gamma_{adc} = \lambda_{a\mu;\rho} \lambda^\mu_d \lambda^\rho_c \; .
\end{equation*}
In other words, the source receives contributions both from the dipole term in the particle stress-energy computed from the spin tensor, and the modified spinning-particle trajectory on the level of the monopole term.

Similarly to Ref.~\cite{Skoupy:2022}, we expand the expression for $C^+_{lmn}$ in the secondary spin $s_\parallel$. However, here the amplitudes and fluxes are expanded with fixed orbital parameters $p$ and $e$ as opposed to fixed orbital frequencies $\Omega_{r,\phi}$. The linear-in-spin part of the amplitude can be written as
\begin{widetext}
\begin{align}\label{eq:deltaC_lmn}
    \delta C^+_{lmn} = - \qty( \frac{\delta \Upsilon^t}{\Upsilon^t_{\rm (g)}} + \frac{\partial_\omega W_{lmn} \delta \omega_{mn}}{W^{\rm (g)}_{lmn}} ) C^{{\rm (g)}+}_{lmn} + \frac{1}{W^{\rm (g)}_{lmn} \Upsilon^t_{\rm (g)}} \int_0^{2\pi} \qty(\delta I^+_{lmn} + i I^{{\rm (g)}+}_{lmn} \delta \psi_{mn} ) e^{i \psi_{mn}^{\rm (g)}(q^r)} \dd q^r \, ,
\end{align}
where
\begin{align}
    \delta I^+_{lmn} &= r_{\rm (g)}^2 \qty( \qty( \frac{2 \delta r}{r_{\rm (g)}} A^{\rm m}_{{\rm (g)}ab} + \delta B^{0}_{ab} ) F^{{\rm (g)}ab}_{lmn} + \qty(B^r_{ab} + A^{\rm m}_{{\rm (g)}ab} \delta r) \pdv{F^{{\rm (g)}ab}_{lmn}}{r} + A^{\rm (m)}_{{\rm (g)}ab} \pdv{F^{{\rm (g)}ab}_{lmn}}{\omega_{mn}} \delta\omega_{mn} ) \, , \\
    \delta \psi_{mn} &= \delta\omega_{mn} \Delta t_{\rm (g)} + \omega^{\rm (g)}_{mn} \delta\Delta t
\end{align}
\end{widetext}
with
\begin{align}
    \delta B^0_{ab} &= \delta A^{\rm m}_{ab} + A^{\rm d}_{ab} + i \qty(\omega B^t_{ab} - m B^\phi_{ab}) \, , \\
    \delta A^{\rm m}_{ab} &= u_\mu u_\nu \partial_r( \lambda^\mu_a \lambda^\nu_b ) \delta r + 2 u_\mu \delta u_\nu \lambda^{(\mu}_a \lambda^{\nu)}_b \, .
\end{align}

Then, the fluxes can be separated into the geodesic and spin part as $\mathcal{F}^{E,J_z} = \mathcal{F}^{E,J_z}_{\rm (g)} + s_\parallel \delta \mathcal{F}^{E,J_z}/M$ to obtain
\begin{subequations}\label{eq:linear_fluxes}
\begin{align}
    \delta \mathcal{F}^E &= \sum_{lmn} \frac{\omega_{mn}\Re{C^{{\rm (g)}+}_{lmn} \overline{\delta C^+_{lmn}}} - \abs{C^{{\rm (g)}+}_{lmn}}^2 \delta\omega_{mn}}{2\pi \omega_{mn}^3} \, , \\
    \delta \mathcal{F}^{J_z} &= \sum_{lmn} m \frac{\omega_{mn}2 \Re{C^{{\rm (g)}+}_{lmn} \overline{\delta C^+_{lmn}}} - 3 \abs{C^{{\rm (g)}+}_{lmn}}^2 \delta\omega_{mn}}{4\pi \omega_{mn}^4} \, .
\end{align}
\end{subequations}

\subsection{PN expansion of the fluxes}

The geodesic amplitude $C^{{\rm (g)}+}_{lmn}$ and the linear-in-spin part $\delta C^+_{lmn}$ can be calculated as a PN series and series in $e$ by substituting the expansions of the trajectory from Section \ref{sec:PNtrajectory}. However, we also need a weak-field and low-velocities expansion of the radial function $R^-_{lmn}$ and Wronskian $W_{lmn}$. This has been done in~\cite{Tagoshi:1994,Tanaka:1996} where these quantities were expanded in $z = \omega r = \order{v}$ and $\epsilon = 2M \omega = \order{v^3}$ (see \cite{Sasaki:2003} for a review). Therefore, after substituting these variables, we obtain the expansion of the function $R^-_{lmn}(r(q^r))$ in $v$ and $e$.

After the expansion in $v$ and $e$, the integrals for the geodesic part $C^{{\rm (g)}+}_{lmn}$ in Eq~\eqref{eq:Cpm_lmn} and for the linear-in-spin part $\delta C^+_{lmn}$ \eqref{eq:deltaC_lmn} consist of a finite Fourier series in $q^r$; therefore, they are trivial to integrate. In this way, we obtain the amplitudes with their linear-in-spin parts from which we calculate the fluxes and their linear-in-spin parts.

In the PN approximation and after expansion in eccentricity, the sums over $l$, $m$, and $n$ in the geodesic fluxes \eqref{eq:fluxes} and their linear-in-spin parts \eqref{eq:linear_fluxes} are finite, since higher terms contribute only to higher order in $v$ and $e$. Unlike the geodesic parts of the $l$, $m$ multipoles of the fluxes, which start at $(l-2)$PN order for even $l+m$ and at $(l-1)$PN order for odd $l+m$, the linear-in-spin parts start at $(l-1/2)$PN order for both even and odd $l+m$. Since the linear-in-spin parts of the fluxes start at 1.5PN order, which corresponds to the spin-orbit coupling, we need to expand them to 3.5PN order NLO to obtain a 5PN expansion. Therefore, the fluxes and their linear-in-spin parts are summed over $2 \leq l \leq 5$ and $-l \leq m \leq l$ to obtain the geodesic fluxes in the 3.5PN order and the linear-in-spin parts to 5PN order. Because the $n$ modes of the fluxes $\mathcal{F}_{lmn}$ behave as $\order{e^{2n}}$ and thanks to the symmetry $\mathcal{F}_{l,m,n} = \mathcal{F}_{l,-m,-n}$, we sum over $n$ in the range $-m < n \leq 5$\footnote{Modes with $n=-m$ contribute to the fluxes with higher PN order, and we do not need to calculate them here.} to obtain expansion to $e^{10}$.

Therefore, when the geodesic fluxes are completed to the 5PN order from, e.g., \cite{Munna:2020}, we obtain the full 5PN energy and angular momentum fluxes from a spinning body orbiting a Schwarzschild black hole up to linear order in spin. Note that during the calculation of the linear-in-spin part, nonzero terms appear in the 1PN position, which cancel out and the series then start at the 1.5PN term. Therefore, because of the subtraction of the leading term, the trajectory must be expanded to one order higher than is the order of the final series. Furthermore, because the horizon fluxes for nonspinning secondary in Schwarzschild spacetime start at 4PN order, the linear-in-spin contribution to the horizon fluxes starts at 5.5PN order. Thus, we do not need to consider them here.

As discussed in \cite{Forseth:2016}, each PN term contains a factor with a certain power of $1-e^2$. When the fluxes are expressed using the parameter $v$, this factor reads as $(1-e^2)^{3/2}$ for all orders and can be factored out \cite{Sago:2015}. The resulting linear-in-spin parts of the energy and angular-momentum fluxes have the form
\begin{widetext}
\begin{align}\label{eq:deltaFEn}
\begin{split}
    & \delta \mathcal{F}^{E} = 
        \mathcal{F}^{E}_{\rm N} \qty(1-e^2)^{3/2} \Bigg[ 
        \delta f_3 v^3 
        + \delta f_5 v^5 
        + \delta f_6 v^6 
        + \delta f_7 v^7 
        + \delta f_8 v^8 
        + \left(\delta f_9 + \delta f_9^{\log v}\left(\gamma - \frac{35 \pi^2}{107} + \log v \right) \right) v^9
        \\
        & \phantom{\delta \mathcal{F}^{E} =}
        + \delta f_{10} v^{10} 
        + \order{v^{11}} 
        \Bigg] \, ,
\end{split}
\end{align}
\begin{align}\label{eq:deltaFJz}
\begin{split}
    & \delta \mathcal{F}^{J_z} = 
        \mathcal{F}^{J_z}_{\rm N} \qty(1-e^2)^{3/2} \Bigg[ \delta g_3 v^3 + \delta g_5 v^5 + \delta g_6 v^6 + \delta g_7 v^7 + \delta g_8 v^8 + \qty(\delta g_9 + \delta g_9^{\log v}\qty(\gamma - \frac{35 \pi^2}{107} + \log v ) ) v^9 
        \\
        &\phantom{\delta \mathcal{F}^{J_z} =}
        + \delta g_{10} v^{10} + \order{v^{11}} \Bigg] \, ,
\end{split}
\end{align}
where
\begin{equation}
    \mathcal{F}^{E}_{\rm N} = \frac{32}{5} \qty(\frac{\mu}{M})^2 v^{10}\,,\quad \mathcal{F}^{J_z}_{\rm N} = \frac{32}{5} \frac{\mu^2}{M} v^{7} \, ,
\end{equation}
are the Newtonian fluxes, and $\gamma$ is the Euler–Mascheroni constant. The $\delta f_i(e), \delta g_i(e)$ are functions of eccentricity similar to the enhancement functions of Peters \& Mathews and can be found in Appendix~\ref{app:PNseries}.
\end{widetext}

Similarly to the geodesic part, we were able to resum the leading term $\delta f_3$ ($\delta g_3$), the 1PN and 2PN contributions $\delta f_5$, $\delta f_7$ ($\delta g_5$, $\delta g_7$) and the logarithmic term $\delta f_9^{\log v}$ ($\delta g_9^{\log v}$) and write them in closed form.

After expansion to $e^{12}$ and factorization of $(1-e^2)^{3/2}$, in the functions $\delta f_3$, $\delta g_3$, $\delta f_9^{\log(v)}$, and $\delta g_9^{\log(v)}$, some of the last terms vanish. In particular, the series $\delta f_3$ ends at $e^6$, $\delta g_3$ at $e^4$, $\delta f_9^{\log(v)}$ at $e^{10}$, and $\delta g_9^{\log(v)}$ at $e^8$. Furthermore, after subtracting terms proportional to $\sqrt{1-e^2}$ from $\delta f_5$, $\delta g_5$, $\delta f_7$, and $\delta g_7$, in the remaining series, similarly, some terms vanish. These series end at $e^8$, $e^6$, $e^{10}$, and $e^8$, respectively. Therefore, we have not verified some of the resummations to \emph{all} orders in eccentricity, but we assume that they are true from the similar behavior of the geodesic part \cite{Munna:2019,Munna:2020a}.

\begin{figure}
    \centering
    \includegraphics[width=0.48\textwidth]{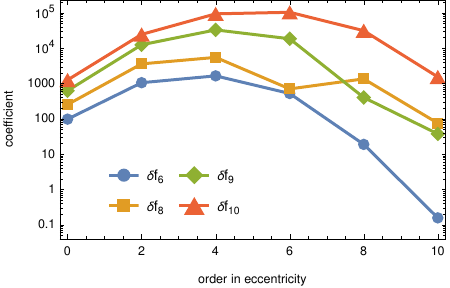}
    \caption{Coefficients in the PN expansion and eccentricity expansion of the linear part of the energy flux $\delta \mathcal{F}^E$ from Eq.~\eqref{eq:deltaFEn}. 
    }
    \label{fig:deltaFE_coefs}
\end{figure}
\begin{figure}
    \centering
    \includegraphics[width=0.48\textwidth]{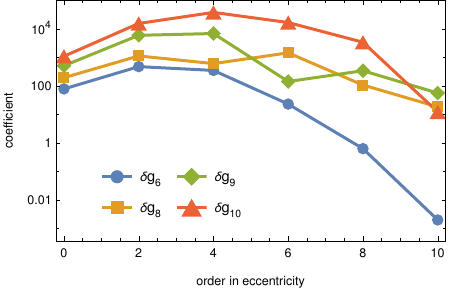}
    \caption{Coefficients in the PN expansion and eccentricity expansion of the linear part of the angular momentum flux $\delta \mathcal{F}^{J_z}$ from Eq.~\eqref{eq:deltaFJz}.}
    \label{fig:deltaFJz_coefs}
\end{figure}

In Figures \ref{fig:deltaFE_coefs} and \ref{fig:deltaFJz_coefs} we plot the coefficients of the PN series of the linear-in-spin parts of the energy and angular momentum flux. Each line shows the coefficients of the series in $e$ for a given PN order. The coefficients seem to decrease with eccentricity for all PN orders, which suggests that the truncation of the eccentricity series does not cause a large error. However, it may be improved by fitting and subtracting unknown terms proportional to $\sqrt{1-e^2}$ or $\log(1+\sqrt{1-e^2})$, which we know to appear in geodesic fluxes \cite{Munna:2019,Munna:2020a}.

Fluxes can also be expressed using the gauge-invariant quantity $x = (M\Omega_\phi)^{2/3}$. Then, they can be linearized as
\begin{align}
    \mathcal{F}(x,e,s_\parallel) = \mathcal{F}_{\rm (g)}(x,e) + s_\parallel \eval{\delta\mathcal{F}}_{x,e}(x,e) \,,
\end{align}
where the linear-in-spin part with fixed $x$ and $e$ can be obtained from the linear-in-spin part with fixed $v$ and $e$ as
\begin{align}
    \eval{\delta\mathcal{F}}_{x,e}(x,e) = \eval{\delta\mathcal{F}}_{v,e}(v_{\rm (g)}(x,e),e) + \pdv{\mathcal{F}_{\rm g}(v,e)}{v} \delta v \,,
\end{align}
where $v_{\rm (g)}(x,e)$ and $\delta v(x,e)$ can be found from $\Omega_\phi$ as an inverse series of the series $x = (M(\Omega_\phi^{\rm (g)}(v(x,e),e) + s_\parallel \delta\Omega_\phi(v(x,e),e)))^{2/3}/M$. The results for $\delta v$ and $\eval{\delta\mathcal{F}}_{x,e}(x,e)$ are given in the Supplemental Material.

After this transformation, the energy flux for zero eccentricity agrees with the results of \citet{Nagar:2019}, where the PN expansion of energy fluxes from spinning bodies on circular orbits of a Schwarzschild black hole was derived.

To verify our results, we compare them with the results of \citet{Henry:2023}, where the energy and angular momentum fluxes from eccentric spinning binaries were calculated using the PN theory. Their results to 3PN and $e^8$ are given as functions of $x$ and the time eccentricity $e_t$, which is used in the quasi-Keplerian parametrization described in Eqs.~(2.26) in \cite{Henry:2023}. Therefore, we had to transform their fluxes to functions of $x$ and $e$ using a relation between $e_t$ and $e$ derived in Appendix \ref{app:quasiKeplerian}. After the transformation, the linear parts of the energy and angular momentum fluxes derived in \cite{Henry:2023} agree with our results up to the 3PN order, $e^8$ and the first order in the mass ratio.

\begin{figure}
    \centering
    \includegraphics[width=0.48\textwidth]{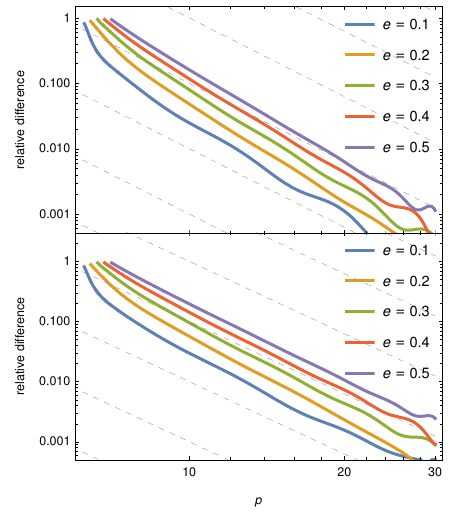}
    \caption{Relative difference between the PN expansion of the linear-in-spin part of the energy (top) and angular momentum (bottom) flux $\delta\mathcal{F}_{\text{PN}}$ and the fully relativistic value of $\delta\mathcal{F}_{\text{num}}$ for different eccentricities. The dashed lines show dependence $\order{p^{-4}}$ which should be the order of the error.}
    \label{fig:error_deltaFEnJz}
\end{figure}

To further validate our results, we compare the PN series with the fully relativistic numerically calculated linear parts of the fluxes calculated in \cite{Skoupy:2022}. We calculate the relative errors
\begin{align}
    \abs{ 1- \frac{\delta \mathcal{F}_{\rm PN}}{\delta \mathcal{F}_{\rm num}} }
\end{align}
for $\delta\mathcal{F}^{E}$ and $\delta\mathcal{F}^{J_z}$ and plot them in Figure~\ref{fig:error_deltaFEnJz} as functions of $p$ for different values of the eccentricity. These plots verify that the relative difference decreases with increasing $p$. For comparison, we also plot the behavior $p^{-4}$ since it is the behavior of the first neglected PN term (because the fluxes are expanded to 3.5PN orders NLO). The relative differences seem to decrease with higher power of $p$ which is probably caused by the smaller magnitude of the 4PN NLO term compared to the 4.5PN NLO term. For higher $p$, the relative difference is dominated by the interpolation error of the numerical fluxes.

\section{Flux-driven inspirals}
\label{sec:inspirals}

Once we obtained the energy and angular momentum fluxes, we can calculate the inspiral, i.e. the evolution of the orbital parameters. As discussed in Section \ref{sec:evolution}, the fluxes of energy and angular momentum are sufficient to calculate the evolution of $p$ and $e$ since $s_\parallel$ is conserved.

\subsection{Analytical integration of quasicircular inspirals}\label{sec:analytical_inspirals}

To obtain a first understanding of the convergence of the PN expansion, it is useful to examine the dynamics analytically. This is achieved by using the PN expansion of Schwarzschild geodesic fluxes as obtained in Refs. \cite{Munna:2020,Munna:2023} and implemented in the \texttt{PostNewtonianSelfForce} Mathematica package \cite{PNSelfForceZenodo} along with the spin fluxes derived here. While it is in principle possible to analytically integrate the dynamics at generic eccentricity, the symbolic computations become prohibitively expensive. For this reason, we restricted ourselves to quasicircular inspirals for the analytical convergence exploration (thus essentially restricting ourselves to the earlier flux formulas of Refs. \cite{Tanaka:1996,Tagoshi:1997jy,Nagar:2019}; see also Ref. \cite{Loutrel:2024} for a similar setup).

In that case, we can evolve the inspiral only in terms of the PN expansion parameter $v$. Furthermore, we can reparametrize the evolution with the azimuthal phase $\phi$
\begin{align}
\begin{split}    
    & \frac{\mathrm{d} v}{\mathrm{d} \phi}\Big|_{e=0} = \frac{1}{2}\sqrt{\frac{M}{p^3}} \left(\frac{\mathrm{d} E}{\mathrm{d} p} \right)^{-1} \!\mathcal{F}^E  \,\Omega^\phi
    \\
    & \phantom{\frac{\mathrm{d} v}{\mathrm{d} \phi}\Big|_{e=0}} = \frac{\mathrm{d} v}{\mathrm{d} \phi}\Bigg|_{\mathrm{(g)}} + \frac{s_\parallel}{M}\frac{\mathrm{d} v}{\mathrm{d} \phi}\Bigg|_{\mathrm{(s)}}\,,
\end{split}
\end{align}
where the relations for the fluxes, frequencies, and $E(p)$ are evaluated at $e=0$ and receive $\mathcal{O}(s_\parallel/M)$ corrections as described above. For the quasicircular inspiral, one could equivalently use the $\mathcal{F}^{J_z}$ flux and a $J_z(p)$ relation due to the identity $E = J_z \Omega^\phi$ at zero eccentricity. 

Then, we get the equation for the evolution of the azimuthal phase as a function of $v$ by $\phi'(v) = (\mathrm{d}v/\mathrm{d}\phi)^{-1}$ as
\begin{widetext}
\begin{align}
    \begin{split}
    & \phi'(v) = 
        \frac{5 M}{32 \mu v^6} \Bigg[ 
            1
            + \frac{743}{336} v^2
            - 4 \pi v^3 
            +  \frac{3}{2} \frac{s_\parallel}{M} v^3
            + \frac{3 058 673 }{1 016 064} v^4
            - \frac{7729 \pi }{672}v^5 
            + \frac{743}{224} \frac{s_\parallel}{M} v^5 
            + \Phi_6 v^6 
            - 6 \pi \frac{s_\parallel}{M} v^6
            \\
            &\phantom{ \phi'(v) = }
            - \frac{15 419 335 \pi }{1 016 064} v^7
            + \frac{3 058 673}{677 376} \frac{s_\parallel}{M} v^7 
            + \Phi_8 v^8 
            - \frac{7729 \pi}{448} \frac{s_\parallel}{M} v^8
            +  \Phi_9 v^9 
            +  \Phi_{s9} \frac{s_\parallel}{M} v^9 
            + \Phi_{10} v^{10} 
            - \frac{15419335 \pi}{677 376} \frac{s_\parallel}{M} v^{10}
        \Bigg],
    \end{split}
\end{align}
where the $\Phi$ coefficients are defined as
\begin{align}
    & \Phi_6 \equiv -\frac{10 817 850 546 611}{93 884 313 600} + \frac{1712}{105} (\gamma+\log v) + \frac{32}{3} \pi^2 + \frac{3424}{105} \log 2 \,,
    \\
    & \Phi_8 \equiv 
        -\frac{2 500 489 942 240 134 443}{3 690 780 136 243 200}
        + \frac{9 203}{210} (\gamma + \log v)
        + \frac{9 049}{252} \pi^2 
        + \frac{50 551}{882} \log 2
        + \frac{47 385}{1 568} \log 3 \,,
    \\
    & \Phi_{9} \equiv 
        \pi\left(\frac{90 036 665 674 763}{187 768 627 200}
        - \frac{6 848}{105} (\gamma + \log v)
        - \frac{64}{3} \pi^2
        - \frac{13 696}{105} \log 2\right) ,
    \\
    & \Phi_{9s} \equiv 
        -\frac{10 270 192 050 611}{62 589 542 400}
        + \frac{856}{35} (\gamma + \log v)
        + 16 \pi^2
        + \frac{1 712}{35} \log 2 \,,
    \\
    \begin{split}
    & \Phi_{10} \equiv
        - \frac{1 417 220 168 422 461 061 151}{505 226 791 983 513 600}
        + \frac{6 470 582 647}{110 020 680}(\gamma + \log v)
        + \frac{578 223 115}{12 192 768} \pi^2
        + \frac{53 992 839 431}{220 041 360} \log 2
        \\ & \phantom{\Phi_{10} =}
        - \frac{5 512 455}{87 808} \log 3 \,.
    \end{split}
\end{align}
\end{widetext}
This relation can then be easily integrated term by term to obtain the change in $\phi$ between two referential values of $v$. We can take the end of the inspiral to be at the innermost stable circular orbit, which is at 
\begin{align}
    v_{\rm ISCO} = \sqrt{\frac{1}{6}} + \frac{1}{18} \frac{s_{\parallel}}{M}\,.
\end{align} 
Furthermore, we want to parametrize the initial condition by a referential initial frequency where the signal enters the band, $\Omega_\phi = \Omega^\phi_{\rm i}$. The perturbative inversion of the relation $\Omega_\phi (v)$ yields $v = \Omega_\phi^{1/3} + s_\parallel \Omega_\phi^{4/3}/(2M) $. As a result, we get the inspiral phase as
\begin{widetext}
\begin{align}
\begin{split} \label{eq:Deltaphi}
    & \Delta \phi_{\rm ISCO}  = 
        \frac{M}{32 \mu \Omi^{5/3}} \Bigg[ 
            1 
            + \frac{3715}{1008} \Omi^{1/3} 
            -10 \pi \Omi 
            + \frac{15 293 365}{1 016 064} \Omi^{4/3} 
            + \Delta_{5/3} \Omi^{5/3}
            + \Delta_{2} \Omi^2
            + \frac{77 096 675 \pi}{2 032 128} \Omi^{7/3}
            \\
            & \phantom{\phi_{\rm ISCO} - \phi_{\rm I} =}
            + \Delta_{8/3} \Omi^{8/3}
            + \Delta_{3} \Omi^{3}
            + \Delta_{10/3} \Omi^{10/3}
        \Bigg]
        + \frac{s_\parallel}{M} \frac{5}{128 \Omi^{2/3}} \Bigg[
            1
            + \Delta_{s 2/3} \Omi^{2/3}
            + 32 \pi \Omi
            - \frac{15 293 365}{1 016 064} \Omi^{4/3}
            \\
            & \phantom{\phi_{\rm ISCO} - \phi_{\rm I} =}
            + \frac{7 729 \pi}{168} \Omi^{5/3}
            + \Delta_{s 2} \Omi^2
            + \frac{3 083 867 \pi}{63 504} \Omi^{7/3}
        \Bigg] + \mathcal{O}(\Omi^{6/3}) \,,
\end{split} 
\end{align}
where we have defined $\Omi \equiv M \Omega^\phi_{\mathrm{i}}$ as the initial frequency in units of $M$.  The $\Delta$ coefficients then read
\begin{align}
\begin{split}
    & \Delta_{5/3} \equiv 
        - \frac{223 791 298 249 051 766 200 631}{163 693 480 602 658 406 400\sqrt{6}}
        + \frac{377 580 814 447}{3 960 744 480 \sqrt{6}} \gamma 
        + \frac{383368458940043 \pi}{5407736463360}
        - \frac{428 \pi \gamma}{189}
        + \frac{28366605835 \pi^2}{438939648 \sqrt{6}}
        \\
        & \phantom{\Delta_{5/3} =}
        - \frac{20 \pi^3}{27}
        + \frac{136763321753}{990186120 \sqrt{6}} \log 2
        + \frac{102239 \pi}{4032}\log 2
        - \frac{5198370032377}{126743823360 \sqrt{6}} \log 3
        + \frac{51643 \pi}{1728} \log 3
         \\
        & \phantom{\Delta_{5/3} =}
        + \frac{38645}{672} \pi \log (\Omi^{1/3})\,,
\end{split} \\
    & \Delta_{2} \equiv 
        \frac{12348611926451}{18776862720}
        - \frac{1712}{21} \left(\gamma + \log(\Omi^{1/3})\right)
        - \frac{160 \pi^2}{3}
        - \frac{3424}{21} \log 2 \,,
    \\
    & \Delta_{8/3} \equiv
        \frac{2554404624135128353}{2214468081745920}
        - \frac{9203}{126} \left(\gamma + \log(\Omi^{1/3})\right)
        - \frac{45 245 \pi^2}{756}
        - \frac{252 755}{2646} \log 2
        - \frac{78 975}{1568} \log 3 \,,
    \\
    & \Delta_3 \equiv \pi \left[
        - \frac{93098188434443}{150214901760}
        + \frac{1712}{21} \left(\gamma + \log(\Omi^{1/3})\right)
        + \frac{80 \pi^2}{3}
        + \frac{3424}{21} \log 2
    \right] \,,
    \\
\begin{split}
    & \Delta_{10/3} \equiv 
        \frac{474387630222958367413}{168408930661171200}
        - \frac{6470582647}{110020680}\left(\gamma + \log(\Omi^{1/3})\right)
        - \frac{578223115 \pi^2}{12192768}
        - \frac{53992839431}{220041360} \log 2
        \\
        & \phantom{\Delta_{10/3} = }
        + \frac{5512455}{87808} \log 3 \,,
\end{split} \\
\begin{split}
    & \Delta_{s 2/3} \equiv
        - \frac{5881840409277979019197}{245540220903987609600}
        - \frac{310713839464837  \pi}{5069752934400 \sqrt{6}}
        + \frac{8765086219 \pi^2}{1975228416}
        - \frac{32}{81} \sqrt{\frac{2}{3}} \pi^3
        \\
        & \phantom{\Delta_{s 2/3} =}
        + \frac{(112087291999 - 7175443968  \sqrt{6} \pi)\gamma}{17823350160}
        + \frac{43923447107}{17823350160} \log 2
        -\frac{1712 \pi }{2835}  \sqrt{\frac{2}{3}} \log\left(\frac{8}{3}\right)
         \\
        & \phantom{\Delta_{s 2/3} =}
        - \frac{5159694245689}{570347205120} \log 3
        - \frac{743}{56} \log(\Omi^{1/3}) \,,
\end{split} \\
    & \Delta_{s2} \equiv 
        \frac{10747149910451}{26824089600}
        - \frac{856}{15} \left(\gamma + \log(\Omi^{1/3})\right)
        - \frac{112 \pi^2}{3} 
        - \frac{1712}{15} \log 2 \,.   
\end{align}
\end{widetext}

Let us now plug in numbers corresponding to LISA EMRIs into this formula to investigate its convergence (we will use the same numbers in Section \ref{subsec:LISAinsp}). We use a primary mass $10^{6}M_{\odot}$, initial frequency of the $l=2,m=2$ mode equal to $f_{m=2} = 2 \Omega^\phi_{\rm i}/(2 \pi) = 1 \rm \, mHz$, and a secondary mass of $\mu = 100 M_{\odot}$, and thus $\epsilon = 10^{-4}$. The spin is also chosen as $s_\parallel = 100 M_\odot$, which corresponds to a maximally spinning and aligned secondary black hole. We evaluate each term separately and summarize the results in Table \ref{tab:PNcirc}. In general alignment with the observations made by \citet{Burke:2023} and \citet{Isoyama:2012bx} for nonspinning secondaries, we see that even though the geodesic adiabatic terms are far from subradian accuracy at 5PN, the spin terms are suppressed by a mass ratio factor and have converged well below radians in this scenario. This supports the hybrid approach which we will use for the evolution of eccentric inspirals in the next Section.

\begin{table}
\caption{Contributions of various terms to the final phase a of a quasicircular EMRI entering band at $1$ mHz at primary mass $M = 10^6 M_\odot$ and secondary mass $\mu = 100 M_\odot$ with a maximally spinning and aligned secondary.}\label{tab:PNcirc}
\begin{tabular}{c|c|c}
    PN ord.   & Geodesic $\Delta \phi$ & Spin $\Delta \phi$  \\  \hline
0   & 325 402           & 0          \\
1    & 74 454.5         & 0          \\
1.5  & -158 135         & 0.629199        \\
2    & 18 877.6         & 0            \\
2.5  & -47 387.3        & -2.2748          \\
3    & 6578.54          & 0.978466            \\
3.5  & 2312.36          & -0.0365018          \\
4    & 2420.03          & 0.087333            \\
4.5  & -1172.24         & -0.0000844637           \\
5    & 688.9            & 0.00572313            \\
\end{tabular}
\end{table}

\subsection{Evolution equations for eccentric inspirals}

To evolve the orbital parameters $p,e$, we must first use Eq.~\eqref{eq:EJdecay} to derive their (average) time derivatives.
From the relation between the constants of motion $E$ and $J_z$ and the $p, e$, the evolution equations can be written as
\begin{align}\label{eq:pdotedot}
    \mqty( \dot{p} \\ \dot{e} ) \equiv \mqty( \dv{p}{t} \\ \dv{e}{t} ) = -\mathbb{J}^{-1} \mqty( \mathcal{F}^{E} \\ \mathcal{F}^{J_z} ) \, ,
\end{align}
where $\mathbb{J}$ is the Jacobian
\begin{align}
    \mathbb{J} = \mqty( \partial_p E & \partial_e E \\ \partial_p J_z & \partial_e J_z ) \,,
\end{align}
which is known analytically and can be found in Appendix \ref{app:jacobian}. Eq.~\eqref{eq:pdotedot} can be expanded in the secondary spin as
\begin{multline}
    \mqty( \dot{p} \\ \dot{e} ) = - \mathbb{J}_{\rm (g)}^{-1} \mqty( \mathcal{F}_{\rm (g)}^{E} \\ \mathcal{F}_{\rm (g)}^{J_z} ) \\ 
    + \frac{s_\parallel}{M} \qty( \mathbb{J}_{\rm (g)}^{-1} \delta \mathbb{J} \mathbb{J}_{\rm (g)}^{-1} \mqty( \mathcal{F}_{\rm (g)}^{E} \\ \mathcal{F}_{\rm (g)}^{J_z} ) - \mathbb{J}_{\rm (g)}^{-1} \mqty( \delta \mathcal{F}^{E} \\ \delta \mathcal{F}^{J_z} ) ) ,
\end{multline}
where we used the relation for the derivative of inverse matrix. The geodesic energy and angular momentum fluxes $\mathcal{F}^{E,J_z}_{\rm (g)}$ can be written using the geodesic evolution of $p$ and $e$ as
\begin{align}
\begin{split}
    & \mqty( \dot{p} \\ \dot{e} ) = \mqty( \dot{p}_{\rm (g)} \\ \dot{e}_{\rm (g)} ) 
    \\
    & \phantom{\mqty( \dot{p} \\ \dot{e} ) =}
    + \frac{s_\parallel}{M} \qty( \mathbb{J}_{\rm (g)}^{-1} \delta \mathbb{J} \mqty( \dot{p}_{\rm (g)} \\ \dot{e}_{\rm (g)} ) - \mathbb{J}_{\rm (g)}^{-1} \mqty( \delta \mathcal{F}^{E} \\ \delta \mathcal{F}^{J_z} ) ) .
\end{split}
\end{align}
Note that the first term is associated with the adiabatic term while the second term contributes only to the postadiabatic term. Therefore, the requirements for the accuracy of the first term are much higher than the requirements for the accuracy of the second term. Thus, we can use the PN expansion of the linear-in-spin parts of the energy and angular momentum fluxes. The geodesic evolution of $p$ and $e$ in the fully relativistic regime was calculated numerically and subsequently interpolated on a grid in the $p$-$e$ plane in \cite{Skoupy:2022}. Therefore, the evolution equations we use in this work read
\begin{multline} \label{eq:pdot_edot_hybrid}
    \mqty( \dot{p} \\ \dot{e} ) = \mqty( \dot{p}_{\rm (g)}^{\rm num} \\ \dot{e}_{\rm (g)}^{\rm num} ) \\ 
    + \frac{s_\parallel}{M} \qty( \mathbb{J}_{\rm (g)}^{-1} \delta \mathbb{J} \mqty( \dot{p}_{\rm (g)}^{\rm num} \\ \dot{e}_{\rm (g)}^{\rm num} ) - \mathbb{J}_{\rm (g)}^{-1} \mqty( \delta \mathcal{F}^{E}_{\rm PN} \\ \delta \mathcal{F}^{J_z}_{\rm PN} ) ) \, ,
\end{multline}
where the superscript ``num'' means fully relativistic results, subscript ``PN'' denotes the PN expansion and the Jacobian and its $s_\parallel$-derivative are fully relativistic as well because they can contain some nontrivial behavior near the last stable orbit. The explicit form of the matrix product $\mathbb{J}_{\rm (g)}^{-1} \delta \mathbb{J}$ can be found in Appendix \ref{app:jacobian}.

After the evolution of $p(t)$ and $e(t)$ is obtained, the inspiral waveform from two-timescale expansion can be calculated as
\begin{align}\label{eq:waveform}
    h = \frac{1}{r} \sum_{lmn} A_{lmn}(t) Y_{lm}(\theta) e^{ - i \Phi_{mn}(t-r^\ast) + i m \phi}\,,
\end{align}
where the amplitude and phase read, respectively,
\begin{align}
    A_{lmn}(t) &= \frac{2 C^+_{lmn}(p(t),e(t))}{\omega_{mn}^2(p(t),e(t))} \,, \\
    \Phi_{mn}(t) &= m \Phi_\phi(t) + n \Phi_r(t) \,, \\
    \Phi_{r,\phi}(t) &= \int_0^t \Omega_{r,\phi}(p(u),e(u),s_\parallel) \dd u  \,.\label{eq:Phases}
\end{align}

\subsection{LISA band inspirals} \label{subsec:LISAinsp}

To verify the validity of the hybrid model \eqref{eq:pdot_edot_hybrid} containing the PN expansions, in this Section we compare inspirals calculated using this model and a fully relativistic model for astrophysically relevant EMRIs that will be possible to detect with LISA. Similarly to Section \ref{sec:analytical_inspirals}, the primary mass is chosen as $M = 10^6 M_\odot$. For this primary mass, the frequency of the dominant mode $l=m=2$, $n=0$ in the innermost stable circular orbit is $f_{20} = \omega_{20}/(2\pi) = 2 \Omega_\phi/(2\pi) = 4.4$~mHz, which is close to the minimum of the LISA noise curve. The mass of the secondary and parallel spin is chosen as $\mu = s_\parallel = 100 M_\odot$. Therefore, the mass ratio is $\epsilon=10^{-4}$, and the secondary corresponds to a maximally spinning Kerr black hole. We evolved the inspirals in a range where the frequency is $2\Omega_\phi/(2\pi) \geq 1$~mHz. 

The inspirals cannot be evolved all the way to the separatrix in our setup. This is caused by the fact that the grid, on which we interpolated the numerical fluxes, starts at a finite distance from the separatrix, and, also, because some quantities linear in the secondary spin diverge there. Therefore, we need to choose a consistent condition to end the inspirals. In our setup, this condition reads as
\begin{align}\label{eq:adiabaticity}
    \frac{\dv{\Omega_r}{t}}{\Omega_r^2} = 10^{-2} \,,
\end{align}
which corresponds to a radial inverse adiabaticity parameter (we draw inspiration from a similar parameter defined in Ref. \cite{Albertini:2022rfe}). This quantity is small for adiabatic inspirals and grows near the separatrix where the two-timescale expansion breaks.

\begin{figure}
    \centering
    \includegraphics[width=0.48\textwidth]{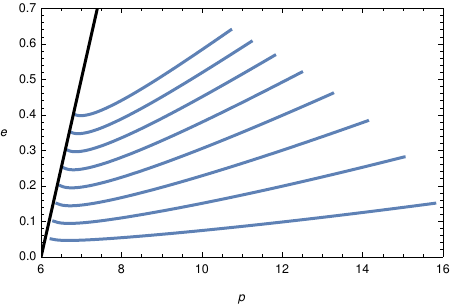}
    \caption{Adiabatic inspirals in the $p$-$e$ plane. The black line shows the separatrix $p=6+2e + \order{s_\parallel}$. The two models are indistinguishable in this plot (see Figure \ref{fig:deltaPhi} for phase differences).}
    \label{fig:inspirals}
\end{figure}

In this setup, we found values of $p$ that satisfy the condition \eqref{eq:adiabaticity} for $e$ between $0.05$ and $0.4$ and evolved the inspirals backward using the fully relativistic model. The evolution was stopped when the condition $f = 2\Omega_\phi<1$~mHz was reached. Then, we used the hybrid model \eqref{eq:pdot_edot_hybrid} to evolve the inspirals from the end points of the previous calculation. In this way, we obtained two sets of evolutions of $p$ and $e$ with different models for comparison. The results in the $p$-$e$ plane are depicted in Fig.~\ref{fig:inspirals}. 

\begin{figure}
    \centering
    \includegraphics[width=0.48\textwidth]{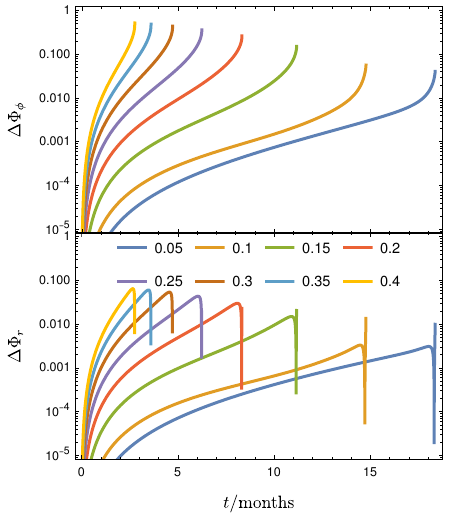}
    \caption{Absolute differences between the azimuthal (top) and radial (bottom) phases obtained from the PN model and fully relativistic model for different final eccentricities.}
    \label{fig:deltaPhi}
\end{figure}

In the next step, we used the analytic formulas for the orbital frequencies $\Omega_r$ and $\Omega_\phi$ to calculate the phases $\Phi_r$ and $\Phi_\phi$ from Eq.~\eqref{eq:Phases}. In Figure~\ref{fig:deltaPhi} we plot the absolute difference between the phases calculated with the hybrid and fully relativistic model. We can see that the phase differences are below unity.

\begin{figure}
    \centering
    \includegraphics[width=0.48\textwidth]{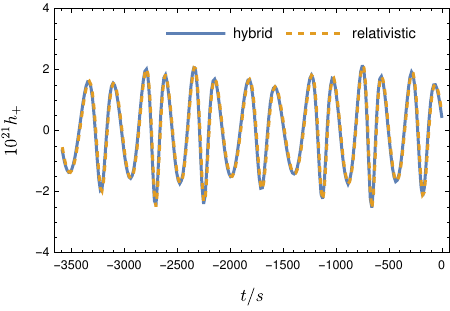}
    \caption{Waveforms of the $+$ polarization of an inspiral ending at $e=0.1$ calculated with the hybrid model (blue) and with the fully relativistic (numerical) model (yellow). The inspiral is observed from the distance of 1 Gpc at the viewing angle $\theta = \pi/3$, $\phi = \pi/4$ in the source frame.}
    \label{fig:waveform01}
\end{figure}

\begin{figure}
    \centering
    \includegraphics[width=0.48\textwidth]{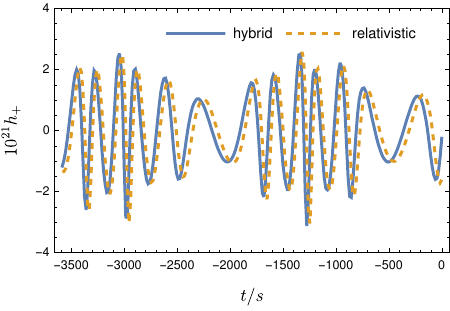}
    \caption{Waveforms of the $+$ polarization of an inspiral ending at $e=0.3$ calculated with the hybrid model (blue) and the fully relativistic (numerical) model (yellow). The distance and viewing angle are the same as in Figure~\ref{fig:waveform01}.}
    \label{fig:waveform03}
\end{figure}

To calculate both the evolution of the orbital parameters and the phases, we used \texttt{NDSolve} function in Mathematica with Adams' method. The resulting time series $p(t)$, $e(t)$, and $\Phi_{r,\phi}(t)$ were then used to calculate the waveform using the FastEMRIWaveforms (FEW) package \cite{Katz:2021,Chua:2020stf,michael_l_katz_2020_4005001,Chua:2018woh}. The distance of the observer was chosen as 1 Gpc and the viewing angle was chosen as $\theta = \pi/3$, $\phi = \pi/4$ in the source frame. FEW calculates the waveforms \eqref{eq:waveform} with the geodesic amplitudes $C^{{\rm (g)}+}_{lmn}$, which introduces $\order{s_\parallel/M}$ error in the amplitudes. However, this is not an issue, since the requirements for the accuracy of the amplitudes are lower than the requirements for the accuracy of the phases \cite{Lindblom:2008cm}. In Figures~\ref{fig:waveform01} and \ref{fig:waveform03} we show a comparison of the waveforms calculated with the two models for two inspirals ending at $e=0.1$ and $e=0.3$. 

\begin{figure}
    \centering
    \includegraphics[width=0.48\textwidth]{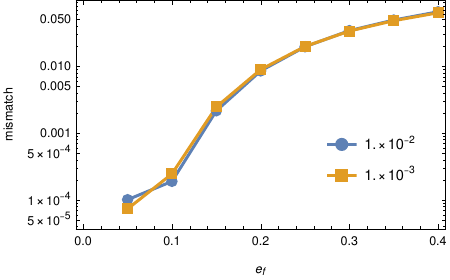}
    \caption{Mismatches between inspirals calculated with the hybrid model and the fully relativistic model for different final eccentricities $e_f$. The blue points show inspirals which end at $\dot{\Omega}_r/\Omega_r^2 = 10^{-2}$ while the yellow points show inspirals ending at $\dot{\Omega}_r/\Omega_r^2 = 10^{-3}$. Small differences between these cases indicate that the mismatches are almost independent of the ending criterion.}
    \label{fig:mismatch}
\end{figure}

From the obtained waveforms we calculated the mismatch between the fully relativistic model and the hybrid model with PN linear-in-spin parts. Mismatch is defined from the overlap $\mathcal{O}$ as
\begin{align}
\begin{split}
    \mathcal{M}(h_1,h_2) 
    & \equiv 1 - \mathcal{O}(h_1,h_2) 
    \\
    & = 1 - \frac{\langle h_1, h_2 \rangle}{\sqrt{\langle h_1, h_1 \rangle \langle h_2, h_2 \rangle}} \,,
\end{split}
\end{align}
where $\langle\cdot,\cdot\rangle$ is the $L^2$ product.
When the two waveforms are identical, the mismatch is zero. We plot the mismatches for different final eccentricities in Figure~\ref{fig:mismatch}. To test whether the ending criterion \eqref{eq:adiabaticity} influences the mismatches, we calculated the inspirals for two ending criterion, namely $\dot{\Omega}_r/\Omega_r^2 = 10^{-2}$ and $10^{-3}$ and compared them in Fig.~\ref{fig:mismatch}. In this plot, we can see that the mismatch is consistent for the two ending criteria and is lower than $10^{-2}$ for the lower final eccentricities.

\section{Discussion and outlooks} \label{sec:Concl}

In the previous sections, we calculated the PN expansions of the trajectories of spinning bodies on eccentric orbits around Schwarzschild black holes. Then, we found the PN expansions of the energy and angular momentum fluxes from the aforementioned orbits. The linear-in-spin parts of the fluxes were then used in a hybrid model, where the subleading secondary-spin effects were analytically approximated by using the PN series. Mismatches between waveforms from the fully relativistic model and our hybrid model showed that for lower eccentricities the models are indistinguishable. This result shows that in some cases the linear-in-spin part of the fluxes can be approximated as an analytical PN series without the need to numerically calculate the fully relativistic contribution. 

However, we only computed the $L^2$ mismatches and did not minimize them with respect to time shifts between the waveforms, which could be easily improved. Nevertheless, the insights gained by these improvements would be minor; to accurately assess the possible biases introduced by the hybrid model across the parameter space, a Fisher-matrix or Markov-chain Monte Carlo analysis such as those carried out by \citet{Burke:2023} and \citet{Piovano:2021iwv} should be performed with this model.

Figure \ref{fig:mismatch} shows that the mismatch is greater for inspirals with higher eccentricity. This could be improved by expanding to higher order in eccentricity or by finding exact (or arbitrary order in eccentricity) formulas, such as in \cite{Munna:2019,Munna:2020a}. However, in Figure \ref{fig:inspirals} we can see that the inspirals with higher eccentricity enter the LISA band in stronger field at lower $p$. Therefore, expanding to higher PN order may also improve accuracy. Nevertheless, the computations at higher PN order increase in complexity. For example, modes with $n=-m$ and higher $l$ modes must be included starting from $5.5$PN and higher for the spin fluxes. Additionally, the horizon fluxes will be needed as well, since they start at 4PN for the geodesic part and at 5.5PN for the linear-in-spin part. To extend the results to higher PN or eccentricity order, more computational resources or more systematic treatment of the Fourier series at each PN eccentricity order would be needed.

Poor convergence of the PN series for higher eccentricities can be caused by the fact that the secondary body reaches a stronger field at the pericenter $r_2 = M p/(1+e)$ even for high $p$ (i.e. small $v$). However, the convergence of the series in $v$ is better than the convergence of the series in $x$ since at fixed $x$ the pericenter approaches zero when $e \rightarrow 1$ and the fluxes diverge there. This is connected to the cancellation of the divergent factors $(1-e^2)^{-k/2}$ appearing in the $x$ series when it is reparametrized by $v$. 

Nevertheless, other nonanalytical terms of the type $(1 - e^2)^{k/2}$ with $k>0$ systematically appear in the flux series. The factorization of such terms on a case-to-case basis allowed us to resum the otherwise infinite $e$-series for a number of terms. However, at high PN orders additional terms with higher $k$ appear and the resummations consequently require more and more terms in the $e$-series to be verified. Additionally, Figs. \ref{fig:deltaFE_coefs} and \ref{fig:deltaFJz_coefs} show that the higher-PN terms are clearly not as well converged as lower-order terms at $e^{10}$. 


On the other hand, the requirement $2\Omega_\phi > 10^{-3}\, \text{Hz}$ we impose for the starting point of inspirals works best only for circular or low-eccentricity inspirals; it may be too crude for the highly eccentric cases. This is because for higher eccentricities, higher $n$ modes are present, thus introducing higher-frequency harmonics into the spectrum, which then enter the LISA band earlier than our cutoff. Thus, a more sophisticated analysis of LISA mismatches of extended waveforms without such simplifications is needed.

Therefore, conclusions about highly eccentric inspirals should not be drawn from the results for quasicircular inspirals. We see this also in Table~\ref{tab:PNcirc}, where the contribution to the phase from the last, 5PN, term is of the order of $10^{-2}$. Such a truncation error would be sufficient for LISA waveforms, but this convergence property unfortunately does not generalize to eccentric inspirals. We can extrapolate our observations using secondary spin even to the hybrid model of \citet{Burke:2023}, where 3PN approximations of second-order fluxes and conservative self-force were used in quasicircular inspirals of nonspinning binaries with encouraging results (see also the earlier work of \citet{Isoyama:2012bx}). We do not expect these encouraging results to generalize to eccentric inspirals. What is more, \textit{we do not expect even 5PN-$e^{10}$ expansions of the second-order fluxes and conservative self-force to be sufficient for LISA parameter estimates of highly eccentric inspirals in hybrid models.} 

How could the results of this paper be further generalized or expanded? One possibility would be to compute metric perturbations sourced by the spinning test particle instead of just curvature perturbations and fluxes. This was done numerically by \citet{Mathews:2022} in a fully relativistic setting, and it would be interesting to obtain PN-expanded analytical counterparts to their work. Another possible extension would be to calculate the PN expansion of energy and angular momentum fluxes from generic orbits of spinning bodies in Kerr spacetime. This could be achieved by expanding the equations of motion obtained from the Hamilton-Jacobi equation \cite{Witzany:2019} in a PN series and solving them order by order. We are already preparing a work in which we solve for the fundamental frequencies of motion of the spinning particle in Kerr in closed form and expand them in a PN series \cite{LeoSashwatInprep} (see also \cite{Gonzo:2024zxo}), but the full trajectories pose more of a technical challenge. 

Nonetheless, generic orbits of spinning test particles in Kerr are parametrized by one additional constant of motion, the R{\"u}diger (Carter-like) constant $K_R$. Hence, to calculate the inspirals, the evolution of this constant must first be derived, similarly to the evolution of $s_\parallel$ presented here in Section \ref{sec:evolution}. Until then, one can only evolve equatorial inspirals as done in \cite{Skoupy:2022}. Another loophole possibility to drive inspirals without the need for the evolution of $K_R$ turns out to be when one restricts to the inspirals of nearly spherical orbits with $e \approx 0$; we are also working on this topic \cite{SpherPapInPrep}.


\begin{acknowledgments}

We are grateful for the support of the Charles U. \textit{Primus} Research Program 23/SCI/017. This work makes use of the Black Hole Perturbation Toolkit.

\end{acknowledgments}

\appendix

\section{Linear-in-spin parts of the PN-expanded fluxes}
\label{app:PNseries}

In this Appendix, we present the results for the linear parts of the energy and angular momentum fluxes as series in the PN parameter $v = \sqrt{1/p}$
and eccentricity $e$ with some terms expressed in closed form.

The linear part of the energy flux reads as
\begin{widetext} 
\begin{align} \label{eq:AppdFE}
\begin{split}
    & \frac{\delta \mathcal{F}^{E}}{\mathcal{F}^{E}_{\rm N} \qty(1-e^2)^{3/2}} = \delta f_3(e) v^3 + \delta f_5(e) v^5 + \delta f_6(e) v^6 + \delta f_7(e) v^7 + \delta f_8(e)v^8 + 
    \\
    & \phantom{\frac{\delta \mathcal{F}^{E}}{\mathcal{F}^{E}_{\rm N} \qty(1-e^2)^{3/2}} =}
    \qty(\delta f_9^0(e) + \delta f_9^{\log(n)}(e) + \delta f_9^{\log(v)}(e)\qty(\gamma - \frac{35 \pi^2}{107} + \log(v) ) ) v^9 + \delta f_{10} v^{10} + \order{v^{11}} ,
\end{split}
\end{align}
where
\begin{align}
    \mathcal{F}^E_{\rm N} = \frac{32}{5} \qty(\frac{\mu}{M})^2 v^{10} 
\end{align}
is the Newtonian flux from circular orbits and $\delta f_i(e)$ are functions of eccentricity which take the form
\begin{align}
    \delta f_3 &= - \qty( \frac{25}{4} + \frac{151 e^2}{4} + \frac{443 e^4}{16} + \frac{355 e^6}{192}) ,
    \\
    \delta f_5 &= \frac{2403}{112}+\frac{16435 e^2}{112}+\frac{6701 e^4}{224}-\frac{78383 e^6}{896}-\frac{108813 e^8}{14336}+\frac{329 e^{10}}{512}+\order{e^{12}} ,
    \\
    \frac{\delta f_6}{\pi} &= - \frac{187 }{6} - \frac{32257  e^2}{96} - \frac{67141  e^4}{128} - \frac{9184435  e^6}{55296} - \frac{663581  e^8}{110592} - \frac{1761277  e^{10}}{35389440} + \order{e^{12}} ,
    \\
    \delta f_7 &= \frac{285211}{4536}+\frac{1027841 e^2}{1134}+\frac{182196563 e^4}{72576}+\frac{187711757 e^6}{145152}-\frac{12275083 e^8}{129024} +\frac{599257 e^{10}}{258048}+\order{e^{12}} ,
    \\
    \frac{\delta f_8}{\pi} &= \frac{62471 }{672}+\frac{1525357  e^2}{1344}+\frac{65526409  e^4}{43008}-\frac{110900285   e^6}{774144} -\frac{15548354173  e^8}{49545216}-\frac{1021275631  e^{10}}{117964800}+\order{e^{12}} ,
    \\
    \delta f_9^0 &= -\frac{29174232523}{34927200}-\frac{7994937281 e^2}{436590}-\frac{389472520471 e^4}{6652800} -\frac{1184172237779 e^6}{22353408}-\frac{15468420403 e^8}{698544} \nonumber\\ &\hphantom{=} -\frac{351984359281 e^{10}}{30105600}+\order{e^{12}} ,
    \\
    \frac{\delta f_{10}}{\pi} &= \frac{86803}{216}+\frac{758931497 e^2}{96768}+\frac{5808116575 e^4}{193536}+\frac{2774132820133 e^6}{83607552} +\frac{6683237111993  e^8}{668860416} \nonumber\\ &\phantom{=} +\frac{524219611964239 e^{10}}{1070176665600}+\order{e^{12}} .
\end{align} 
The various logarithmic terms are then given as
\begin{align}     
    \delta f_9^{\log(n)} &\equiv  \delta f_9^{\log(2)} \log(2) + \delta f_9^{\log(3)} \log(3) + \delta f_9^{\log(5)} \log(5) + \delta f_9^{\log(7)} \log(7) \\
    \delta f_9^{\log(v)} &= \frac{15943}{105}+\frac{37985 e^2}{14}+\frac{39804 e^4}{5}+\frac{172163 e^6}{28}+\frac{1088725 e^8}{896}+\frac{257121 e^{10}}{8960} \\
    \begin{split}
    \delta f_9^{\log(2)} &= \frac{19153}{63}+\frac{243853 e^2}{630}+\frac{44397296 e^4}{315}-\frac{3518430389 e^6}{2268}+\frac{4038595503577 e^8}{362880} -\frac{1099970344876951 e^{10}}{18144000}
    \\ &\phantom{=} +\order{e^{12}} ,
    \end{split}
    \\
    \delta f_9^{\log(3)} &= \frac{702027 e^2}{140}-\frac{147191661 e^4}{2240}+\frac{427378437 e^6}{1120}+\frac{591272646381 e^8}{573440} \nonumber\\ &\phantom{=} -\frac{1574559880088247 e^{10}}{57344000}+\order{e^{12}} ,
    \\
    \delta f_9^{\log(5)} &= \frac{31138671875 e^6}{72576}-\frac{50939732421875 e^8}{9289728}+\frac{56782099609375 e^{10}}{1769472}+\order{e^{12}} ,
    \\
    \delta f_9^{\log(7)} &= \frac{56067099797765 e^{10}}{5308416}+\order{e^{12}} .
\end{align}
The term $\delta f_5$ can be resummed in eccentricity as
\begin{align}
    \delta f_5 &= \frac{1731}{112} + \frac{15399 e^2}{112} + \frac{11811 e^4}{224} - \frac{81743 e^6}{896} - \frac{139613 e^8}{14336} + 6(1-e^2)^{3/2} \qty( 1 + \frac{73}{24} e^2 + \frac{37}{96} e^4 )\,,
\end{align}
which is consistent with the results of \citet{Henry:2023}. We managed to resum also the term $\delta f_7$ in the form 
\begin{align}
    \delta f_7 &= -\frac{473}{648}+\frac{848905 e^2}{1296}+\frac{193342649 e^4}{72576}+\frac{237963833 e^6}{145152}-\frac{25289371 e^8}{129024}-\frac{10496681 e^{10}}{258048} \nonumber \\ &\phantom{=} + (1-e^2)^{3/2} \qty(\frac{1781}{28}+\frac{38839 e^2}{112}+\frac{21935 e^4}{64}+\frac{15179 e^6}{448})\,.
\end{align}

The angular momentum fluxes can be expressed as
\begin{align} \label{eq:AppdFJz}
\begin{split}
    & \frac{\delta \mathcal{F}^{J_z}}{\mathcal{F}^{J_z}_{\rm N} \qty(1-e^2)^{3/2}} = 
        \delta g_3(e) v^3 + \delta g_5(e) v^5 + \delta g_6(e) v^6 + \delta g_7(e) v^7 + \delta g_8(e) v^8 +  
        \\
        & \phantom{\frac{\delta \mathcal{F}^{J_z}}{\mathcal{F}^{J_z}_{\rm N} \qty(1-e^2)^{3/2}} = }
         \qty(\delta g_9^0 + \delta g_9^{\log(n)}(e) + \delta g_9^{\log(v)}\qty(\gamma - \frac{35 \pi^2}{107} + \log(v) ) ) v^9  + \delta g_{10} v^{10} + \order{v^{11}},
\end{split}
\end{align}
where 
\begin{align}
    \mathcal{F}^{J_z}_{\rm N} = \frac{32}{5} \frac{\mu^2}{M} v^7
\end{align}
is the Newtonian flux from circular orbits and the functions $\delta g_i(e)$ read as
\begin{align}
    \delta g_3 &= -\frac{19}{4}-\frac{683 e^2}{48}-\frac{403 e^4}{96} ,
    \\ 
    \delta g_5 &= \frac{3559}{224}+\frac{30509 e^2}{672}-\frac{135161 e^4}{5376}-\frac{1251 e^6}{64}+\frac{15 e^8}{32}+\frac{99 e^{10}}{512} + \order{e^{12}} ,
    \\ 
    \frac{\delta g_6}{\pi} &= - \frac{151 }{6} - \frac{3671  e^2}{24} - \frac{21607  e^4}{192} - \frac{102199  e^6}{13824} + \frac{88799  e^8}{442368} - \frac{27779  e^{10}}{44236800} + \order{e^{12}} ,
    \\ 
    \delta g_7 &= \frac{1006711}{18144}+\frac{2535215 e^2}{5184}+\frac{1709749 e^4}{2304}+\frac{1212805 e^6}{13824}-\frac{795497 e^8}{18432} +\frac{38149 e^{10}}{4096} + \order{e^{12}} ,
    \\ 
    \frac{\delta g_8}{\pi} &= \frac{100369 }{1344}+\frac{296213  e^2}{896}-\frac{7426709  e^4}{28672}-\frac{620472319  e^6}{1548288}-\frac{728183081  e^8}{33030144} +\frac{42902045527  e^{10}}{4954521600} + \order{e^{12}} ,
    \\
    \begin{split}
    \delta g_9^0 &= -\frac{19353142307}{27941760}-\frac{258040969517 e^2}{27941760}-\frac{39004110703 e^4}{2540160} -\frac{7526116434163 e^6}{1117670400}-\frac{19004463748957 e^8}{5960908800} \\ &\phantom{=} -\frac{953163710537 e^{10}}{425779200}+\order{e^{12}} ,
    \end{split}
    \\
    \begin{split}
    \frac{\delta g_{10}}{\pi} &= \frac{1068677}{3024}+\frac{6760111 e^2}{1344}+\frac{113252009 e^4}{9216}+\frac{56647762139 e^6}{10450944} -\frac{104974004267 e^8}{95551488}+\frac{174401161771 e^{10}}{44590694400}
    \\ & \phantom{=} +\order{e^{12}}.
    \end{split}
\end{align}
The various logarithmic terms are then 
\begin{align}
    \delta g_9^{\log(n)} &\equiv \delta g_9^{\log(2)} \log(2) + \delta g_9^{\log(3)} \log(3) + \delta g_9^{\log(5)} \log(5) + \delta g_9^{\log(7)} \log(7) ,
    \\
    \delta g_9^{\log(v)} &= \frac{2675}{21}+\frac{347429 e^2}{252}+\frac{1090651 e^4}{504}+\frac{6934349 e^6}{10080}+\frac{105823 e^8}{4480} ,
    \\
    \begin{split}
    \delta g_9^{\log(2)} &= \frac{80357}{315}-\frac{615143 e^2}{1260}+\frac{181175503 e^4}{2520}-\frac{63567501707 e^6}{90720} +\frac{18959825303 e^8}{4480}-\frac{3138557026727 e^{10}}{162000}
    \\ & \phantom{=} +\order{e^{12}} ,
    \end{split}
    \\
    \delta g_9^{\log(3)} &= \frac{1794069 e^2}{560}-\frac{21138813 e^4}{560}+\frac{196801569 e^6}{1024}+\frac{136487855331 e^8}{573440} -\frac{1881487783393587 e^{10}}{229376000}+\order{e^{12}} ,
    \\
    \delta g_9^{\log(5)} &= \frac{14503515625 e^6}{82944}-\frac{877191015625 e^8}{442368}+\frac{503789897265625 e^{10}}{49545216}+O\left(e^{12}\right) ,
    \\
    \delta g_9^{\log(7)} &= \frac{8289498036460823 e^{10}}{2654208000}+\order{e^{12}} .
\end{align}
The second and fourth term in Eq. \eqref{eq:AppdFJz} can be again resummed as 
\begin{align}
    \delta g_5 &= \frac{2215}{224}+\frac{33029 e^2}{672}-\frac{104921 e^4}{5376}-\frac{1401 e^6}{64} + 6 (1-e^2)^{3/2} \qty( 1 + \frac{7}{8}e^2 ) , \\
    \begin{split}
    \delta g_7 &= \frac{56743}{18144}+\frac{2111585 e^2}{5184}+\frac{2036089 e^4}{2304}+\frac{1977175 e^6}{13824}-\frac{1549517 e^8}{18432} 
    \\
    & \phantom{=} + (1-e^2)^{3/2} \qty(\frac{733}{14}+\frac{35897 e^2}{224}+\frac{2215 e^4}{28}) .
    \end{split}
\end{align}
\end{widetext}

The linear parts of the fluxes as functions of alternate $x-e$ parametrization are given in the Supplemental Material \cite{SupMat}.

\section{Comparison with Phys. Rev. D 108, 104016 (2023)}
\label{app:quasiKeplerian}

In this Appendix, we show the derivation of the transformation between time eccentricity $e_t$ and Darwin eccentricity $e$ which is needed for the comparison between our results and the results of \cite{Henry:2023}. 

In the quasi-Keplerian parametrization and harmonic coordinates $(t_\text{H}, r_\text{H}, \phi_\text{H}) = (t, r - M, \phi)$, the orbit is given as \cite{Henry:2023}
\begin{align}
    r_H &= a_r ( 1 - e_r \cos u ) \,, \\
    \Omega_r t_H &= u - e_t \sin u + f_{v-u}(v-u) + f_v \sin v \,, \\
    \frac{2\pi}{\Phi} \phi_H &= v + g_{2v} \sin 2v + g_{3v} \sin 3v \,, \\
    \tan \frac{v}{2} &= \sqrt{\frac{1 + e_\phi}{1-e_\phi}} \tan \frac{u}{2} \,,
\end{align}
where $a_r$ is the semi-major axis, $u$ is the eccentric anomaly, $\Phi$ is the total phase between two successive periastron passages, $v$ is the true anomaly and $f_{v-u}$, $f_{v}$, $g_{2v}$, and $g_{3v}$ are functions given in \cite{Henry:2023}. 

To derive the transformation, we first find the relation between $(a_r, e_r)$ and $(p, e)$ parametrization from the expression for the turning points $r_{1,2}$ and $r^{\text{H}}_{1,2}$:
\begin{align}
\begin{split}
    &r_{\text{H} 1,2} = a_r (1 \pm e_r) = r_{1,2} - M 
    \\ & \phantom{r_{\text{H} 1,2}} = M p/(1 \mp e) - M \,.
\end{split}
\end{align}
The parameters $a_r$, $x=(M\Omega_\phi)^{2/3}$, $e_r$, and $e_t$ are given in the supplemental material of \cite{Henry:2023} as functions of $\Tilde{E} = -(E - M c^2)/\mu$ and $h = L/(GM\mu)$. By inverting the PN series to obtain $\Tilde{E}$ and $h$, we were able to express the time eccentricity using the Darwin eccentricity $e$ and the PN parameter $x$ as
\begin{align}
    e_t^2/e^2 &= \qty(e_t^2/e^2)_{\rm (g)} + \frac{s_\parallel}{M} \delta(e_t^2/e^2) \, , \\
    \begin{split}
    \delta(e_t^2/e^2) &= \frac{2 x^{3/2}}{\sqrt{1-e^2}} + \frac{6( e^2 - 2 + 2 \sqrt{1-e^2}) x^{5/2}}{(1-e^2)^{3/2}} 
    \\ & \phantom{=} + \order{x^{7/2}} \, .
    \end{split}
\end{align}
The geodesic part $\qty(e_t^2/e^2)_{\rm (g)}$ can be found in Eq.~(4.38) of \cite{Forseth:2016}.

Alternatively, one can solve the equation for $t$ as a function of the eccentric anomaly $u$ and collect all the terms that generate $e_t$, as was done in \cite{Munna:2020b}, however, this process is long and difficult and we leave it for future work.

\section{Evolution of the orbital parameters}
\label{app:jacobian}

In this Appendix we present the formulas for the evolution of the orbital parameters $p$ and $e$ used in the hybrid model in Eq.~\eqref{eq:pdot_edot_hybrid}. The elements of the geodesic part of the inverse Jacobian
\begin{align}
    \mathbb{J}^{-1}_{\rm (g)} = \mqty( \partial_E p & \partial_{J_z} p \\ \partial_E e & \partial_{J_z} e ) 
\end{align}
read as
\begin{align}
    \partial_E p &= \frac{-2 p^{3/2} \sqrt{P_3 P_2} }{P_1} \, , \\
    \partial_{J_z} p &= \frac{2(p-4)^2 \sqrt{P_3}}{P_1} \, , \\
    \partial_{E} e &= \frac{(p-6-2e^2) \sqrt{p P_2 P_3}}{e P_1} \, , \\
    \partial_{J_z} e &= - \frac{(1-e^2)((p-2)(p-6)+4e^2) \sqrt{P_3}}{p e P_1} \, ,
\end{align}
where we introduced the polynomials
\begin{align}
    P_1 &= (p-6)^2 - 4e^2 \,,\\
    P_2 &= (p-2)^2 - 4e^2 \,,\\
    P_3 &= p-3-e^2 \,.
\end{align}
Note that the polynomial $P_1$ vanishes at the separatrix $p = 6+2e$; therefore the inverse Jacobian diverges there.

We can factor out some terms from the matrix product $\mathbb{J}_{\rm (g)}^{-1} \delta \mathbb{J}$ and express it in the form
\begin{align}
    \mathbb{J}_{\rm (g)}^{-1} \delta \mathbb{J} = \frac{1}{P_1 P_3 \sqrt{P_2 p^3}} \mathbb{M} \,,
\end{align}
where
\begin{widetext}
\begin{align}
    \mathbb{M}_{1,1} &= -4 e^6 p+e^4 (p (p (p+8)-36)+96)-4 e^2 \left(p \left(4 p^2-30 p+83\right)-48\right)-(p-6) (p-2) (p (4 p-15)+24) \,, \\
    \mathbb{M}_{1,2} &= e p \left(-\left(\left(e^2-21\right) p^3\right)-8 \left(e^2+25\right) p^2+4 \left(e^4+10 e^2+165\right) p-768\right) , \\
    \begin{split}
    \mathbb{M}_{2,1} &= \frac{1-e^2}{4 e p} \big(16 e^6 (p-3)-4 e^4 ((p-4) p (p+14)+60)+e^2 (p (p (p (5 p-12)-96)+336)-144) \\ &\phantom{=} +3 (p-6)^2 (p-2)^2 \big)  \,,
    \end{split}\\
    \mathbb{M}_{2,2} &= \left(e^2-1\right) \left(4 e^4 p-e^2 \left(p^3+16 p^2-120 p+96\right)+2 p^4-13 p^3-24 p^2+228 p-288\right) .
\end{align}
\end{widetext}

\bibliography{paper}

\begin{thebibliography}{77}%
\makeatletter
\providecommand \@ifxundefined [1]{%
 \@ifx{#1\undefined}
}%
\providecommand \@ifnum [1]{%
 \ifnum #1\expandafter \@firstoftwo
 \else \expandafter \@secondoftwo
 \fi
}%
\providecommand \@ifx [1]{%
 \ifx #1\expandafter \@firstoftwo
 \else \expandafter \@secondoftwo
 \fi
}%
\providecommand \natexlab [1]{#1}%
\providecommand \enquote  [1]{``#1''}%
\providecommand \bibnamefont  [1]{#1}%
\providecommand \bibfnamefont [1]{#1}%
\providecommand \citenamefont [1]{#1}%
\providecommand \href@noop [0]{\@secondoftwo}%
\providecommand \href [0]{\begingroup \@sanitize@url \@href}%
\providecommand \@href[1]{\@@startlink{#1}\@@href}%
\providecommand \@@href[1]{\endgroup#1\@@endlink}%
\providecommand \@sanitize@url [0]{\catcode `\\12\catcode `\$12\catcode `\&12\catcode `\#12\catcode `\^12\catcode `\_12\catcode `\%12\relax}%
\providecommand \@@startlink[1]{}%
\providecommand \@@endlink[0]{}%
\providecommand \url  [0]{\begingroup\@sanitize@url \@url }%
\providecommand \@url [1]{\endgroup\@href {#1}{\urlprefix }}%
\providecommand \urlprefix  [0]{URL }%
\providecommand \Eprint [0]{\href }%
\providecommand \doibase [0]{https://doi.org/}%
\providecommand \selectlanguage [0]{\@gobble}%
\providecommand \bibinfo  [0]{\@secondoftwo}%
\providecommand \bibfield  [0]{\@secondoftwo}%
\providecommand \translation [1]{[#1]}%
\providecommand \BibitemOpen [0]{}%
\providecommand \bibitemStop [0]{}%
\providecommand \bibitemNoStop [0]{.\EOS\space}%
\providecommand \EOS [0]{\spacefactor3000\relax}%
\providecommand \BibitemShut  [1]{\csname bibitem#1\endcsname}%
\let\auto@bib@innerbib\@empty
\bibitem [{\citenamefont {{Amaro-Seoane}}\ \emph {et~al.}(2017)\citenamefont {{Amaro-Seoane}}, \citenamefont {{Audley}}, \citenamefont {{Babak}}, \citenamefont {{Baker}}, \citenamefont {{Barausse}}, \citenamefont {{Bender}}, \citenamefont {{Berti}}, \citenamefont {{Binetruy}}, \citenamefont {{Born}}, \citenamefont {{Bortoluzzi}}, \citenamefont {{Camp}}, \citenamefont {{Caprini}} \emph {et~al.}}]{LISA}%
  \BibitemOpen
  \bibfield  {author} {\bibinfo {author} {\bibfnamefont {P.}~\bibnamefont {{Amaro-Seoane}}}, \bibinfo {author} {\bibfnamefont {H.}~\bibnamefont {{Audley}}}, \bibinfo {author} {\bibfnamefont {S.}~\bibnamefont {{Babak}}}, \bibinfo {author} {\bibfnamefont {J.}~\bibnamefont {{Baker}}}, \bibinfo {author} {\bibfnamefont {E.}~\bibnamefont {{Barausse}}}, \bibinfo {author} {\bibfnamefont {P.}~\bibnamefont {{Bender}}}, \bibinfo {author} {\bibfnamefont {E.}~\bibnamefont {{Berti}}}, \bibinfo {author} {\bibfnamefont {P.}~\bibnamefont {{Binetruy}}}, \bibinfo {author} {\bibfnamefont {M.}~\bibnamefont {{Born}}}, \bibinfo {author} {\bibfnamefont {D.}~\bibnamefont {{Bortoluzzi}}}, \bibinfo {author} {\bibfnamefont {J.}~\bibnamefont {{Camp}}}, \bibinfo {author} {\bibfnamefont {C.}~\bibnamefont {{Caprini}}}, \emph {et~al.},\ }\bibfield  {title} {\bibinfo {title} {{Laser Interferometer Space Antenna}},\ }\href@noop {} {\bibfield  {journal} {\bibinfo  {journal} {arXiv e-prints}\ ,\ \bibinfo {eid} {arXiv:1702.00786}} (\bibinfo
  {year} {2017})},\ \Eprint {https://arxiv.org/abs/1702.00786} {arXiv:1702.00786 [astro-ph.IM]} \BibitemShut {NoStop}%
\bibitem [{\citenamefont {{Colpi}}\ \emph {et~al.}(2024)\citenamefont {{Colpi}}, \citenamefont {{Danzmann}}, \citenamefont {{Hewitson}} \emph {et~al.}}]{LISA:2024}%
  \BibitemOpen
  \bibfield  {author} {\bibinfo {author} {\bibfnamefont {M.}~\bibnamefont {{Colpi}}}, \bibinfo {author} {\bibfnamefont {K.}~\bibnamefont {{Danzmann}}}, \bibinfo {author} {\bibfnamefont {M.}~\bibnamefont {{Hewitson}}}, \emph {et~al.},\ }\bibfield  {title} {\bibinfo {title} {{LISA Definition Study Report}},\ }\href {https://doi.org/10.48550/arXiv.2402.07571} {\bibfield  {journal} {\bibinfo  {journal} {arXiv e-prints}\ ,\ \bibinfo {eid} {arXiv:2402.07571}} (\bibinfo {year} {2024})},\ \Eprint {https://arxiv.org/abs/2402.07571} {arXiv:2402.07571 [astro-ph.CO]} \BibitemShut {NoStop}%
\bibitem [{\citenamefont {{Luo}}\ \emph {et~al.}(2016)\citenamefont {{Luo}}, \citenamefont {{Chen}}, \citenamefont {{Duan}}, \citenamefont {{Gong}}, \citenamefont {{Hu}}, \citenamefont {{Ji}}, \citenamefont {{Liu}}, \citenamefont {{Mei}}, \citenamefont {{Milyukov}}, \citenamefont {{Sazhin}}, \citenamefont {{Shao}}, \citenamefont {{Toth}}, \citenamefont {{Tu}}, \citenamefont {{Wang}}, \citenamefont {{Wang}}, \citenamefont {{Yeh}}, \citenamefont {{Zhan}}, \citenamefont {{Zhang}}, \citenamefont {{Zharov}},\ and\ \citenamefont {{Zhou}}}]{TianQin}%
  \BibitemOpen
  \bibfield  {author} {\bibinfo {author} {\bibfnamefont {J.}~\bibnamefont {{Luo}}}, \bibinfo {author} {\bibfnamefont {L.-S.}\ \bibnamefont {{Chen}}}, \bibinfo {author} {\bibfnamefont {H.-Z.}\ \bibnamefont {{Duan}}}, \bibinfo {author} {\bibfnamefont {Y.-G.}\ \bibnamefont {{Gong}}}, \bibinfo {author} {\bibfnamefont {S.}~\bibnamefont {{Hu}}}, \bibinfo {author} {\bibfnamefont {J.}~\bibnamefont {{Ji}}}, \bibinfo {author} {\bibfnamefont {Q.}~\bibnamefont {{Liu}}}, \bibinfo {author} {\bibfnamefont {J.}~\bibnamefont {{Mei}}}, \bibinfo {author} {\bibfnamefont {V.}~\bibnamefont {{Milyukov}}}, \bibinfo {author} {\bibfnamefont {M.}~\bibnamefont {{Sazhin}}}, \bibinfo {author} {\bibfnamefont {C.-G.}\ \bibnamefont {{Shao}}}, \bibinfo {author} {\bibfnamefont {V.~T.}\ \bibnamefont {{Toth}}}, \bibinfo {author} {\bibfnamefont {H.-B.}\ \bibnamefont {{Tu}}}, \bibinfo {author} {\bibfnamefont {Y.}~\bibnamefont {{Wang}}}, \bibinfo {author} {\bibfnamefont {Y.}~\bibnamefont {{Wang}}}, \bibinfo {author} {\bibfnamefont {H.-C.}\
  \bibnamefont {{Yeh}}}, \bibinfo {author} {\bibfnamefont {M.-S.}\ \bibnamefont {{Zhan}}}, \bibinfo {author} {\bibfnamefont {Y.}~\bibnamefont {{Zhang}}}, \bibinfo {author} {\bibfnamefont {V.}~\bibnamefont {{Zharov}}},\ and\ \bibinfo {author} {\bibfnamefont {Z.-B.}\ \bibnamefont {{Zhou}}},\ }\bibfield  {title} {\bibinfo {title} {{TianQin: a space-borne gravitational wave detector}},\ }\href {https://doi.org/10.1088/0264-9381/33/3/035010} {\bibfield  {journal} {\bibinfo  {journal} {Classical and Quantum Gravity}\ }\textbf {\bibinfo {volume} {33}},\ \bibinfo {eid} {035010} (\bibinfo {year} {2016})},\ \Eprint {https://arxiv.org/abs/1512.02076} {arXiv:1512.02076 [astro-ph.IM]} \BibitemShut {NoStop}%
\bibitem [{\citenamefont {{Ruan}}\ \emph {et~al.}(2020)\citenamefont {{Ruan}}, \citenamefont {{Guo}}, \citenamefont {{Cai}},\ and\ \citenamefont {{Zhang}}}]{Taiji}%
  \BibitemOpen
  \bibfield  {author} {\bibinfo {author} {\bibfnamefont {W.-H.}\ \bibnamefont {{Ruan}}}, \bibinfo {author} {\bibfnamefont {Z.-K.}\ \bibnamefont {{Guo}}}, \bibinfo {author} {\bibfnamefont {R.-G.}\ \bibnamefont {{Cai}}},\ and\ \bibinfo {author} {\bibfnamefont {Y.-Z.}\ \bibnamefont {{Zhang}}},\ }\bibfield  {title} {\bibinfo {title} {{Taiji program: Gravitational-wave sources}},\ }\href {https://doi.org/10.1142/S0217751X2050075X} {\bibfield  {journal} {\bibinfo  {journal} {International Journal of Modern Physics A}\ }\textbf {\bibinfo {volume} {35}},\ \bibinfo {eid} {2050075} (\bibinfo {year} {2020})}\BibitemShut {NoStop}%
\bibitem [{\citenamefont {{Babak}}\ \emph {et~al.}(2017)\citenamefont {{Babak}}, \citenamefont {{Gair}}, \citenamefont {{Sesana}}, \citenamefont {{Barausse}}, \citenamefont {{Sopuerta}}, \citenamefont {{Berry}}, \citenamefont {{Berti}}, \citenamefont {{Amaro-Seoane}}, \citenamefont {{Petiteau}},\ and\ \citenamefont {{Klein}}}]{Babak:2017}%
  \BibitemOpen
  \bibfield  {author} {\bibinfo {author} {\bibfnamefont {S.}~\bibnamefont {{Babak}}}, \bibinfo {author} {\bibfnamefont {J.}~\bibnamefont {{Gair}}}, \bibinfo {author} {\bibfnamefont {A.}~\bibnamefont {{Sesana}}}, \bibinfo {author} {\bibfnamefont {E.}~\bibnamefont {{Barausse}}}, \bibinfo {author} {\bibfnamefont {C.~F.}\ \bibnamefont {{Sopuerta}}}, \bibinfo {author} {\bibfnamefont {C.~P.~L.}\ \bibnamefont {{Berry}}}, \bibinfo {author} {\bibfnamefont {E.}~\bibnamefont {{Berti}}}, \bibinfo {author} {\bibfnamefont {P.}~\bibnamefont {{Amaro-Seoane}}}, \bibinfo {author} {\bibfnamefont {A.}~\bibnamefont {{Petiteau}}},\ and\ \bibinfo {author} {\bibfnamefont {A.}~\bibnamefont {{Klein}}},\ }\bibfield  {title} {\bibinfo {title} {{Science with the space-based interferometer LISA. V. Extreme mass-ratio inspirals}},\ }\href {https://doi.org/10.1103/PhysRevD.95.103012} {\bibfield  {journal} {\bibinfo  {journal} {\prd}\ }\textbf {\bibinfo {volume} {95}},\ \bibinfo {eid} {103012} (\bibinfo {year} {2017})},\ \Eprint
  {https://arxiv.org/abs/1703.09722} {arXiv:1703.09722 [gr-qc]} \BibitemShut {NoStop}%
\bibitem [{\citenamefont {Barack}\ and\ \citenamefont {Cutler}(2007)}]{Barack:2006pq}%
  \BibitemOpen
  \bibfield  {author} {\bibinfo {author} {\bibfnamefont {L.}~\bibnamefont {Barack}}\ and\ \bibinfo {author} {\bibfnamefont {C.}~\bibnamefont {Cutler}},\ }\bibfield  {title} {\bibinfo {title} {{Using LISA EMRI sources to test off-Kerr deviations in the geometry of massive black holes}},\ }\href {https://doi.org/10.1103/PhysRevD.75.042003} {\bibfield  {journal} {\bibinfo  {journal} {Phys. Rev. D}\ }\textbf {\bibinfo {volume} {75}},\ \bibinfo {pages} {042003} (\bibinfo {year} {2007})},\ \Eprint {https://arxiv.org/abs/gr-qc/0612029} {arXiv:gr-qc/0612029} \BibitemShut {NoStop}%
\bibitem [{\citenamefont {Arun}\ \emph {et~al.}(2022)\citenamefont {Arun} \emph {et~al.}}]{LISAFunWG:2022}%
  \BibitemOpen
  \bibfield  {author} {\bibinfo {author} {\bibfnamefont {K.~G.}\ \bibnamefont {Arun}} \emph {et~al.} (\bibinfo {collaboration} {LISA}),\ }\bibfield  {title} {\bibinfo {title} {{New horizons for fundamental physics with LISA}},\ }\href {https://doi.org/10.1007/s41114-022-00036-9} {\bibfield  {journal} {\bibinfo  {journal} {Living Rev. Rel.}\ }\textbf {\bibinfo {volume} {25}},\ \bibinfo {pages} {4} (\bibinfo {year} {2022})},\ \Eprint {https://arxiv.org/abs/2205.01597} {arXiv:2205.01597 [gr-qc]} \BibitemShut {NoStop}%
\bibitem [{\citenamefont {Seoane}\ \emph {et~al.}(2023)\citenamefont {Seoane} \emph {et~al.}}]{LISAAstroWG:2022}%
  \BibitemOpen
  \bibfield  {author} {\bibinfo {author} {\bibfnamefont {P.~A.}\ \bibnamefont {Seoane}} \emph {et~al.} (\bibinfo {collaboration} {LISA}),\ }\bibfield  {title} {\bibinfo {title} {{Astrophysics with the Laser Interferometer Space Antenna}},\ }\href {https://doi.org/10.1007/s41114-022-00041-y} {\bibfield  {journal} {\bibinfo  {journal} {Living Rev. Rel.}\ }\textbf {\bibinfo {volume} {26}},\ \bibinfo {pages} {2} (\bibinfo {year} {2023})},\ \Eprint {https://arxiv.org/abs/2203.06016} {arXiv:2203.06016 [gr-qc]} \BibitemShut {NoStop}%
\bibitem [{\citenamefont {{LISA Consortium Waveform Working Group}}\ \emph {et~al.}(2023)\citenamefont {{LISA Consortium Waveform Working Group}} \emph {et~al.}}]{LISAWavWG:2023}%
  \BibitemOpen
  \bibfield  {author} {\bibinfo {author} {\bibnamefont {{LISA Consortium Waveform Working Group}}} \emph {et~al.},\ }\bibfield  {title} {\bibinfo {title} {{Waveform Modelling for the Laser Interferometer Space Antenna}},\ }\href {https://doi.org/10.48550/arXiv.2311.01300} {\bibfield  {journal} {\bibinfo  {journal} {arXiv e-prints}\ ,\ \bibinfo {eid} {arXiv:2311.01300}} (\bibinfo {year} {2023})},\ \Eprint {https://arxiv.org/abs/2311.01300} {arXiv:2311.01300 [gr-qc]} \BibitemShut {NoStop}%
\bibitem [{\citenamefont {Pound}\ and\ \citenamefont {Wardell}(2020)}]{Pound:2021}%
  \BibitemOpen
  \bibfield  {author} {\bibinfo {author} {\bibfnamefont {A.}~\bibnamefont {Pound}}\ and\ \bibinfo {author} {\bibfnamefont {B.}~\bibnamefont {Wardell}},\ }\bibinfo {title} {Black hole perturbation theory and gravitational self-force},\ in\ \href {https://doi.org/10.1007/978-981-15-4702-7_38-1} {\emph {\bibinfo {booktitle} {Handbook of Gravitational Wave Astronomy}}},\ \bibinfo {editor} {edited by\ \bibinfo {editor} {\bibfnamefont {C.}~\bibnamefont {Bambi}}, \bibinfo {editor} {\bibfnamefont {S.}~\bibnamefont {Katsanevas}},\ and\ \bibinfo {editor} {\bibfnamefont {K.~D.}\ \bibnamefont {Kokkotas}}}\ (\bibinfo  {publisher} {Springer Singapore},\ \bibinfo {address} {Singapore},\ \bibinfo {year} {2020})\ pp.\ \bibinfo {pages} {1--119}\BibitemShut {NoStop}%
\bibitem [{\citenamefont {{Hinderer}}\ and\ \citenamefont {{Flanagan}}(2008)}]{Hinderer:2008}%
  \BibitemOpen
  \bibfield  {author} {\bibinfo {author} {\bibfnamefont {T.}~\bibnamefont {{Hinderer}}}\ and\ \bibinfo {author} {\bibfnamefont {{\'E}.~{\'E}.}\ \bibnamefont {{Flanagan}}},\ }\bibfield  {title} {\bibinfo {title} {{Two-timescale analysis of extreme mass ratio inspirals in Kerr spacetime: Orbital motion}},\ }\href {https://doi.org/10.1103/PhysRevD.78.064028} {\bibfield  {journal} {\bibinfo  {journal} {\prd}\ }\textbf {\bibinfo {volume} {78}},\ \bibinfo {eid} {064028} (\bibinfo {year} {2008})},\ \Eprint {https://arxiv.org/abs/0805.3337} {arXiv:0805.3337 [gr-qc]} \BibitemShut {NoStop}%
\bibitem [{\citenamefont {{Mino}}(2003)}]{Mino:2003}%
  \BibitemOpen
  \bibfield  {author} {\bibinfo {author} {\bibfnamefont {Y.}~\bibnamefont {{Mino}}},\ }\bibfield  {title} {\bibinfo {title} {{Perturbative approach to an orbital evolution around a supermassive black hole}},\ }\href {https://doi.org/10.1103/PhysRevD.67.084027} {\bibfield  {journal} {\bibinfo  {journal} {\prd}\ }\textbf {\bibinfo {volume} {67}},\ \bibinfo {eid} {084027} (\bibinfo {year} {2003})},\ \Eprint {https://arxiv.org/abs/gr-qc/0302075} {arXiv:gr-qc/0302075 [gr-qc]} \BibitemShut {NoStop}%
\bibitem [{\citenamefont {{Sago}}\ \emph {et~al.}(2006)\citenamefont {{Sago}}, \citenamefont {{Tanaka}}, \citenamefont {{Hikida}}, \citenamefont {{Ganz}},\ and\ \citenamefont {{Nakano}}}]{Sago:2006}%
  \BibitemOpen
  \bibfield  {author} {\bibinfo {author} {\bibfnamefont {N.}~\bibnamefont {{Sago}}}, \bibinfo {author} {\bibfnamefont {T.}~\bibnamefont {{Tanaka}}}, \bibinfo {author} {\bibfnamefont {W.}~\bibnamefont {{Hikida}}}, \bibinfo {author} {\bibfnamefont {K.}~\bibnamefont {{Ganz}}},\ and\ \bibinfo {author} {\bibfnamefont {H.}~\bibnamefont {{Nakano}}},\ }\bibfield  {title} {\bibinfo {title} {{Adiabatic Evolution of Orbital Parameters in Kerr Spacetime}},\ }\href {https://doi.org/10.1143/PTP.115.873} {\bibfield  {journal} {\bibinfo  {journal} {Progress of Theoretical Physics}\ }\textbf {\bibinfo {volume} {115}},\ \bibinfo {pages} {873} (\bibinfo {year} {2006})},\ \Eprint {https://arxiv.org/abs/gr-qc/0511151} {arXiv:gr-qc/0511151 [gr-qc]} \BibitemShut {NoStop}%
\bibitem [{\citenamefont {{Akcay}}\ \emph {et~al.}(2020)\citenamefont {{Akcay}}, \citenamefont {{Dolan}}, \citenamefont {{Kavanagh}}, \citenamefont {{Moxon}}, \citenamefont {{Warburton}},\ and\ \citenamefont {{Wardell}}}]{Akcay:2020}%
  \BibitemOpen
  \bibfield  {author} {\bibinfo {author} {\bibfnamefont {S.}~\bibnamefont {{Akcay}}}, \bibinfo {author} {\bibfnamefont {S.~R.}\ \bibnamefont {{Dolan}}}, \bibinfo {author} {\bibfnamefont {C.}~\bibnamefont {{Kavanagh}}}, \bibinfo {author} {\bibfnamefont {J.}~\bibnamefont {{Moxon}}}, \bibinfo {author} {\bibfnamefont {N.}~\bibnamefont {{Warburton}}},\ and\ \bibinfo {author} {\bibfnamefont {B.}~\bibnamefont {{Wardell}}},\ }\bibfield  {title} {\bibinfo {title} {{Dissipation in extreme mass-ratio binaries with a spinning secondary}},\ }\href {https://doi.org/10.1103/PhysRevD.102.064013} {\bibfield  {journal} {\bibinfo  {journal} {Phys. Rev. D}\ }\textbf {\bibinfo {volume} {102}},\ \bibinfo {eid} {064013} (\bibinfo {year} {2020})},\ \Eprint {https://arxiv.org/abs/1912.09461} {arXiv:1912.09461 [gr-qc]} \BibitemShut {NoStop}%
\bibitem [{\citenamefont {{Warburton}}\ \emph {et~al.}(2021)\citenamefont {{Warburton}}, \citenamefont {{Pound}}, \citenamefont {{Wardell}}, \citenamefont {{Miller}},\ and\ \citenamefont {{Durkan}}}]{Warburton:2021}%
  \BibitemOpen
  \bibfield  {author} {\bibinfo {author} {\bibfnamefont {N.}~\bibnamefont {{Warburton}}}, \bibinfo {author} {\bibfnamefont {A.}~\bibnamefont {{Pound}}}, \bibinfo {author} {\bibfnamefont {B.}~\bibnamefont {{Wardell}}}, \bibinfo {author} {\bibfnamefont {J.}~\bibnamefont {{Miller}}},\ and\ \bibinfo {author} {\bibfnamefont {L.}~\bibnamefont {{Durkan}}},\ }\bibfield  {title} {\bibinfo {title} {{Gravitational-Wave Energy Flux for Compact Binaries through Second Order in the Mass Ratio}},\ }\href {https://doi.org/10.1103/PhysRevLett.127.151102} {\bibfield  {journal} {\bibinfo  {journal} {\prl}\ }\textbf {\bibinfo {volume} {127}},\ \bibinfo {eid} {151102} (\bibinfo {year} {2021})},\ \Eprint {https://arxiv.org/abs/2107.01298} {arXiv:2107.01298 [gr-qc]} \BibitemShut {NoStop}%
\bibitem [{\citenamefont {Miller}\ and\ \citenamefont {Pound}(2021)}]{Miller:2020bft}%
  \BibitemOpen
  \bibfield  {author} {\bibinfo {author} {\bibfnamefont {J.}~\bibnamefont {Miller}}\ and\ \bibinfo {author} {\bibfnamefont {A.}~\bibnamefont {Pound}},\ }\bibfield  {title} {\bibinfo {title} {{Two-timescale evolution of extreme-mass-ratio inspirals: waveform generation scheme for quasicircular orbits in Schwarzschild spacetime}},\ }\href {https://doi.org/10.1103/PhysRevD.103.064048} {\bibfield  {journal} {\bibinfo  {journal} {Phys. Rev. D}\ }\textbf {\bibinfo {volume} {103}},\ \bibinfo {pages} {064048} (\bibinfo {year} {2021})},\ \Eprint {https://arxiv.org/abs/2006.11263} {arXiv:2006.11263 [gr-qc]} \BibitemShut {NoStop}%
\bibitem [{\citenamefont {{Grant}}(2024)}]{Grant:2024ivt}%
  \BibitemOpen
  \bibfield  {author} {\bibinfo {author} {\bibfnamefont {A.~M.}\ \bibnamefont {{Grant}}},\ }\bibfield  {title} {\bibinfo {title} {{Flux-balance laws for spinning bodies under the gravitational self-force}},\ }\href {https://doi.org/10.48550/arXiv.2406.10343} {\bibfield  {journal} {\bibinfo  {journal} {arXiv e-prints}\ ,\ \bibinfo {eid} {arXiv:2406.10343}} (\bibinfo {year} {2024})},\ \Eprint {https://arxiv.org/abs/2406.10343} {arXiv:2406.10343 [gr-qc]} \BibitemShut {NoStop}%
\bibitem [{\citenamefont {{Blanchet}}(2014)}]{Blanchet:2014}%
  \BibitemOpen
  \bibfield  {author} {\bibinfo {author} {\bibfnamefont {L.}~\bibnamefont {{Blanchet}}},\ }\bibfield  {title} {\bibinfo {title} {{Gravitational Radiation from Post-Newtonian Sources and Inspiralling Compact Binaries}},\ }\href {https://doi.org/10.12942/lrr-2014-2} {\bibfield  {journal} {\bibinfo  {journal} {Living Reviews in Relativity}\ }\textbf {\bibinfo {volume} {17}},\ \bibinfo {eid} {2} (\bibinfo {year} {2014})},\ \Eprint {https://arxiv.org/abs/1310.1528} {arXiv:1310.1528 [gr-qc]} \BibitemShut {NoStop}%
\bibitem [{\citenamefont {{Henry}}\ and\ \citenamefont {{Khalil}}(2023)}]{Henry:2023}%
  \BibitemOpen
  \bibfield  {author} {\bibinfo {author} {\bibfnamefont {Q.}~\bibnamefont {{Henry}}}\ and\ \bibinfo {author} {\bibfnamefont {M.}~\bibnamefont {{Khalil}}},\ }\bibfield  {title} {\bibinfo {title} {{Spin effects in gravitational waveforms and fluxes for binaries on eccentric orbits to the third post-Newtonian order}},\ }\href {https://doi.org/10.1103/PhysRevD.108.104016} {\bibfield  {journal} {\bibinfo  {journal} {\prd}\ }\textbf {\bibinfo {volume} {108}},\ \bibinfo {eid} {104016} (\bibinfo {year} {2023})},\ \Eprint {https://arxiv.org/abs/2308.13606} {arXiv:2308.13606 [gr-qc]} \BibitemShut {NoStop}%
\bibitem [{\citenamefont {Peters}\ and\ \citenamefont {Mathews}(1963)}]{Peters:1963}%
  \BibitemOpen
  \bibfield  {author} {\bibinfo {author} {\bibfnamefont {P.~C.}\ \bibnamefont {Peters}}\ and\ \bibinfo {author} {\bibfnamefont {J.}~\bibnamefont {Mathews}},\ }\bibfield  {title} {\bibinfo {title} {Gravitational radiation from point masses in a keplerian orbit},\ }\href {https://doi.org/10.1103/PhysRev.131.435} {\bibfield  {journal} {\bibinfo  {journal} {Phys. Rev.}\ }\textbf {\bibinfo {volume} {131}},\ \bibinfo {pages} {435} (\bibinfo {year} {1963})}\BibitemShut {NoStop}%
\bibitem [{\citenamefont {Blanchet}\ \emph {et~al.}(2023)\citenamefont {Blanchet}, \citenamefont {Faye}, \citenamefont {Henry}, \citenamefont {Larrouturou},\ and\ \citenamefont {Trestini}}]{Blanchet:2023bwj}%
  \BibitemOpen
  \bibfield  {author} {\bibinfo {author} {\bibfnamefont {L.}~\bibnamefont {Blanchet}}, \bibinfo {author} {\bibfnamefont {G.}~\bibnamefont {Faye}}, \bibinfo {author} {\bibfnamefont {Q.}~\bibnamefont {Henry}}, \bibinfo {author} {\bibfnamefont {F.}~\bibnamefont {Larrouturou}},\ and\ \bibinfo {author} {\bibfnamefont {D.}~\bibnamefont {Trestini}},\ }\bibfield  {title} {\bibinfo {title} {{Gravitational-Wave Phasing of Quasicircular Compact Binary Systems to the Fourth-and-a-Half Post-Newtonian Order}},\ }\href {https://doi.org/10.1103/PhysRevLett.131.121402} {\bibfield  {journal} {\bibinfo  {journal} {Phys. Rev. Lett.}\ }\textbf {\bibinfo {volume} {131}},\ \bibinfo {pages} {121402} (\bibinfo {year} {2023})},\ \Eprint {https://arxiv.org/abs/2304.11185} {arXiv:2304.11185 [gr-qc]} \BibitemShut {NoStop}%
\bibitem [{\citenamefont {{Mino}}\ \emph {et~al.}(1997)\citenamefont {{Mino}}, \citenamefont {{Sasaki}}, \citenamefont {{Shibata}}, \citenamefont {{Tagoshi}},\ and\ \citenamefont {{Tanaka}}}]{Mino:1997}%
  \BibitemOpen
  \bibfield  {author} {\bibinfo {author} {\bibfnamefont {Y.}~\bibnamefont {{Mino}}}, \bibinfo {author} {\bibfnamefont {M.}~\bibnamefont {{Sasaki}}}, \bibinfo {author} {\bibfnamefont {M.}~\bibnamefont {{Shibata}}}, \bibinfo {author} {\bibfnamefont {H.}~\bibnamefont {{Tagoshi}}},\ and\ \bibinfo {author} {\bibfnamefont {T.}~\bibnamefont {{Tanaka}}},\ }\bibfield  {title} {\bibinfo {title} {{Chapter 1. Black Hole Perturbation}},\ }\href {https://doi.org/10.1143/PTPS.128.1} {\bibfield  {journal} {\bibinfo  {journal} {Progress of Theoretical Physics Supplement}\ }\textbf {\bibinfo {volume} {128}},\ \bibinfo {pages} {1} (\bibinfo {year} {1997})},\ \Eprint {https://arxiv.org/abs/gr-qc/9712057} {arXiv:gr-qc/9712057 [gr-qc]} \BibitemShut {NoStop}%
\bibitem [{\citenamefont {{Sasaki}}\ and\ \citenamefont {{Tagoshi}}(2003)}]{Sasaki:2003}%
  \BibitemOpen
  \bibfield  {author} {\bibinfo {author} {\bibfnamefont {M.}~\bibnamefont {{Sasaki}}}\ and\ \bibinfo {author} {\bibfnamefont {H.}~\bibnamefont {{Tagoshi}}},\ }\bibfield  {title} {\bibinfo {title} {{Analytic Black Hole Perturbation Approach to Gravitational Radiation}},\ }\href {https://doi.org/10.12942/lrr-2003-6} {\bibfield  {journal} {\bibinfo  {journal} {Living Reviews in Relativity}\ }\textbf {\bibinfo {volume} {6}},\ \bibinfo {eid} {6} (\bibinfo {year} {2003})},\ \Eprint {https://arxiv.org/abs/gr-qc/0306120} {arXiv:gr-qc/0306120 [gr-qc]} \BibitemShut {NoStop}%
\bibitem [{\citenamefont {Bini}\ \emph {et~al.}(2019)\citenamefont {Bini}, \citenamefont {Damour},\ and\ \citenamefont {Geralico}}]{Bini:2019nra}%
  \BibitemOpen
  \bibfield  {author} {\bibinfo {author} {\bibfnamefont {D.}~\bibnamefont {Bini}}, \bibinfo {author} {\bibfnamefont {T.}~\bibnamefont {Damour}},\ and\ \bibinfo {author} {\bibfnamefont {A.}~\bibnamefont {Geralico}},\ }\bibfield  {title} {\bibinfo {title} {{Novel approach to binary dynamics: application to the fifth post-Newtonian level}},\ }\href {https://doi.org/10.1103/PhysRevLett.123.231104} {\bibfield  {journal} {\bibinfo  {journal} {Phys. Rev. Lett.}\ }\textbf {\bibinfo {volume} {123}},\ \bibinfo {pages} {231104} (\bibinfo {year} {2019})},\ \Eprint {https://arxiv.org/abs/1909.02375} {arXiv:1909.02375 [gr-qc]} \BibitemShut {NoStop}%
\bibitem [{\citenamefont {Bl\"umlein}\ \emph {et~al.}(2021)\citenamefont {Bl\"umlein}, \citenamefont {Maier}, \citenamefont {Marquard},\ and\ \citenamefont {Sch\"afer}}]{Blumlein:2020pyo}%
  \BibitemOpen
  \bibfield  {author} {\bibinfo {author} {\bibfnamefont {J.}~\bibnamefont {Bl\"umlein}}, \bibinfo {author} {\bibfnamefont {A.}~\bibnamefont {Maier}}, \bibinfo {author} {\bibfnamefont {P.}~\bibnamefont {Marquard}},\ and\ \bibinfo {author} {\bibfnamefont {G.}~\bibnamefont {Sch\"afer}},\ }\bibfield  {title} {\bibinfo {title} {{The fifth-order post-Newtonian Hamiltonian dynamics of two-body systems from an effective field theory approach: potential contributions}},\ }\href {https://doi.org/10.1016/j.nuclphysb.2021.115352} {\bibfield  {journal} {\bibinfo  {journal} {Nucl. Phys. B}\ }\textbf {\bibinfo {volume} {965}},\ \bibinfo {pages} {115352} (\bibinfo {year} {2021})},\ \Eprint {https://arxiv.org/abs/2010.13672} {arXiv:2010.13672 [gr-qc]} \BibitemShut {NoStop}%
\bibitem [{\citenamefont {Bl\"umlein}\ \emph {et~al.}(2022)\citenamefont {Bl\"umlein}, \citenamefont {Maier}, \citenamefont {Marquard},\ and\ \citenamefont {Sch\"afer}}]{Blumlein:2021txe}%
  \BibitemOpen
  \bibfield  {author} {\bibinfo {author} {\bibfnamefont {J.}~\bibnamefont {Bl\"umlein}}, \bibinfo {author} {\bibfnamefont {A.}~\bibnamefont {Maier}}, \bibinfo {author} {\bibfnamefont {P.}~\bibnamefont {Marquard}},\ and\ \bibinfo {author} {\bibfnamefont {G.}~\bibnamefont {Sch\"afer}},\ }\bibfield  {title} {\bibinfo {title} {{The fifth-order post-Newtonian Hamiltonian dynamics of two-body systems from an effective field theory approach}},\ }\href {https://doi.org/10.1016/j.nuclphysb.2022.115900} {\bibfield  {journal} {\bibinfo  {journal} {Nucl. Phys. B}\ }\textbf {\bibinfo {volume} {983}},\ \bibinfo {pages} {115900} (\bibinfo {year} {2022})},\ \bibinfo {note} {[Erratum: Nucl.Phys.B 985, 115991 (2022)]},\ \Eprint {https://arxiv.org/abs/2110.13822} {arXiv:2110.13822 [gr-qc]} \BibitemShut {NoStop}%
\bibitem [{\citenamefont {{Bini}}\ and\ \citenamefont {{Damour}}(2024)}]{Bini:2024tft}%
  \BibitemOpen
  \bibfield  {author} {\bibinfo {author} {\bibfnamefont {D.}~\bibnamefont {{Bini}}}\ and\ \bibinfo {author} {\bibfnamefont {T.}~\bibnamefont {{Damour}}},\ }\bibfield  {title} {\bibinfo {title} {{Fourth post-Minkowskian local-in-time conservative dynamics of binary systems}},\ }\href {https://doi.org/10.1103/PhysRevD.110.064005} {\bibfield  {journal} {\bibinfo  {journal} {\prd}\ }\textbf {\bibinfo {volume} {110}},\ \bibinfo {eid} {064005} (\bibinfo {year} {2024})},\ \Eprint {https://arxiv.org/abs/2406.04878} {arXiv:2406.04878 [gr-qc]} \BibitemShut {NoStop}%
\bibitem [{\citenamefont {Buonanno}\ and\ \citenamefont {Damour}(1999)}]{Buonanno:1998gg}%
  \BibitemOpen
  \bibfield  {author} {\bibinfo {author} {\bibfnamefont {A.}~\bibnamefont {Buonanno}}\ and\ \bibinfo {author} {\bibfnamefont {T.}~\bibnamefont {Damour}},\ }\bibfield  {title} {\bibinfo {title} {{Effective one-body approach to general relativistic two-body dynamics}},\ }\href {https://doi.org/10.1103/PhysRevD.59.084006} {\bibfield  {journal} {\bibinfo  {journal} {Phys. Rev. D}\ }\textbf {\bibinfo {volume} {59}},\ \bibinfo {pages} {084006} (\bibinfo {year} {1999})},\ \Eprint {https://arxiv.org/abs/gr-qc/9811091} {arXiv:gr-qc/9811091} \BibitemShut {NoStop}%
\bibitem [{\citenamefont {Albertini}\ \emph {et~al.}(2022)\citenamefont {Albertini}, \citenamefont {Nagar}, \citenamefont {Pound}, \citenamefont {Warburton}, \citenamefont {Wardell}, \citenamefont {Durkan},\ and\ \citenamefont {Miller}}]{Albertini:2022rfe}%
  \BibitemOpen
  \bibfield  {author} {\bibinfo {author} {\bibfnamefont {A.}~\bibnamefont {Albertini}}, \bibinfo {author} {\bibfnamefont {A.}~\bibnamefont {Nagar}}, \bibinfo {author} {\bibfnamefont {A.}~\bibnamefont {Pound}}, \bibinfo {author} {\bibfnamefont {N.}~\bibnamefont {Warburton}}, \bibinfo {author} {\bibfnamefont {B.}~\bibnamefont {Wardell}}, \bibinfo {author} {\bibfnamefont {L.}~\bibnamefont {Durkan}},\ and\ \bibinfo {author} {\bibfnamefont {J.}~\bibnamefont {Miller}},\ }\bibfield  {title} {\bibinfo {title} {{Comparing second-order gravitational self-force, numerical relativity, and effective one body waveforms from inspiralling, quasicircular, and nonspinning black hole binaries}},\ }\href {https://doi.org/10.1103/PhysRevD.106.084061} {\bibfield  {journal} {\bibinfo  {journal} {Phys. Rev. D}\ }\textbf {\bibinfo {volume} {106}},\ \bibinfo {pages} {084061} (\bibinfo {year} {2022})},\ \Eprint {https://arxiv.org/abs/2208.01049} {arXiv:2208.01049 [gr-qc]} \BibitemShut {NoStop}%
\bibitem [{\citenamefont {van~de Meent}\ \emph {et~al.}(2023)\citenamefont {van~de Meent}, \citenamefont {Buonanno}, \citenamefont {Mihaylov}, \citenamefont {Ossokine}, \citenamefont {Pompili}, \citenamefont {Warburton}, \citenamefont {Pound}, \citenamefont {Wardell}, \citenamefont {Durkan},\ and\ \citenamefont {Miller}}]{vandeMeent:2023ols}%
  \BibitemOpen
  \bibfield  {author} {\bibinfo {author} {\bibfnamefont {M.}~\bibnamefont {van~de Meent}}, \bibinfo {author} {\bibfnamefont {A.}~\bibnamefont {Buonanno}}, \bibinfo {author} {\bibfnamefont {D.~P.}\ \bibnamefont {Mihaylov}}, \bibinfo {author} {\bibfnamefont {S.}~\bibnamefont {Ossokine}}, \bibinfo {author} {\bibfnamefont {L.}~\bibnamefont {Pompili}}, \bibinfo {author} {\bibfnamefont {N.}~\bibnamefont {Warburton}}, \bibinfo {author} {\bibfnamefont {A.}~\bibnamefont {Pound}}, \bibinfo {author} {\bibfnamefont {B.}~\bibnamefont {Wardell}}, \bibinfo {author} {\bibfnamefont {L.}~\bibnamefont {Durkan}},\ and\ \bibinfo {author} {\bibfnamefont {J.}~\bibnamefont {Miller}},\ }\bibfield  {title} {\bibinfo {title} {{Enhancing the SEOBNRv5 effective-one-body waveform model with second-order gravitational self-force fluxes}},\ }\href {https://doi.org/10.1103/PhysRevD.108.124038} {\bibfield  {journal} {\bibinfo  {journal} {Phys. Rev. D}\ }\textbf {\bibinfo {volume} {108}},\ \bibinfo {pages} {124038} (\bibinfo {year} {2023})},\
  \Eprint {https://arxiv.org/abs/2303.18026} {arXiv:2303.18026 [gr-qc]} \BibitemShut {NoStop}%
\bibitem [{\citenamefont {Barausse}\ and\ \citenamefont {Buonanno}(2010)}]{Barausse:2009xi}%
  \BibitemOpen
  \bibfield  {author} {\bibinfo {author} {\bibfnamefont {E.}~\bibnamefont {Barausse}}\ and\ \bibinfo {author} {\bibfnamefont {A.}~\bibnamefont {Buonanno}},\ }\bibfield  {title} {\bibinfo {title} {{An Improved effective-one-body Hamiltonian for spinning black-hole binaries}},\ }\href {https://doi.org/10.1103/PhysRevD.81.084024} {\bibfield  {journal} {\bibinfo  {journal} {Phys. Rev. D}\ }\textbf {\bibinfo {volume} {81}},\ \bibinfo {pages} {084024} (\bibinfo {year} {2010})},\ \Eprint {https://arxiv.org/abs/0912.3517} {arXiv:0912.3517 [gr-qc]} \BibitemShut {NoStop}%
\bibitem [{\citenamefont {{Nagar}}\ \emph {et~al.}(2019)\citenamefont {{Nagar}}, \citenamefont {{Messina}}, \citenamefont {{Kavanagh}}, \citenamefont {{Lukes-Gerakopoulos}}, \citenamefont {{Warburton}}, \citenamefont {{Bernuzzi}},\ and\ \citenamefont {{Harms}}}]{Nagar:2019}%
  \BibitemOpen
  \bibfield  {author} {\bibinfo {author} {\bibfnamefont {A.}~\bibnamefont {{Nagar}}}, \bibinfo {author} {\bibfnamefont {F.}~\bibnamefont {{Messina}}}, \bibinfo {author} {\bibfnamefont {C.}~\bibnamefont {{Kavanagh}}}, \bibinfo {author} {\bibfnamefont {G.}~\bibnamefont {{Lukes-Gerakopoulos}}}, \bibinfo {author} {\bibfnamefont {N.}~\bibnamefont {{Warburton}}}, \bibinfo {author} {\bibfnamefont {S.}~\bibnamefont {{Bernuzzi}}},\ and\ \bibinfo {author} {\bibfnamefont {E.}~\bibnamefont {{Harms}}},\ }\bibfield  {title} {\bibinfo {title} {{Factorization and resummation: A new paradigm to improve gravitational wave amplitudes. III. The spinning test-body terms}},\ }\href {https://doi.org/10.1103/PhysRevD.100.104056} {\bibfield  {journal} {\bibinfo  {journal} {\prd}\ }\textbf {\bibinfo {volume} {100}},\ \bibinfo {eid} {104056} (\bibinfo {year} {2019})},\ \Eprint {https://arxiv.org/abs/1907.12233} {arXiv:1907.12233 [gr-qc]} \BibitemShut {NoStop}%
\bibitem [{\citenamefont {{Albertini}}\ \emph {et~al.}(2024)\citenamefont {{Albertini}}, \citenamefont {{Nagar}}, \citenamefont {{Mathews}},\ and\ \citenamefont {{Lukes-Gerakopoulos}}}]{Albertini:2024rrs}%
  \BibitemOpen
  \bibfield  {author} {\bibinfo {author} {\bibfnamefont {A.}~\bibnamefont {{Albertini}}}, \bibinfo {author} {\bibfnamefont {A.}~\bibnamefont {{Nagar}}}, \bibinfo {author} {\bibfnamefont {J.}~\bibnamefont {{Mathews}}},\ and\ \bibinfo {author} {\bibfnamefont {G.}~\bibnamefont {{Lukes-Gerakopoulos}}},\ }\bibfield  {title} {\bibinfo {title} {{Comparing second-order gravitational self-force and effective-one-body waveforms from inspiralling, quasicircular black hole binaries with a nonspinning primary and a spinning secondary}},\ }\href {https://doi.org/10.1103/PhysRevD.110.044034} {\bibfield  {journal} {\bibinfo  {journal} {\prd}\ }\textbf {\bibinfo {volume} {110}},\ \bibinfo {eid} {044034} (\bibinfo {year} {2024})},\ \Eprint {https://arxiv.org/abs/2406.04108} {arXiv:2406.04108 [gr-qc]} \BibitemShut {NoStop}%
\bibitem [{\citenamefont {Poisson}(1993)}]{Poisson:1993}%
  \BibitemOpen
  \bibfield  {author} {\bibinfo {author} {\bibfnamefont {E.}~\bibnamefont {Poisson}},\ }\bibfield  {title} {\bibinfo {title} {Gravitational radiation from a particle in circular orbit around a black hole. i. analytical results for the nonrotating case},\ }\href {https://doi.org/10.1103/PhysRevD.47.1497} {\bibfield  {journal} {\bibinfo  {journal} {Phys. Rev. D}\ }\textbf {\bibinfo {volume} {47}},\ \bibinfo {pages} {1497} (\bibinfo {year} {1993})}\BibitemShut {NoStop}%
\bibitem [{\citenamefont {Tagoshi}\ and\ \citenamefont {Sasaki}(1994)}]{Tagoshi:1994}%
  \BibitemOpen
  \bibfield  {author} {\bibinfo {author} {\bibfnamefont {H.}~\bibnamefont {Tagoshi}}\ and\ \bibinfo {author} {\bibfnamefont {M.}~\bibnamefont {Sasaki}},\ }\bibfield  {title} {\bibinfo {title} {{Post-Newtonian Expansion of Gravitational Waves from a Particle in Circular Orbit around a Schwarzschild Black Hole}},\ }\href {https://doi.org/10.1143/ptp/92.4.745} {\bibfield  {journal} {\bibinfo  {journal} {Progress of Theoretical Physics}\ }\textbf {\bibinfo {volume} {92}},\ \bibinfo {pages} {745} (\bibinfo {year} {1994})},\ \Eprint {https://arxiv.org/abs/https://academic.oup.com/ptp/article-pdf/92/4/745/5358195/92-4-745.pdf} {https://academic.oup.com/ptp/article-pdf/92/4/745/5358195/92-4-745.pdf} \BibitemShut {NoStop}%
\bibitem [{\citenamefont {{Poisson}}\ and\ \citenamefont {{Sasaki}}(1995)}]{Poisson:1995}%
  \BibitemOpen
  \bibfield  {author} {\bibinfo {author} {\bibfnamefont {E.}~\bibnamefont {{Poisson}}}\ and\ \bibinfo {author} {\bibfnamefont {M.}~\bibnamefont {{Sasaki}}},\ }\bibfield  {title} {\bibinfo {title} {{Gravitational radiation from a particle in circular orbit around a black hole. V. Black-hole absorption and tail corrections}},\ }\href {https://doi.org/10.1103/PhysRevD.51.5753} {\bibfield  {journal} {\bibinfo  {journal} {\prd}\ }\textbf {\bibinfo {volume} {51}},\ \bibinfo {pages} {5753} (\bibinfo {year} {1995})},\ \Eprint {https://arxiv.org/abs/gr-qc/9412027} {arXiv:gr-qc/9412027 [gr-qc]} \BibitemShut {NoStop}%
\bibitem [{\citenamefont {{Tanaka}}\ \emph {et~al.}(1996{\natexlab{a}})\citenamefont {{Tanaka}}, \citenamefont {{Tagoshi}},\ and\ \citenamefont {{Sasaki}}}]{Tanaka:1996}%
  \BibitemOpen
  \bibfield  {author} {\bibinfo {author} {\bibfnamefont {T.}~\bibnamefont {{Tanaka}}}, \bibinfo {author} {\bibfnamefont {H.}~\bibnamefont {{Tagoshi}}},\ and\ \bibinfo {author} {\bibfnamefont {M.}~\bibnamefont {{Sasaki}}},\ }\bibfield  {title} {\bibinfo {title} {{Gravitational Waves by a Particle in Circular Orbits around a Schwarzschild Black Hole --- 5.5 Post-Newtonian Formula ---}},\ }\href {https://doi.org/10.1143/PTP.96.1087} {\bibfield  {journal} {\bibinfo  {journal} {Progress of Theoretical Physics}\ }\textbf {\bibinfo {volume} {96}},\ \bibinfo {pages} {1087} (\bibinfo {year} {1996}{\natexlab{a}})},\ \Eprint {https://arxiv.org/abs/gr-qc/9701050} {arXiv:gr-qc/9701050 [gr-qc]} \BibitemShut {NoStop}%
\bibitem [{\citenamefont {Shibata}\ \emph {et~al.}(1995)\citenamefont {Shibata}, \citenamefont {Sasaki}, \citenamefont {Tagoshi},\ and\ \citenamefont {Tanaka}}]{Shibata:1995}%
  \BibitemOpen
  \bibfield  {author} {\bibinfo {author} {\bibfnamefont {M.}~\bibnamefont {Shibata}}, \bibinfo {author} {\bibfnamefont {M.}~\bibnamefont {Sasaki}}, \bibinfo {author} {\bibfnamefont {H.}~\bibnamefont {Tagoshi}},\ and\ \bibinfo {author} {\bibfnamefont {T.}~\bibnamefont {Tanaka}},\ }\bibfield  {title} {\bibinfo {title} {Gravitational waves from a particle orbiting around a rotating black hole: Post-newtonian expansion},\ }\href {https://doi.org/10.1103/PhysRevD.51.1646} {\bibfield  {journal} {\bibinfo  {journal} {Phys. Rev. D}\ }\textbf {\bibinfo {volume} {51}},\ \bibinfo {pages} {1646} (\bibinfo {year} {1995})}\BibitemShut {NoStop}%
\bibitem [{\citenamefont {{Tagoshi}}\ \emph {et~al.}(1996)\citenamefont {{Tagoshi}}, \citenamefont {{Shibata}}, \citenamefont {{Tanaka}},\ and\ \citenamefont {{Sasaki}}}]{Tagoshi:1996}%
  \BibitemOpen
  \bibfield  {author} {\bibinfo {author} {\bibfnamefont {H.}~\bibnamefont {{Tagoshi}}}, \bibinfo {author} {\bibfnamefont {M.}~\bibnamefont {{Shibata}}}, \bibinfo {author} {\bibfnamefont {T.}~\bibnamefont {{Tanaka}}},\ and\ \bibinfo {author} {\bibfnamefont {M.}~\bibnamefont {{Sasaki}}},\ }\bibfield  {title} {\bibinfo {title} {{Post-Newtonian expansion of gravitational waves from a particle in circular orbit around a rotating black hole: Up to O(v$^{8}$) beyond the quadrupole formula}},\ }\href {https://doi.org/10.1103/PhysRevD.54.1439} {\bibfield  {journal} {\bibinfo  {journal} {\prd}\ }\textbf {\bibinfo {volume} {54}},\ \bibinfo {pages} {1439} (\bibinfo {year} {1996})},\ \Eprint {https://arxiv.org/abs/gr-qc/9603028} {arXiv:gr-qc/9603028 [gr-qc]} \BibitemShut {NoStop}%
\bibitem [{\citenamefont {Tagoshi}\ \emph {et~al.}(1997)\citenamefont {Tagoshi}, \citenamefont {Mano},\ and\ \citenamefont {Takasugi}}]{Tagoshi:1997jy}%
  \BibitemOpen
  \bibfield  {author} {\bibinfo {author} {\bibfnamefont {H.}~\bibnamefont {Tagoshi}}, \bibinfo {author} {\bibfnamefont {S.}~\bibnamefont {Mano}},\ and\ \bibinfo {author} {\bibfnamefont {E.}~\bibnamefont {Takasugi}},\ }\bibfield  {title} {\bibinfo {title} {{PostNewtonian expansion of gravitational waves from a particle in circular orbits around a rotating black hole: Effects of black hole absorption}},\ }\href {https://doi.org/10.1143/PTP.98.829} {\bibfield  {journal} {\bibinfo  {journal} {Prog. Theor. Phys.}\ }\textbf {\bibinfo {volume} {98}},\ \bibinfo {pages} {829} (\bibinfo {year} {1997})},\ \Eprint {https://arxiv.org/abs/gr-qc/9711072} {arXiv:gr-qc/9711072} \BibitemShut {NoStop}%
\bibitem [{\citenamefont {{Fujita}}(2012)}]{Fujita:2012}%
  \BibitemOpen
  \bibfield  {author} {\bibinfo {author} {\bibfnamefont {R.}~\bibnamefont {{Fujita}}},\ }\bibfield  {title} {\bibinfo {title} {{Gravitational Waves from a Particle in Circular Orbits around a Schwarzschild Black Hole to the 22nd Post-Newtonian Order}},\ }\href {https://doi.org/10.1143/PTP.128.971} {\bibfield  {journal} {\bibinfo  {journal} {Progress of Theoretical Physics}\ }\textbf {\bibinfo {volume} {128}},\ \bibinfo {pages} {971} (\bibinfo {year} {2012})},\ \Eprint {https://arxiv.org/abs/1211.5535} {arXiv:1211.5535 [gr-qc]} \BibitemShut {NoStop}%
\bibitem [{\citenamefont {{Fujita}}(2015)}]{Fujita:2015}%
  \BibitemOpen
  \bibfield  {author} {\bibinfo {author} {\bibfnamefont {R.}~\bibnamefont {{Fujita}}},\ }\bibfield  {title} {\bibinfo {title} {{Gravitational waves from a particle in circular orbits around a rotating black hole to the 11th post-Newtonian order}},\ }\href {https://doi.org/10.1093/ptep/ptv012} {\bibfield  {journal} {\bibinfo  {journal} {Progress of Theoretical and Experimental Physics}\ }\textbf {\bibinfo {volume} {2015}},\ \bibinfo {eid} {033E01} (\bibinfo {year} {2015})},\ \Eprint {https://arxiv.org/abs/1412.5689} {arXiv:1412.5689 [gr-qc]} \BibitemShut {NoStop}%
\bibitem [{\citenamefont {{Tanaka}}\ \emph {et~al.}(1996{\natexlab{b}})\citenamefont {{Tanaka}}, \citenamefont {{Mino}}, \citenamefont {{Sasaki}},\ and\ \citenamefont {{Shibata}}}]{Tanaka:1996b}%
  \BibitemOpen
  \bibfield  {author} {\bibinfo {author} {\bibfnamefont {T.}~\bibnamefont {{Tanaka}}}, \bibinfo {author} {\bibfnamefont {Y.}~\bibnamefont {{Mino}}}, \bibinfo {author} {\bibfnamefont {M.}~\bibnamefont {{Sasaki}}},\ and\ \bibinfo {author} {\bibfnamefont {M.}~\bibnamefont {{Shibata}}},\ }\bibfield  {title} {\bibinfo {title} {{Gravitational waves from a spinning particle in circular orbits around a rotating black hole}},\ }\href {https://doi.org/10.1103/PhysRevD.54.3762} {\bibfield  {journal} {\bibinfo  {journal} {\prd}\ }\textbf {\bibinfo {volume} {54}},\ \bibinfo {pages} {3762} (\bibinfo {year} {1996}{\natexlab{b}})},\ \Eprint {https://arxiv.org/abs/gr-qc/9602038} {arXiv:gr-qc/9602038 [gr-qc]} \BibitemShut {NoStop}%
\bibitem [{\citenamefont {{Ganz}}\ \emph {et~al.}(2007)\citenamefont {{Ganz}}, \citenamefont {{Hikida}}, \citenamefont {{Nakano}}, \citenamefont {{Sago}},\ and\ \citenamefont {{Tanaka}}}]{Ganz:2007}%
  \BibitemOpen
  \bibfield  {author} {\bibinfo {author} {\bibfnamefont {K.}~\bibnamefont {{Ganz}}}, \bibinfo {author} {\bibfnamefont {W.}~\bibnamefont {{Hikida}}}, \bibinfo {author} {\bibfnamefont {H.}~\bibnamefont {{Nakano}}}, \bibinfo {author} {\bibfnamefont {N.}~\bibnamefont {{Sago}}},\ and\ \bibinfo {author} {\bibfnamefont {T.}~\bibnamefont {{Tanaka}}},\ }\bibfield  {title} {\bibinfo {title} {{Adiabatic Evolution of Three `Constants' of Motion for Greatly Inclined Orbits in Kerr Spacetime}},\ }\href {https://doi.org/10.1143/PTP.117.1041} {\bibfield  {journal} {\bibinfo  {journal} {Progress of Theoretical Physics}\ }\textbf {\bibinfo {volume} {117}},\ \bibinfo {pages} {1041} (\bibinfo {year} {2007})},\ \Eprint {https://arxiv.org/abs/gr-qc/0702054} {arXiv:gr-qc/0702054 [gr-qc]} \BibitemShut {NoStop}%
\bibitem [{\citenamefont {{Sago}}\ and\ \citenamefont {{Fujita}}(2015)}]{Sago:2015}%
  \BibitemOpen
  \bibfield  {author} {\bibinfo {author} {\bibfnamefont {N.}~\bibnamefont {{Sago}}}\ and\ \bibinfo {author} {\bibfnamefont {R.}~\bibnamefont {{Fujita}}},\ }\bibfield  {title} {\bibinfo {title} {{Calculation of radiation reaction effect on orbital parameters in Kerr spacetime}},\ }\href {https://doi.org/10.1093/ptep/ptv092} {\bibfield  {journal} {\bibinfo  {journal} {Progress of Theoretical and Experimental Physics}\ }\textbf {\bibinfo {volume} {2015}},\ \bibinfo {eid} {073E03} (\bibinfo {year} {2015})},\ \Eprint {https://arxiv.org/abs/1505.01600} {arXiv:1505.01600 [gr-qc]} \BibitemShut {NoStop}%
\bibitem [{\citenamefont {{Isoyama}}\ \emph {et~al.}(2022)\citenamefont {{Isoyama}}, \citenamefont {{Fujita}}, \citenamefont {{Chua}}, \citenamefont {{Nakano}}, \citenamefont {{Pound}},\ and\ \citenamefont {{Sago}}}]{Isoyama:2022}%
  \BibitemOpen
  \bibfield  {author} {\bibinfo {author} {\bibfnamefont {S.}~\bibnamefont {{Isoyama}}}, \bibinfo {author} {\bibfnamefont {R.}~\bibnamefont {{Fujita}}}, \bibinfo {author} {\bibfnamefont {A.~J.~K.}\ \bibnamefont {{Chua}}}, \bibinfo {author} {\bibfnamefont {H.}~\bibnamefont {{Nakano}}}, \bibinfo {author} {\bibfnamefont {A.}~\bibnamefont {{Pound}}},\ and\ \bibinfo {author} {\bibfnamefont {N.}~\bibnamefont {{Sago}}},\ }\bibfield  {title} {\bibinfo {title} {{Adiabatic Waveforms from Extreme-Mass-Ratio Inspirals: An Analytical Approach}},\ }\href {https://doi.org/10.1103/PhysRevLett.128.231101} {\bibfield  {journal} {\bibinfo  {journal} {\prl}\ }\textbf {\bibinfo {volume} {128}},\ \bibinfo {eid} {231101} (\bibinfo {year} {2022})},\ \Eprint {https://arxiv.org/abs/2111.05288} {arXiv:2111.05288 [gr-qc]} \BibitemShut {NoStop}%
\bibitem [{\citenamefont {{Forseth}}\ \emph {et~al.}(2016)\citenamefont {{Forseth}}, \citenamefont {{Evans}},\ and\ \citenamefont {{Hopper}}}]{Forseth:2016}%
  \BibitemOpen
  \bibfield  {author} {\bibinfo {author} {\bibfnamefont {E.}~\bibnamefont {{Forseth}}}, \bibinfo {author} {\bibfnamefont {C.~R.}\ \bibnamefont {{Evans}}},\ and\ \bibinfo {author} {\bibfnamefont {S.}~\bibnamefont {{Hopper}}},\ }\bibfield  {title} {\bibinfo {title} {{Eccentric-orbit extreme-mass-ratio inspiral gravitational wave energy fluxes to 7PN order}},\ }\href {https://doi.org/10.1103/PhysRevD.93.064058} {\bibfield  {journal} {\bibinfo  {journal} {\prd}\ }\textbf {\bibinfo {volume} {93}},\ \bibinfo {eid} {064058} (\bibinfo {year} {2016})},\ \Eprint {https://arxiv.org/abs/1512.03051} {arXiv:1512.03051 [gr-qc]} \BibitemShut {NoStop}%
\bibitem [{\citenamefont {{Munna}}\ and\ \citenamefont {{Evans}}(2019)}]{Munna:2019}%
  \BibitemOpen
  \bibfield  {author} {\bibinfo {author} {\bibfnamefont {C.}~\bibnamefont {{Munna}}}\ and\ \bibinfo {author} {\bibfnamefont {C.~R.}\ \bibnamefont {{Evans}}},\ }\bibfield  {title} {\bibinfo {title} {{Eccentric-orbit extreme-mass-ratio-inspiral radiation: Analytic forms of leading-logarithm and subleading-logarithm flux terms at high PN orders}},\ }\href {https://doi.org/10.1103/PhysRevD.100.104060} {\bibfield  {journal} {\bibinfo  {journal} {\prd}\ }\textbf {\bibinfo {volume} {100}},\ \bibinfo {eid} {104060} (\bibinfo {year} {2019})},\ \Eprint {https://arxiv.org/abs/1909.05877} {arXiv:1909.05877 [gr-qc]} \BibitemShut {NoStop}%
\bibitem [{\citenamefont {{Munna}}\ and\ \citenamefont {{Evans}}(2020)}]{Munna:2020a}%
  \BibitemOpen
  \bibfield  {author} {\bibinfo {author} {\bibfnamefont {C.}~\bibnamefont {{Munna}}}\ and\ \bibinfo {author} {\bibfnamefont {C.~R.}\ \bibnamefont {{Evans}}},\ }\bibfield  {title} {\bibinfo {title} {{Eccentric-orbit extreme-mass-ratio-inspiral radiation. II. 1PN correction to leading-logarithm and subleading-logarithm flux sequences and the entire perturbative 4PN flux}},\ }\href {https://doi.org/10.1103/PhysRevD.102.104006} {\bibfield  {journal} {\bibinfo  {journal} {\prd}\ }\textbf {\bibinfo {volume} {102}},\ \bibinfo {eid} {104006} (\bibinfo {year} {2020})},\ \Eprint {https://arxiv.org/abs/2009.01254} {arXiv:2009.01254 [gr-qc]} \BibitemShut {NoStop}%
\bibitem [{\citenamefont {{Munna}}\ \emph {et~al.}(2020)\citenamefont {{Munna}}, \citenamefont {{Evans}}, \citenamefont {{Hopper}},\ and\ \citenamefont {{Forseth}}}]{Munna:2020}%
  \BibitemOpen
  \bibfield  {author} {\bibinfo {author} {\bibfnamefont {C.}~\bibnamefont {{Munna}}}, \bibinfo {author} {\bibfnamefont {C.~R.}\ \bibnamefont {{Evans}}}, \bibinfo {author} {\bibfnamefont {S.}~\bibnamefont {{Hopper}}},\ and\ \bibinfo {author} {\bibfnamefont {E.}~\bibnamefont {{Forseth}}},\ }\bibfield  {title} {\bibinfo {title} {{Determination of new coefficients in the angular momentum and energy fluxes at infinity to 9PN order for eccentric Schwarzschild extreme-mass-ratio inspirals using mode-by-mode fitting}},\ }\href {https://doi.org/10.1103/PhysRevD.102.024047} {\bibfield  {journal} {\bibinfo  {journal} {\prd}\ }\textbf {\bibinfo {volume} {102}},\ \bibinfo {eid} {024047} (\bibinfo {year} {2020})},\ \Eprint {https://arxiv.org/abs/2005.03044} {arXiv:2005.03044 [gr-qc]} \BibitemShut {NoStop}%
\bibitem [{\citenamefont {{Munna}}(2020)}]{Munna:2020b}%
  \BibitemOpen
  \bibfield  {author} {\bibinfo {author} {\bibfnamefont {C.}~\bibnamefont {{Munna}}},\ }\bibfield  {title} {\bibinfo {title} {{Analytic post-Newtonian expansion of the energy and angular momentum radiated to infinity by eccentric-orbit nonspinning extreme-mass-ratio inspirals to the 19th order}},\ }\href {https://doi.org/10.1103/PhysRevD.102.124001} {\bibfield  {journal} {\bibinfo  {journal} {\prd}\ }\textbf {\bibinfo {volume} {102}},\ \bibinfo {eid} {124001} (\bibinfo {year} {2020})},\ \Eprint {https://arxiv.org/abs/2008.10622} {arXiv:2008.10622 [gr-qc]} \BibitemShut {NoStop}%
\bibitem [{\citenamefont {{Munna}}\ \emph {et~al.}(2023)\citenamefont {{Munna}}, \citenamefont {{Evans}},\ and\ \citenamefont {{Forseth}}}]{Munna:2023}%
  \BibitemOpen
  \bibfield  {author} {\bibinfo {author} {\bibfnamefont {C.}~\bibnamefont {{Munna}}}, \bibinfo {author} {\bibfnamefont {C.~R.}\ \bibnamefont {{Evans}}},\ and\ \bibinfo {author} {\bibfnamefont {E.}~\bibnamefont {{Forseth}}},\ }\bibfield  {title} {\bibinfo {title} {{Tidal heating and torquing of the primary black hole in eccentric-orbit, nonspinning, extreme-mass-ratio inspirals to 22PN order}},\ }\href {https://doi.org/10.1103/PhysRevD.108.044039} {\bibfield  {journal} {\bibinfo  {journal} {\prd}\ }\textbf {\bibinfo {volume} {108}},\ \bibinfo {eid} {044039} (\bibinfo {year} {2023})},\ \Eprint {https://arxiv.org/abs/2306.12481} {arXiv:2306.12481 [gr-qc]} \BibitemShut {NoStop}%
\bibitem [{\citenamefont {{Burke}}\ \emph {et~al.}(2024)\citenamefont {{Burke}}, \citenamefont {{Piovano}}, \citenamefont {{Warburton}}, \citenamefont {{Lynch}}, \citenamefont {{Speri}}, \citenamefont {{Kavanagh}}, \citenamefont {{Wardell}}, \citenamefont {{Pound}}, \citenamefont {{Durkan}},\ and\ \citenamefont {{Miller}}}]{Burke:2023}%
  \BibitemOpen
  \bibfield  {author} {\bibinfo {author} {\bibfnamefont {O.}~\bibnamefont {{Burke}}}, \bibinfo {author} {\bibfnamefont {G.~A.}\ \bibnamefont {{Piovano}}}, \bibinfo {author} {\bibfnamefont {N.}~\bibnamefont {{Warburton}}}, \bibinfo {author} {\bibfnamefont {P.}~\bibnamefont {{Lynch}}}, \bibinfo {author} {\bibfnamefont {L.}~\bibnamefont {{Speri}}}, \bibinfo {author} {\bibfnamefont {C.}~\bibnamefont {{Kavanagh}}}, \bibinfo {author} {\bibfnamefont {B.}~\bibnamefont {{Wardell}}}, \bibinfo {author} {\bibfnamefont {A.}~\bibnamefont {{Pound}}}, \bibinfo {author} {\bibfnamefont {L.}~\bibnamefont {{Durkan}}},\ and\ \bibinfo {author} {\bibfnamefont {J.}~\bibnamefont {{Miller}}},\ }\bibfield  {title} {\bibinfo {title} {{Assessing the importance of first postadiabatic terms for small-mass-ratio binaries}},\ }\href {https://doi.org/10.1103/PhysRevD.109.124048} {\bibfield  {journal} {\bibinfo  {journal} {\prd}\ }\textbf {\bibinfo {volume} {109}},\ \bibinfo {eid} {124048} (\bibinfo {year} {2024})},\ \Eprint
  {https://arxiv.org/abs/2310.08927} {arXiv:2310.08927 [gr-qc]} \BibitemShut {NoStop}%
\bibitem [{\citenamefont {Harms}\ \emph {et~al.}(2016)\citenamefont {Harms}, \citenamefont {Lukes-Gerakopoulos}, \citenamefont {Bernuzzi},\ and\ \citenamefont {Nagar}}]{Harms:2016ctx}%
  \BibitemOpen
  \bibfield  {author} {\bibinfo {author} {\bibfnamefont {E.}~\bibnamefont {Harms}}, \bibinfo {author} {\bibfnamefont {G.}~\bibnamefont {Lukes-Gerakopoulos}}, \bibinfo {author} {\bibfnamefont {S.}~\bibnamefont {Bernuzzi}},\ and\ \bibinfo {author} {\bibfnamefont {A.}~\bibnamefont {Nagar}},\ }\bibfield  {title} {\bibinfo {title} {{Spinning test body orbiting around a Schwarzschild black hole: Circular dynamics and gravitational-wave fluxes}},\ }\href {https://doi.org/10.1103/PhysRevD.94.104010} {\bibfield  {journal} {\bibinfo  {journal} {Phys. Rev. D}\ }\textbf {\bibinfo {volume} {94}},\ \bibinfo {pages} {104010} (\bibinfo {year} {2016})},\ \Eprint {https://arxiv.org/abs/1609.00356} {arXiv:1609.00356 [gr-qc]} \BibitemShut {NoStop}%
\bibitem [{\citenamefont {Lukes-Gerakopoulos}\ \emph {et~al.}(2017)\citenamefont {Lukes-Gerakopoulos}, \citenamefont {Harms}, \citenamefont {Bernuzzi},\ and\ \citenamefont {Nagar}}]{Lukes-Gerakopoulos:2017vkj}%
  \BibitemOpen
  \bibfield  {author} {\bibinfo {author} {\bibfnamefont {G.}~\bibnamefont {Lukes-Gerakopoulos}}, \bibinfo {author} {\bibfnamefont {E.}~\bibnamefont {Harms}}, \bibinfo {author} {\bibfnamefont {S.}~\bibnamefont {Bernuzzi}},\ and\ \bibinfo {author} {\bibfnamefont {A.}~\bibnamefont {Nagar}},\ }\bibfield  {title} {\bibinfo {title} {{Spinning test-body orbiting around a Kerr black hole: circular dynamics and gravitational-wave fluxes}},\ }\href {https://doi.org/10.1103/PhysRevD.96.064051} {\bibfield  {journal} {\bibinfo  {journal} {Phys. Rev. D}\ }\textbf {\bibinfo {volume} {96}},\ \bibinfo {pages} {064051} (\bibinfo {year} {2017})},\ \Eprint {https://arxiv.org/abs/1707.07537} {arXiv:1707.07537 [gr-qc]} \BibitemShut {NoStop}%
\bibitem [{\citenamefont {{Skoup{\'y}}}\ and\ \citenamefont {{Lukes-Gerakopoulos}}(2021)}]{Skoupy:2021}%
  \BibitemOpen
  \bibfield  {author} {\bibinfo {author} {\bibfnamefont {V.}~\bibnamefont {{Skoup{\'y}}}}\ and\ \bibinfo {author} {\bibfnamefont {G.}~\bibnamefont {{Lukes-Gerakopoulos}}},\ }\bibfield  {title} {\bibinfo {title} {{Spinning test body orbiting around a Kerr black hole: Eccentric equatorial orbits and their asymptotic gravitational-wave fluxes}},\ }\href {https://doi.org/10.1103/PhysRevD.103.104045} {\bibfield  {journal} {\bibinfo  {journal} {\prd}\ }\textbf {\bibinfo {volume} {103}},\ \bibinfo {eid} {104045} (\bibinfo {year} {2021})},\ \Eprint {https://arxiv.org/abs/2102.04819} {arXiv:2102.04819 [gr-qc]} \BibitemShut {NoStop}%
\bibitem [{\citenamefont {{Mathews}}\ \emph {et~al.}(2022)\citenamefont {{Mathews}}, \citenamefont {{Pound}},\ and\ \citenamefont {{Wardell}}}]{Mathews:2022}%
  \BibitemOpen
  \bibfield  {author} {\bibinfo {author} {\bibfnamefont {J.}~\bibnamefont {{Mathews}}}, \bibinfo {author} {\bibfnamefont {A.}~\bibnamefont {{Pound}}},\ and\ \bibinfo {author} {\bibfnamefont {B.}~\bibnamefont {{Wardell}}},\ }\bibfield  {title} {\bibinfo {title} {{Self-force calculations with a spinning secondary}},\ }\href {https://doi.org/10.1103/PhysRevD.105.084031} {\bibfield  {journal} {\bibinfo  {journal} {\prd}\ }\textbf {\bibinfo {volume} {105}},\ \bibinfo {eid} {084031} (\bibinfo {year} {2022})},\ \Eprint {https://arxiv.org/abs/2112.13069} {arXiv:2112.13069 [gr-qc]} \BibitemShut {NoStop}%
\bibitem [{\citenamefont {{Skoup{\'y}}}\ \emph {et~al.}(2023)\citenamefont {{Skoup{\'y}}}, \citenamefont {{Lukes-Gerakopoulos}}, \citenamefont {{Drummond}},\ and\ \citenamefont {{Hughes}}}]{Skoupy:2023}%
  \BibitemOpen
  \bibfield  {author} {\bibinfo {author} {\bibfnamefont {V.}~\bibnamefont {{Skoup{\'y}}}}, \bibinfo {author} {\bibfnamefont {G.}~\bibnamefont {{Lukes-Gerakopoulos}}}, \bibinfo {author} {\bibfnamefont {L.~V.}\ \bibnamefont {{Drummond}}},\ and\ \bibinfo {author} {\bibfnamefont {S.~A.}\ \bibnamefont {{Hughes}}},\ }\bibfield  {title} {\bibinfo {title} {{Asymptotic gravitational-wave fluxes from a spinning test body on generic orbits around a Kerr black hole}},\ }\href {https://doi.org/10.1103/PhysRevD.108.044041} {\bibfield  {journal} {\bibinfo  {journal} {\prd}\ }\textbf {\bibinfo {volume} {108}},\ \bibinfo {eid} {044041} (\bibinfo {year} {2023})},\ \Eprint {https://arxiv.org/abs/2303.16798} {arXiv:2303.16798 [gr-qc]} \BibitemShut {NoStop}%
\bibitem [{\citenamefont {{Witzany}}\ and\ \citenamefont {{Piovano}}(2024)}]{Witzany:2023}%
  \BibitemOpen
  \bibfield  {author} {\bibinfo {author} {\bibfnamefont {V.}~\bibnamefont {{Witzany}}}\ and\ \bibinfo {author} {\bibfnamefont {G.~A.}\ \bibnamefont {{Piovano}}},\ }\bibfield  {title} {\bibinfo {title} {{Analytic Solutions for the Motion of Spinning Particles near Spherically Symmetric Black Holes and Exotic Compact Objects}},\ }\href {https://doi.org/10.1103/PhysRevLett.132.171401} {\bibfield  {journal} {\bibinfo  {journal} {\prl}\ }\textbf {\bibinfo {volume} {132}},\ \bibinfo {eid} {171401} (\bibinfo {year} {2024})},\ \Eprint {https://arxiv.org/abs/2308.00021} {arXiv:2308.00021 [gr-qc]} \BibitemShut {NoStop}%
\bibitem [{Sup()}]{SupMat}%
  \BibitemOpen
  \href@noop {} {\bibinfo {title} {{See Supplemental Material at \url{http://link.aps.org/supplemental/10.1103/PhysRevD.110.084061} or arXiv ancillary file for a Mathematica notebook containing the PN expansions of the trajectory and the gravitational-wave fluxes.}}}\BibitemShut {Stop}%
\bibitem [{\citenamefont {{Katz}}\ \emph {et~al.}(2021)\citenamefont {{Katz}}, \citenamefont {{Chua}}, \citenamefont {{Speri}}, \citenamefont {{Warburton}},\ and\ \citenamefont {{Hughes}}}]{Katz:2021}%
  \BibitemOpen
  \bibfield  {author} {\bibinfo {author} {\bibfnamefont {M.~L.}\ \bibnamefont {{Katz}}}, \bibinfo {author} {\bibfnamefont {A.~J.~K.}\ \bibnamefont {{Chua}}}, \bibinfo {author} {\bibfnamefont {L.}~\bibnamefont {{Speri}}}, \bibinfo {author} {\bibfnamefont {N.}~\bibnamefont {{Warburton}}},\ and\ \bibinfo {author} {\bibfnamefont {S.~A.}\ \bibnamefont {{Hughes}}},\ }\bibfield  {title} {\bibinfo {title} {{Fast extreme-mass-ratio-inspiral waveforms: New tools for millihertz gravitational-wave data analysis}},\ }\href {https://doi.org/10.1103/PhysRevD.104.064047} {\bibfield  {journal} {\bibinfo  {journal} {\prd}\ }\textbf {\bibinfo {volume} {104}},\ \bibinfo {eid} {064047} (\bibinfo {year} {2021})},\ \Eprint {https://arxiv.org/abs/2104.04582} {arXiv:2104.04582 [gr-qc]} \BibitemShut {NoStop}%
\bibitem [{\citenamefont {Chua}\ \emph {et~al.}(2021)\citenamefont {Chua}, \citenamefont {Katz}, \citenamefont {Warburton},\ and\ \citenamefont {Hughes}}]{Chua:2020stf}%
  \BibitemOpen
  \bibfield  {author} {\bibinfo {author} {\bibfnamefont {A.~J.~K.}\ \bibnamefont {Chua}}, \bibinfo {author} {\bibfnamefont {M.~L.}\ \bibnamefont {Katz}}, \bibinfo {author} {\bibfnamefont {N.}~\bibnamefont {Warburton}},\ and\ \bibinfo {author} {\bibfnamefont {S.~A.}\ \bibnamefont {Hughes}},\ }\bibfield  {title} {\bibinfo {title} {{Rapid generation of fully relativistic extreme-mass-ratio-inspiral waveform templates for LISA data analysis}},\ }\href {https://doi.org/10.1103/PhysRevLett.126.051102} {\bibfield  {journal} {\bibinfo  {journal} {Phys. Rev. Lett.}\ }\textbf {\bibinfo {volume} {126}},\ \bibinfo {pages} {051102} (\bibinfo {year} {2021})},\ \Eprint {https://arxiv.org/abs/2008.06071} {arXiv:2008.06071 [gr-qc]} \BibitemShut {NoStop}%
\bibitem [{\citenamefont {Katz}\ \emph {et~al.}(2020)\citenamefont {Katz}, \citenamefont {Chua}, \citenamefont {Warburton},\ and\ \citenamefont {Hughes.}}]{michael_l_katz_2020_4005001}%
  \BibitemOpen
  \bibfield  {author} {\bibinfo {author} {\bibfnamefont {M.~L.}\ \bibnamefont {Katz}}, \bibinfo {author} {\bibfnamefont {A.~J.~K.}\ \bibnamefont {Chua}}, \bibinfo {author} {\bibfnamefont {N.}~\bibnamefont {Warburton}},\ and\ \bibinfo {author} {\bibfnamefont {S.~A.}\ \bibnamefont {Hughes.}},\ }\href {https://doi.org/10.5281/zenodo.4005001} {\bibinfo {title} {{BlackHolePerturbationToolkit/FastEMRIWaveforms: Official Release}}} (\bibinfo {year} {2020})\BibitemShut {NoStop}%
\bibitem [{\citenamefont {Chua}\ \emph {et~al.}(2019)\citenamefont {Chua}, \citenamefont {Galley},\ and\ \citenamefont {Vallisneri}}]{Chua:2018woh}%
  \BibitemOpen
  \bibfield  {author} {\bibinfo {author} {\bibfnamefont {A.~J.}\ \bibnamefont {Chua}}, \bibinfo {author} {\bibfnamefont {C.~R.}\ \bibnamefont {Galley}},\ and\ \bibinfo {author} {\bibfnamefont {M.}~\bibnamefont {Vallisneri}},\ }\bibfield  {title} {\bibinfo {title} {{Reduced-order modeling with artificial neurons for gravitational-wave inference}},\ }\href {https://doi.org/10.1103/PhysRevLett.122.211101} {\bibfield  {journal} {\bibinfo  {journal} {Phys. Rev. Lett.}\ }\textbf {\bibinfo {volume} {122}},\ \bibinfo {pages} {211101} (\bibinfo {year} {2019})},\ \Eprint {https://arxiv.org/abs/1811.05491} {arXiv:1811.05491 [astro-ph.IM]} \BibitemShut {NoStop}%
\bibitem [{\citenamefont {Rudiger}(1981)}]{Rudiger:1981c}%
  \BibitemOpen
  \bibfield  {author} {\bibinfo {author} {\bibfnamefont {R.}~\bibnamefont {Rudiger}},\ }\bibfield  {title} {\bibinfo {title} {Conserved quantities of spinning test particles in general relativity. i},\ }\href {https://doi.org/10.1098/rspa.1981.0046} {\bibfield  {journal} {\bibinfo  {journal} {Proceedings of the Royal Society of London A: Mathematical, Physical and Engineering Sciences}\ }\textbf {\bibinfo {volume} {375}},\ \bibinfo {pages} {185} (\bibinfo {year} {1981})}\BibitemShut {NoStop}%
\bibitem [{\citenamefont {{Witzany}}\ \emph {et~al.}(2019)\citenamefont {{Witzany}}, \citenamefont {{Steinhoff}},\ and\ \citenamefont {{Lukes-Gerakopoulos}}}]{Witzany:2019}%
  \BibitemOpen
  \bibfield  {author} {\bibinfo {author} {\bibfnamefont {V.}~\bibnamefont {{Witzany}}}, \bibinfo {author} {\bibfnamefont {J.}~\bibnamefont {{Steinhoff}}},\ and\ \bibinfo {author} {\bibfnamefont {G.}~\bibnamefont {{Lukes-Gerakopoulos}}},\ }\bibfield  {title} {\bibinfo {title} {{Hamiltonians and canonical coordinates for spinning particles in curved space-time}},\ }\href {https://doi.org/10.1088/1361-6382/ab002f} {\bibfield  {journal} {\bibinfo  {journal} {Classical and Quantum Gravity}\ }\textbf {\bibinfo {volume} {36}},\ \bibinfo {eid} {075003} (\bibinfo {year} {2019})},\ \Eprint {https://arxiv.org/abs/1808.06582} {arXiv:1808.06582 [gr-qc]} \BibitemShut {NoStop}%
\bibitem [{\citenamefont {Detweiler}\ and\ \citenamefont {Whiting}(2003)}]{Detweiler:2002mi}%
  \BibitemOpen
  \bibfield  {author} {\bibinfo {author} {\bibfnamefont {S.~L.}\ \bibnamefont {Detweiler}}\ and\ \bibinfo {author} {\bibfnamefont {B.~F.}\ \bibnamefont {Whiting}},\ }\bibfield  {title} {\bibinfo {title} {{Selfforce via a Green's function decomposition}},\ }\href {https://doi.org/10.1103/PhysRevD.67.024025} {\bibfield  {journal} {\bibinfo  {journal} {Phys. Rev. D}\ }\textbf {\bibinfo {volume} {67}},\ \bibinfo {pages} {024025} (\bibinfo {year} {2003})},\ \Eprint {https://arxiv.org/abs/gr-qc/0202086} {arXiv:gr-qc/0202086} \BibitemShut {NoStop}%
\bibitem [{\citenamefont {Harte}(2012)}]{Harte:2011ku}%
  \BibitemOpen
  \bibfield  {author} {\bibinfo {author} {\bibfnamefont {A.~I.}\ \bibnamefont {Harte}},\ }\bibfield  {title} {\bibinfo {title} {{Mechanics of extended masses in general relativity}},\ }\href {https://doi.org/10.1088/0264-9381/29/5/055012} {\bibfield  {journal} {\bibinfo  {journal} {Class. Quant. Grav.}\ }\textbf {\bibinfo {volume} {29}},\ \bibinfo {pages} {055012} (\bibinfo {year} {2012})},\ \Eprint {https://arxiv.org/abs/1103.0543} {arXiv:1103.0543 [gr-qc]} \BibitemShut {NoStop}%
\bibitem [{\citenamefont {{Skoup{\'y}}}\ and\ \citenamefont {{Lukes-Gerakopoulos}}(2022)}]{Skoupy:2022}%
  \BibitemOpen
  \bibfield  {author} {\bibinfo {author} {\bibfnamefont {V.}~\bibnamefont {{Skoup{\'y}}}}\ and\ \bibinfo {author} {\bibfnamefont {G.}~\bibnamefont {{Lukes-Gerakopoulos}}},\ }\bibfield  {title} {\bibinfo {title} {{Adiabatic equatorial inspirals of a spinning body into a Kerr black hole}},\ }\href {https://doi.org/10.1103/PhysRevD.105.084033} {\bibfield  {journal} {\bibinfo  {journal} {\prd}\ }\textbf {\bibinfo {volume} {105}},\ \bibinfo {eid} {084033} (\bibinfo {year} {2022})},\ \Eprint {https://arxiv.org/abs/2201.07044} {arXiv:2201.07044 [gr-qc]} \BibitemShut {NoStop}%
\bibitem [{\citenamefont {Warburton}\ \emph {et~al.}(2023)\citenamefont {Warburton}, \citenamefont {Wardell}, \citenamefont {Munna},\ and\ \citenamefont {Kavanagh}}]{PNSelfForceZenodo}%
  \BibitemOpen
  \bibfield  {author} {\bibinfo {author} {\bibfnamefont {N.}~\bibnamefont {Warburton}}, \bibinfo {author} {\bibfnamefont {B.}~\bibnamefont {Wardell}}, \bibinfo {author} {\bibfnamefont {C.}~\bibnamefont {Munna}},\ and\ \bibinfo {author} {\bibfnamefont {C.}~\bibnamefont {Kavanagh}},\ }\href {https://doi.org/10.5281/zenodo.8112975} {\bibinfo {title} {Postnewtonianselfforce}} (\bibinfo {year} {2023})\BibitemShut {NoStop}%
\bibitem [{\citenamefont {{Loutrel}}\ \emph {et~al.}(2024)\citenamefont {{Loutrel}}, \citenamefont {{Mukherjee}}, \citenamefont {{Maselli}},\ and\ \citenamefont {{Pani}}}]{Loutrel:2024}%
  \BibitemOpen
  \bibfield  {author} {\bibinfo {author} {\bibfnamefont {N.}~\bibnamefont {{Loutrel}}}, \bibinfo {author} {\bibfnamefont {S.}~\bibnamefont {{Mukherjee}}}, \bibinfo {author} {\bibfnamefont {A.}~\bibnamefont {{Maselli}}},\ and\ \bibinfo {author} {\bibfnamefont {P.}~\bibnamefont {{Pani}}},\ }\bibfield  {title} {\bibinfo {title} {{Analytical model of precessing binaries using post-Newtonian theory in the extreme mass-ratio limit: General formalism}},\ }\href {https://doi.org/10.1103/PhysRevD.110.024006} {\bibfield  {journal} {\bibinfo  {journal} {\prd}\ }\textbf {\bibinfo {volume} {110}},\ \bibinfo {eid} {024006} (\bibinfo {year} {2024})},\ \Eprint {https://arxiv.org/abs/2402.08883} {arXiv:2402.08883 [gr-qc]} \BibitemShut {NoStop}%
\bibitem [{\citenamefont {Isoyama}\ \emph {et~al.}(2013)\citenamefont {Isoyama}, \citenamefont {Fujita}, \citenamefont {Sago}, \citenamefont {Tagoshi},\ and\ \citenamefont {Tanaka}}]{Isoyama:2012bx}%
  \BibitemOpen
  \bibfield  {author} {\bibinfo {author} {\bibfnamefont {S.}~\bibnamefont {Isoyama}}, \bibinfo {author} {\bibfnamefont {R.}~\bibnamefont {Fujita}}, \bibinfo {author} {\bibfnamefont {N.}~\bibnamefont {Sago}}, \bibinfo {author} {\bibfnamefont {H.}~\bibnamefont {Tagoshi}},\ and\ \bibinfo {author} {\bibfnamefont {T.}~\bibnamefont {Tanaka}},\ }\bibfield  {title} {\bibinfo {title} {{Impact of the second-order self-forces on the dephasing of the gravitational waves from quasicircular extreme mass-ratio inspirals}},\ }\href {https://doi.org/10.1103/PhysRevD.87.024010} {\bibfield  {journal} {\bibinfo  {journal} {Phys. Rev. D}\ }\textbf {\bibinfo {volume} {87}},\ \bibinfo {pages} {024010} (\bibinfo {year} {2013})},\ \Eprint {https://arxiv.org/abs/1210.2569} {arXiv:1210.2569 [gr-qc]} \BibitemShut {NoStop}%
\bibitem [{\citenamefont {Lindblom}\ \emph {et~al.}(2008)\citenamefont {Lindblom}, \citenamefont {Owen},\ and\ \citenamefont {Brown}}]{Lindblom:2008cm}%
  \BibitemOpen
  \bibfield  {author} {\bibinfo {author} {\bibfnamefont {L.}~\bibnamefont {Lindblom}}, \bibinfo {author} {\bibfnamefont {B.~J.}\ \bibnamefont {Owen}},\ and\ \bibinfo {author} {\bibfnamefont {D.~A.}\ \bibnamefont {Brown}},\ }\bibfield  {title} {\bibinfo {title} {{Model Waveform Accuracy Standards for Gravitational Wave Data Analysis}},\ }\href {https://doi.org/10.1103/PhysRevD.78.124020} {\bibfield  {journal} {\bibinfo  {journal} {Phys. Rev. D}\ }\textbf {\bibinfo {volume} {78}},\ \bibinfo {pages} {124020} (\bibinfo {year} {2008})},\ \Eprint {https://arxiv.org/abs/0809.3844} {arXiv:0809.3844 [gr-qc]} \BibitemShut {NoStop}%
\bibitem [{\citenamefont {Piovano}\ \emph {et~al.}(2021)\citenamefont {Piovano}, \citenamefont {Brito}, \citenamefont {Maselli},\ and\ \citenamefont {Pani}}]{Piovano:2021iwv}%
  \BibitemOpen
  \bibfield  {author} {\bibinfo {author} {\bibfnamefont {G.~A.}\ \bibnamefont {Piovano}}, \bibinfo {author} {\bibfnamefont {R.}~\bibnamefont {Brito}}, \bibinfo {author} {\bibfnamefont {A.}~\bibnamefont {Maselli}},\ and\ \bibinfo {author} {\bibfnamefont {P.}~\bibnamefont {Pani}},\ }\bibfield  {title} {\bibinfo {title} {{Assessing the detectability of the secondary spin in extreme mass-ratio inspirals with fully relativistic numerical waveforms}},\ }\href {https://doi.org/10.1103/PhysRevD.104.124019} {\bibfield  {journal} {\bibinfo  {journal} {Phys. Rev. D}\ }\textbf {\bibinfo {volume} {104}},\ \bibinfo {pages} {124019} (\bibinfo {year} {2021})},\ \Eprint {https://arxiv.org/abs/2105.07083} {arXiv:2105.07083 [gr-qc]} \BibitemShut {NoStop}%
\bibitem [{\citenamefont {{Witzany}}\ \emph {et~al.}(2024)\citenamefont {{Witzany}}, \citenamefont {{Skoup{\'y}}}, \citenamefont {{Stein}},\ and\ \citenamefont {{Tanay}}}]{LeoSashwatInprep}%
  \BibitemOpen
  \bibfield  {author} {\bibinfo {author} {\bibfnamefont {V.}~\bibnamefont {{Witzany}}}, \bibinfo {author} {\bibfnamefont {V.}~\bibnamefont {{Skoup{\'y}}}}, \bibinfo {author} {\bibfnamefont {L.~C.}\ \bibnamefont {{Stein}}},\ and\ \bibinfo {author} {\bibfnamefont {S.}~\bibnamefont {{Tanay}}},\ }\bibfield  {title} {\bibinfo {title} {{Actions of spinning compact binaries: Spinning particle in Kerr matched to dynamics at 1.5 post-Newtonian order}},\ }\href {https://doi.org/10.48550/arXiv.2411.09742} {\bibfield  {journal} {\bibinfo  {journal} {arXiv e-prints}\ ,\ \bibinfo {eid} {arXiv:2411.09742}} (\bibinfo {year} {2024})},\ \Eprint {https://arxiv.org/abs/2411.09742} {arXiv:2411.09742 [gr-qc]} \BibitemShut {NoStop}%
\bibitem [{\citenamefont {Gonzo}\ and\ \citenamefont {Shi}(2024)}]{Gonzo:2024zxo}%
  \BibitemOpen
  \bibfield  {author} {\bibinfo {author} {\bibfnamefont {R.}~\bibnamefont {Gonzo}}\ and\ \bibinfo {author} {\bibfnamefont {C.}~\bibnamefont {Shi}},\ }\bibfield  {title} {\bibinfo {title} {{Scattering and bound observables for spinning particles in Kerr spacetime with generic spin orientations}},\ }\href@noop {} {\  (\bibinfo {year} {2024})},\ \Eprint {https://arxiv.org/abs/2405.09687} {arXiv:2405.09687 [hep-th]} \BibitemShut {NoStop}%
\bibitem [{\citenamefont {Skoupý}\ \emph {et~al.}(2024)\citenamefont {Skoupý}, \citenamefont {Piovano},\ and\ \citenamefont {Witzany}}]{SpherPapInPrep}%
  \BibitemOpen
  \bibfield  {author} {\bibinfo {author} {\bibfnamefont {V.}~\bibnamefont {Skoupý}}, \bibinfo {author} {\bibfnamefont {G.}~\bibnamefont {Piovano}},\ and\ \bibinfo {author} {\bibfnamefont {V.}~\bibnamefont {Witzany}},\ }\bibfield  {title} {\bibinfo {title} {{In preparation}},\ }\href@noop {} {\  (\bibinfo {year} {2024})}\BibitemShut {NoStop}%
\end{thebibliography}%

\end{document}